\documentclass[final, 1p, times]{elsarticle}

\usepackage{graphicx}
\usepackage{amssymb}
\usepackage{amsmath}
\usepackage{amsfonts}
\usepackage{bbm}
\usepackage{braket}
\usepackage{hyperref}
\usepackage{siunitx}

\usepackage{lineno}
\usepackage{xcolor}
\journal{Annals of Physics Special Issue: Localisation 2020}

\begin{document}

\begin{frontmatter}


\title{The microscopic picture of the integer quantum Hall regime}



\author[1]{Rudolf A.\ R\"{o}mer}
\ead{R.Roemer@warwick.ac.uk}
\address[1]{Department of Physics, University of Warwick, Coventry, CV4 7AL, UK}

\author[2]{Josef Oswald}
\ead{Josef.Oswald@unileoben.ac.at}
\address[2]{Institut f\"{u}r Physik, Montanuniversit\"{a}t Leoben, Franz-Josef-Strasse 18, 8700 Leoben, Austria}

%%%%%%%%%%%%%%%%%%%%%%%%%%%%%%%%%%%%%%%%%%%%%%%%%%%%%%%%%%%%%%%%%%%%%%%%%
\begin{abstract}
Computer modelling of the integer quantum Hall effect based on self-consistent Hartee-Fock calculations has now reached an astonishing level of maturity. Spatially-resolved studies of the electron density at near macroscopic system sizes of up to $\SI{1}{\mu m^2}$ reveal self-organized clusters of locally fully filled and locally fully depleted Landau levels depending on which spin polarization is favoured. The behaviour results, for strong disorders, in an exchange-interaction induced $g$-factor enhancement and, ultimately, gives rise to narrow transport channels, including the celebrated narrow edge channels. For weak disorder, we find that bubble and stripes phases emerge with characteristics that predict experimental results very well. Hence the HF approach has become a convenient numerical basis to \emph{quantitatively} study the quantum Hall effects, superseding previous more qualitative approaches.
\end{abstract}
%%%%%%%%%%%%%%%%%%%%%%%%%%%%%%%%%%%%%%%%%%%%%%%%%%%%%%%%%%%%%%%%%%%%%%%%%

%%%%%%%%%%%%%%%%%%%%%%%%%%%%%%%%%%%%%%%%%%%%%%%%%%%%%%%%%%%%%%%%%%%%%%%%%
%%%%%%%%%%%%%%%%%%%%%%%%%%%%%%%%%%%%%%%%%%%%%%%%%%%%%%%%%%%%%%%%%%%%%%%%%

\end{frontmatter}

%\linenumbers

\section{Introduction}
%%%%%%%%%%%%%%%%%%%%%%%%%%%%%%%%%%%%%%%%%%%%%%%%%%%%%%%%%%%%%%%%%%%%%%%%%

%the 3 pillars of the QHE

Our current understanding of the (integer) quantum Hall effects rests on three main pillars. The ground work for them was laid in the 1980s and 1990s and the scientific papers connected with each pillar continue to be cited by the quantum Hall community and continue to inspire new experiments within the quantum Hall setting and beyond. 

The first pillar represents the famous \emph{scaling theory} of the integer quantum Hall effect (IQHE) \cite{Huckestein1995ScalingEffect}. It is based on the so-called localization picture of the IQHE and has been underpinned experimentally and theoretically many times, for example by providing a consistent explanation of the existence of plateaus in the transversal transport characteristics as well as leading to the estimation of critical exponents for transitions between plateaus \cite{Chakraborty1995QuantumBasics}. The theoretical basis for this approach are narrow quantum channels of non-interacting electrons that are created at the Fermi level in strongly disordered electron systems. The paradigmatic model for this scaling scenario is provided by the famous Chalker-Coddington model \cite{ChaC88}. Equally well, a tight-binding Anderson model with onsite disorder and Peierls phases modelling the perpendicular magnetic field gives similar answers for universal properties \cite{Puschmann2019IntegerLattice}. 

The existence of narrow transport channels had to be assumed without detailed miscroscopic justifications. %scaling theory.
The second pillar of the IQHE provides this detail by also including electron-electron interactions at the single-particle level. This delivers a much more realistic screening behavior in the smoothly disordered bulk and edge potentials. The main result is that the narrow channel picture has to be modified and, instead, the associated screening behavior creates wide channels at the Fermi level \cite{Chklovskii1992ELECTROSTATICSCHANNELS}. At this point wide, so-called \emph{compressible and incompressible}, stripes have been born. These became a real focus of IQHE research and the existence of compressible and incompressible stripes has been confirmed experimentally by local probe measurements albeit not their spatial widths.

Going beyond screening effects leads to the third pillar of IQHE physics. First theoretical approaches used completely disorder-free systems and successfully delivered, for example,  insights into the surprising spontaneous appearance of ordered charge-density waves observed at lower disorder and weaker magnetic fields. These density modulations appears as stripes or bubbles in high mobility samples at larger filling factors \cite{Fogler1996d}.

More than 20 years after introduction of these seminal discoveries and more then 40 years after the discovery of the IQHE itself, one might expect that by now all those milestones have merged into a well understood unifying picture of the IQHE as a whole. However, a look at the literature delivers an quite unexpected and different situation: each of the "pillar" papers cited above has about 500, and some times even more citations. However, they share in citations much less often. Pillar 1 and 2 papers share only 15 papers citing them both, pillar 2 and 3 only 8, while pillar 1 and 3 have just 6 such joint citations. All three pillar papers together share only two joint citations \cite{Werner2020SizeRegime}, namely, an older paper on using percolation-type physics to explain QH effects in anti-dot lattices \cite{Gusev1998PercolationPotential} and a recent review in Russian \cite{Dolgopolov2014IntegerPhenomena}. Indeed, analysing for shared Phys.\ Rev.\ Lett.\ citations --- as a proxy for generally accepted importance --- gives an even bleaker picture: only pillar 1 and pillar 2 paper share 2 papers in  Phys.\ Rev.\ Lett.\ citing them both. This is clearly a surprising and somewhat disconcerting finding; von Klitzing himself mentions ``a microscopic picture of the quantum Hall effect for real devices with electrical contacts and finite current flow is still missing'' \cite{vonKlitzing2019}.

Surely it should be possible to explain the experimental observations of IQHE physics within a unified understanding. We believe that we have recently made great progress towards such an understanding. In a series of papers, using a self-consistently converged numerical Hartree-Fock approach, together with an effective non-equilibrium model for the finite current flow, we have been able to model (i) the IQHE plateau-plateau transitions, (ii) the appearance of bulk and edge channels as well as (iii) the emergence of the bubble and stripe phases. Within this model, the only parameters changing across all three areas were the strength of the magnetic field $B$ and the characteristic parameters of the smooth disorder potential such a dopant density and potential strength.

In the following, we shall provide some of the details of our approach, focusing on hitherto unpublished results and information which at best appeared in supplemental materials accompanying the original publications. This paper should hence usefully be read as a companion source to those previous publications. 

%%%%%%%%%%%%%%%%%%%%%%%%%%%%%%%%%%%%%%%%%%%%%%%%%%%%%%%%%%%%%%%%%%%%%%%%%
\section{Method}
%%%%%%%%%%%%%%%%%%%%%%%%%%%%%%%%%%%%%%%%%%%%%%%%%%%%%%%%%%%%%%%%%%%%%%%%%

In order to model a high-mobility heterostructure in the QH regime, we
consider a two-dimensional electron system (2DES) in the $(x,y)$-plane subject to a perpendicular magnetic
field $\vec{B} = B\vec{e}_z$. The Hamiltonian 
\begin{equation}
H^\sigma_{\rm 2DES} =
           h^\sigma + V_\text{Coulomb} =
           \frac{(\vec{p}-e\vec{A})^2}{2m^*} +
           \frac{\sigma g^* \mu_{B} B}{2} +
           V_\text{impurities}(\vec{r}) +
           V_\text{Coulomb}(\vec{r},\vec{r}'),
\label{eq-hamiltonian}
\end{equation}
describes spin degree of freedom, $\sigma = \pm 1$, a smooth
random potential $V_\text{impurities}$ modeling the effect of the electron-impurity
interaction, the electron-electron interaction $V_\text{Coulomb}$ and with $m^*$, $g^*$, and $\mu_{B}$ the parameters for effective electron mass,
$g$-factor, and Bohr magneton, respectively.
The electron-impurity
interaction is modeled electrostatically and describes a remote
impurity density separated from the plane of the 2DES by a spacer-layer
of thickness $d$, as found for instance in modulation-doped GaAs-GaAlAs
heterojunctions. This creates a random,
spatially correlated potential with a typical length scale $d$ within the plane of the 2DES. We use $N_{\rm I}$ Gaussian-type "impurities", randomly distributed at $\vec{r}_s$, with random strengths $w_s \in [-W,W]$, and a fixed width $d$. The areal density of impurities is given by $n_{\rm I} = N_{\rm I}/L^2$ \cite{Soh07}.
For the system's many-body state, $|\Phi\rangle$, we use the ansatz  \cite{Aok79,MacA86} of an anti-symmetrized product of single
particle wave-functions, chosen as a linear combination of Landau states \cite{Soh07}.
The number of flux quanta
piercing the 2DES is given by $N_\phi=L^2/2\pi l_{\rm c}^2$, yielding a total number of $M = N_{\rm LL} N_\phi$ states per spin direction. The filling of the system is characterized by the filling factor $\nu = N_{\rm e}/N_\phi$, with $N_{\rm e}$ the number of electrons in the system and areal density $n_{\rm e} = N_{\rm e}/L^2$. In terms of $N_{\mathrm{e},\uparrow}$ spin-up and $N_{\mathrm{e},\uparrow}$ spin-down electrons, we can hence write $\nu= \nu_\uparrow + \nu_\downarrow$ and $N_\mathrm{e}= N_{\mathrm{e},\uparrow} + N_{\mathrm{e},\downarrow}$. The total LL density is given by $n_0 = eB/h$ and $l_{\rm c} = \sqrt{\hbar/eB}$ the magnetic length.
%

%%%%%%%%%%%%%%%%%%%%%%%%%%%%%%%%%%%%%%%%%%%%%%%%%%%%%%%%%%%%%%%%%%%%%%%%%
\section{Results}
%%%%%%%%%%%%%%%%%%%%%%%%%%%%%%%%%%%%%%%%%%%%%%%%%%%%%%%%%%%%%%%%%%%%%%%%%

%%%%%%%%%%%%%%%%%%%%%%%%%%%%%%%%%%%%%%%%%%%%%%%%%%%%%%%%%%%%%%%%%%%%%%%%%
\subsection{Transport results}
%%%%%%%%%%%%%%%%%%%%%%%%%%%%%%%%%%%%%%%%%%%%%%%%%%%%%%%%%%%%%%%%%%%%%%%%%

\begin{figure}[tb]
\begin{center}
    (a)\includegraphics[width=0.95\textwidth]{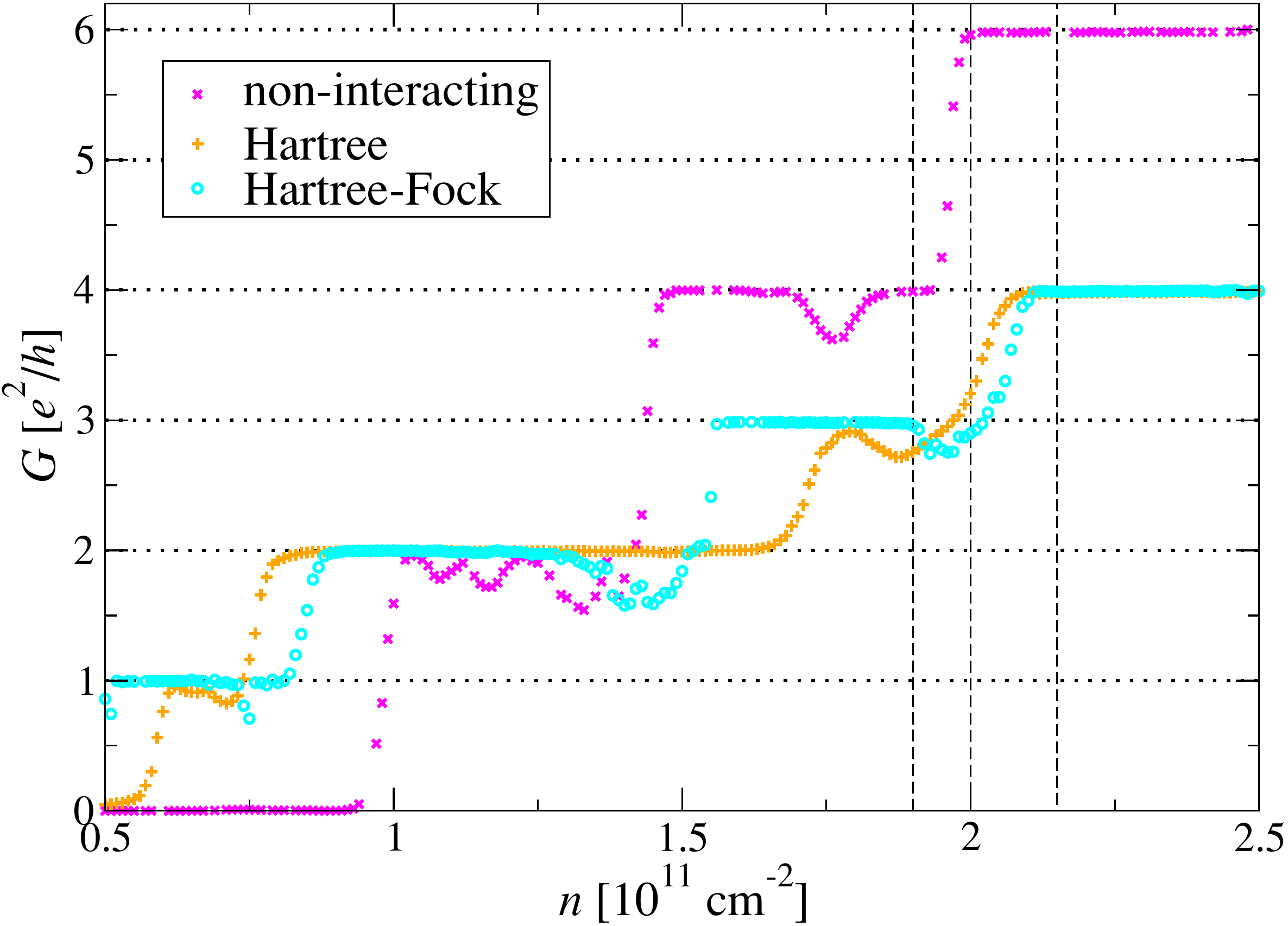}
    \end{center}%\\
    (b)\includegraphics[width=0.5\textwidth,clip]{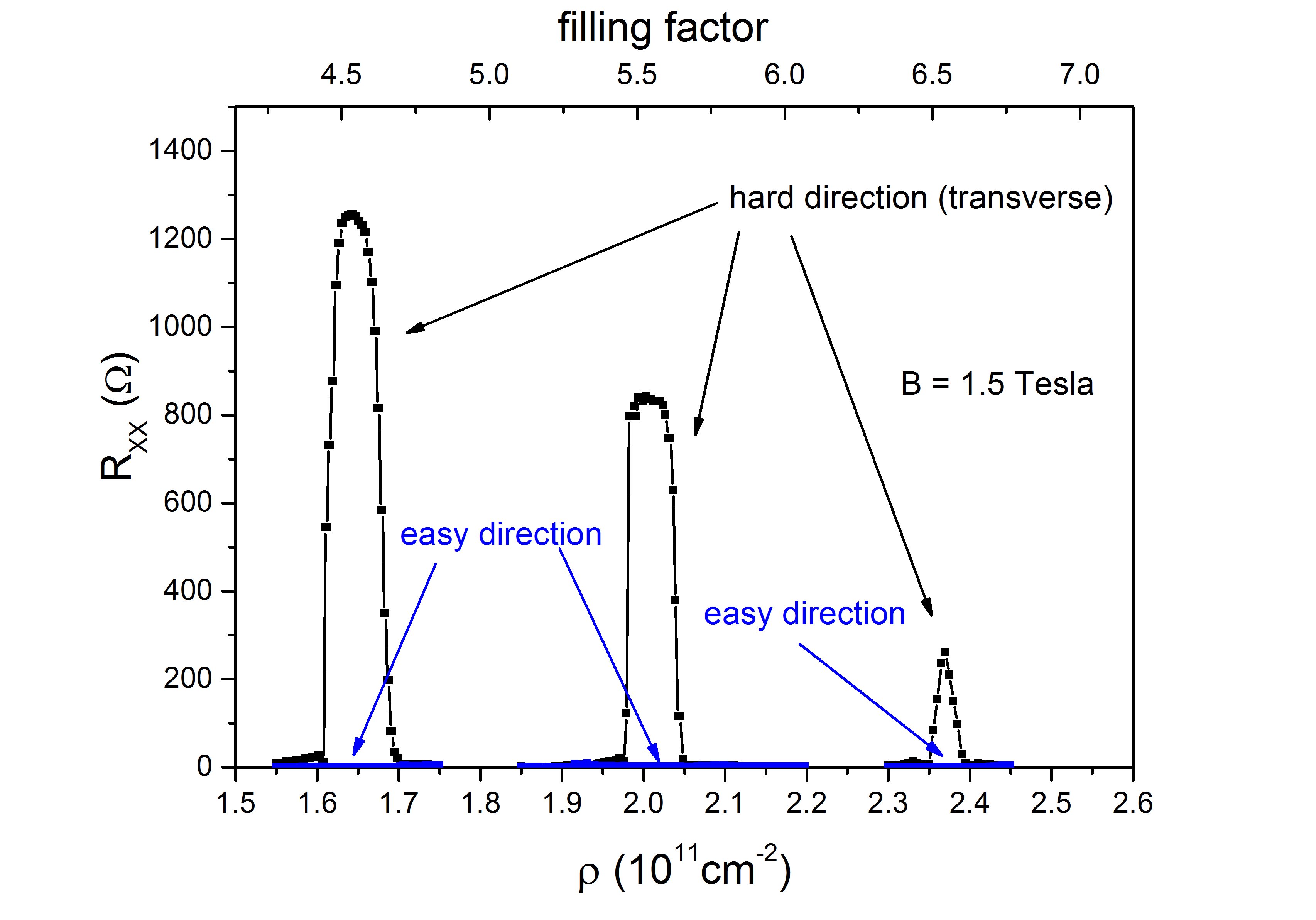} 
    \includegraphics[width=0.5\textwidth,clip]{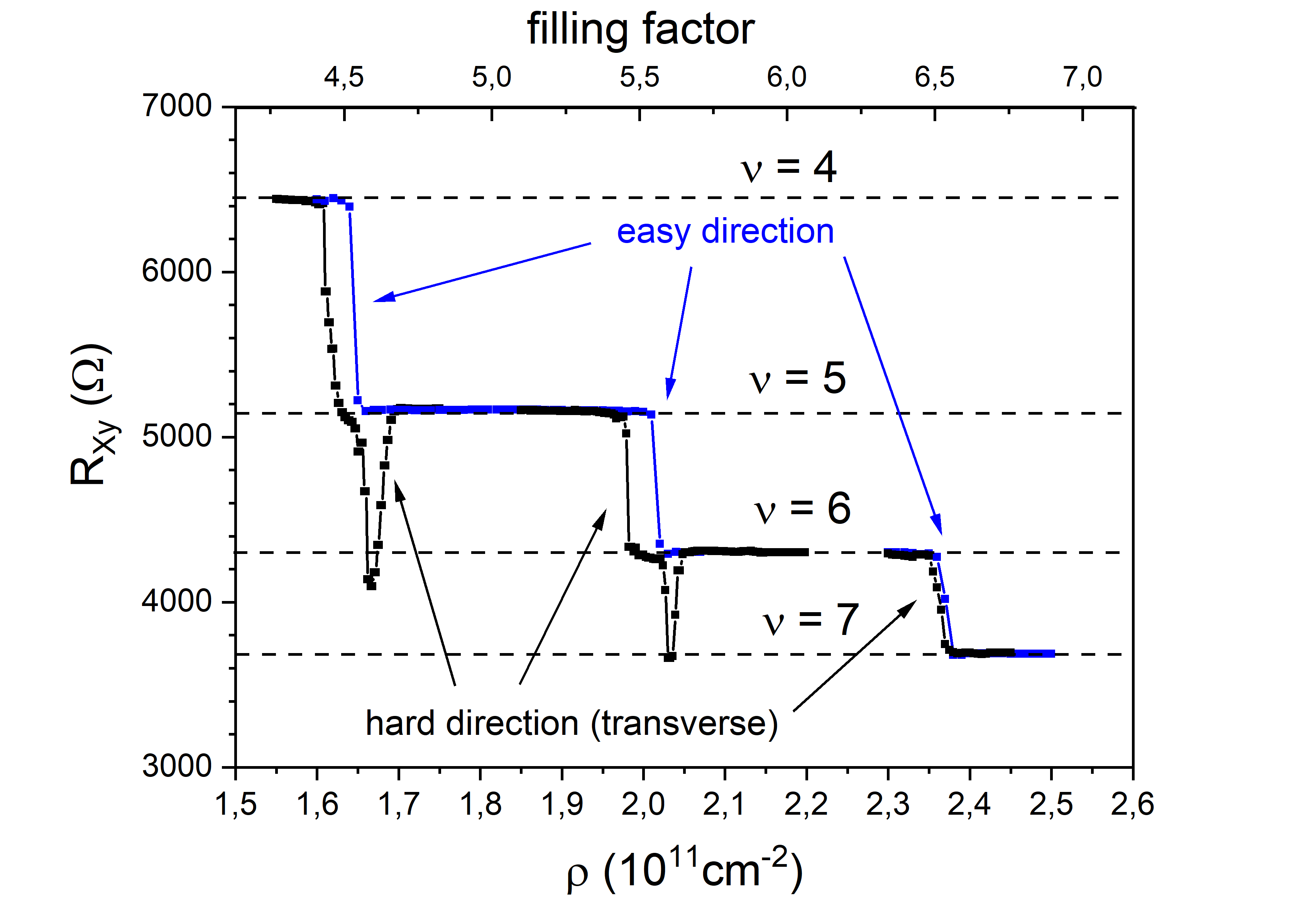}
\caption{
\label{fig-transport}
%\label{fig-Gxx_ns_00-HH-HF}
%\label{fig-conductance-density-00-HH-HF}
(a) Two-point conductance $G$ versus carrier density $n$ at fixed magnetic field of $B=3$ T for the non-interacting ($\times$), the Hartree-interacting ($+$) and the Hartree-Fock-interacting model ($\circ$) at strong disorder. The horizontal dotted lines indicate integer multiples of $e^2/h$ while the vertical dashed lines indicate the three density values of $n= 1.9, 2.0, 2.15$ ($\times 10^{11}$ cm$^{-2}$).
(b) Longitudinal resistence $R_{xx}$ and Hall resistance $R_{xy}$ versus density for a HF interacting system at weak disorder, exhibiting transport anisotropies due to the formation of stripe phases. Easy and hard directions are indicated by different colors.}
\end{figure}

The hallmark of the IQHE is of course the quantization observed for the transveral conductance/resistance of a 2DES at finite perpendicular $B$ field with steps in multiples of $e^2/h$ ($h/e^2$) \cite{KliDP80}. Such behaviour can already very well be modelled using non-interacting approaches \cite{Chakraborty1995QuantumBasics}. In Ref.\ \cite{Sohrmann2008a}, it was shown previously that, with HF interactions, a straightforward application of the Kubo formula  also captures this integer quantization, but has large fluctuations in the vicinity of the plateau-plateau transitions due to finite-size effects. In Fig.\ 
\ref{fig-transport}(a) we show transport results also obtained from our HF calculations, but now constructed after coupling with a non-equilibrium network model (NNM) \cite{Oswald2006CircuitRegime,SohOR09} which allows to reach larger effective sizes. While the results in Fig.\ \ref{fig-transport} exhibit much less fluctuations around the transitions, it must be emphasized that this is due to effectively averaging over many more nodes in the network. Furthermore, for the implementation of the procedure, an ad-hoc assumption has to be made about how to go from the computed charge densities in the HF approach to the transmittivity of a dissipative saddle-point in the NNM \cite{SohOR09,PolS95,Osw16,OswaldPRB2017}. Effectively, this rules out the determination of the critical properties of the QH transition \cite{Huckestein1995ScalingEffect} from the NNM. 
In order to generate transport data, a very large number of step by step calculations are required \cite{SohOR09} in the NNM as well as in the HF procedure. Hence, for keeping within the available computing time, the transport simulations have been performed for sample sizes around $500 \times \SI{500}{nm^2}$ while the underlying HF simulations were performed for a typical IQHE set-up with relatively strong disorder potential fluctuations of $V|\text{impurities}\in \SI{10}{mV}$.
%%%%%%%%%%%%%%%%%%%%%%%%%%%%%%%%%%%%%%%%%%%%%%%%%%%%%%%%%%%%%%%%%%%%
%%%%%%%%%%%%%%%%%%%%%%%%%%%%%%%%%%%%%%%%%%%%%%%%%%%%%%%%%%%%%%%%%%%%
As shown in Fig.\ \ref{fig-transport} (b), the approach can also very well capture the difference in transport between hard and easy directions in the stripe phase of the QH system \cite{OswR17,OswaldPRB2017}. The only change of substance in the simulation parameters is a reduction of the fluctuations of $V|\text{impurities}$ to range from $\SI{0.43}{mV}$ to  $\SI{-0.50}{mV}$. Hence the self-consistent HF approach demonstrates its powers in capturing the large variety of transport properties observed across the realm of IQHE physics. In the following, we shall show that this power also extends to spatially resolved properties, again for larger disorders, cp.\ sections \ref{sec-CD}, and bubble and stripe phases in sections \ref{sec-SB} and \ref{sec-KS}.

%%%%%%%%%%%%%%%%%%%%%%%%%%%%%%%%%%%%%%%%%%%%%%%%%%%%%%%%%%%%%%%%%%%%%%%%%
\subsection{Charge densities}
\label{sec-CD}
%%%%%%%%%%%%%%%%%%%%%%%%%%%%%%%%%%%%%%%%%%%%%%%%%%%%%%%%%%%%%%%%%%%%%%%%%

\begin{figure*}[b]
%
% (b)
% \includegraphics[width=0.45\textwidth]{HF-CD_B291_n200_S1030_sp1.eps}
% \includegraphics[width=0.45\textwidth]{HF-CD_B291_n200_S1030_sp2.eps}
% %
%(a)
$\nu_\uparrow$\includegraphics[width=0.45\textwidth]{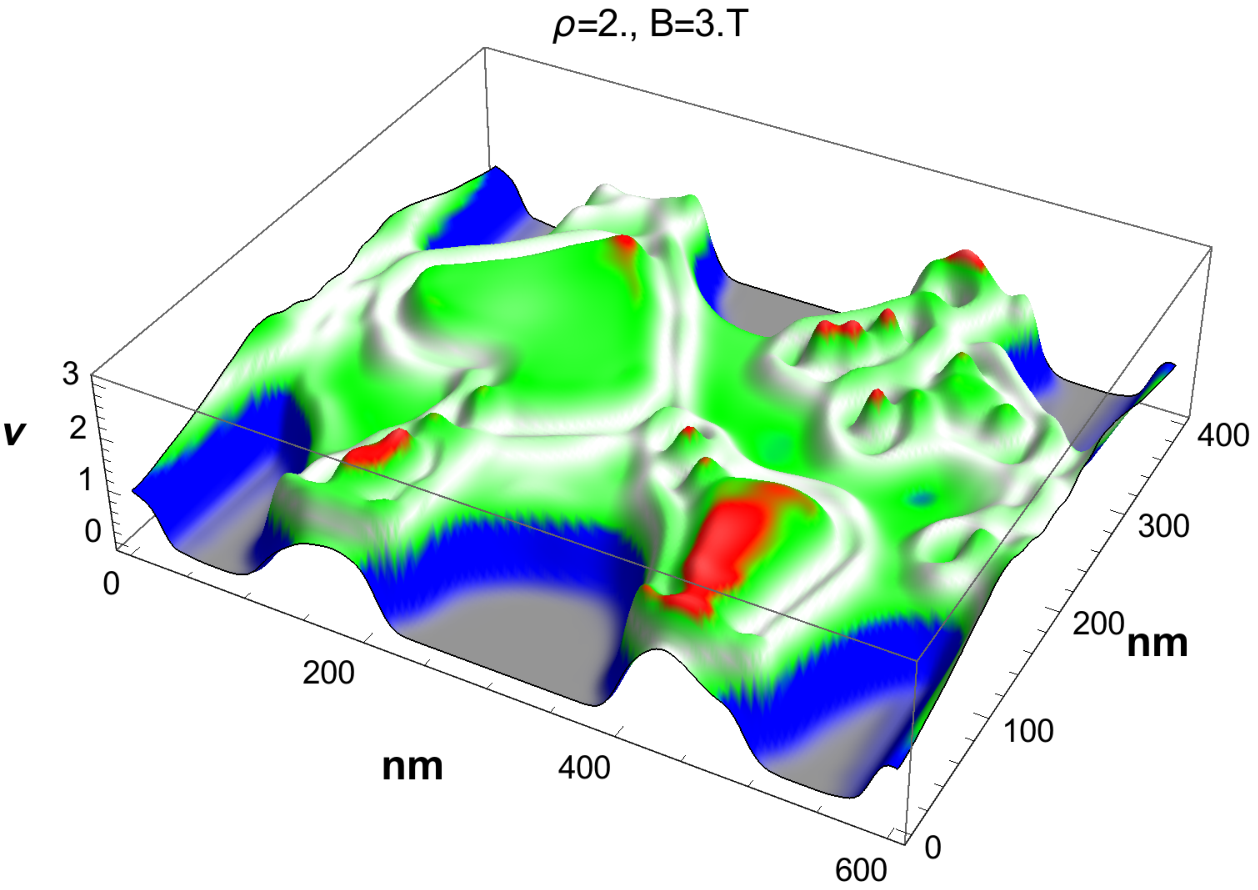} $\nu_\downarrow$\includegraphics[width=0.45\textwidth]{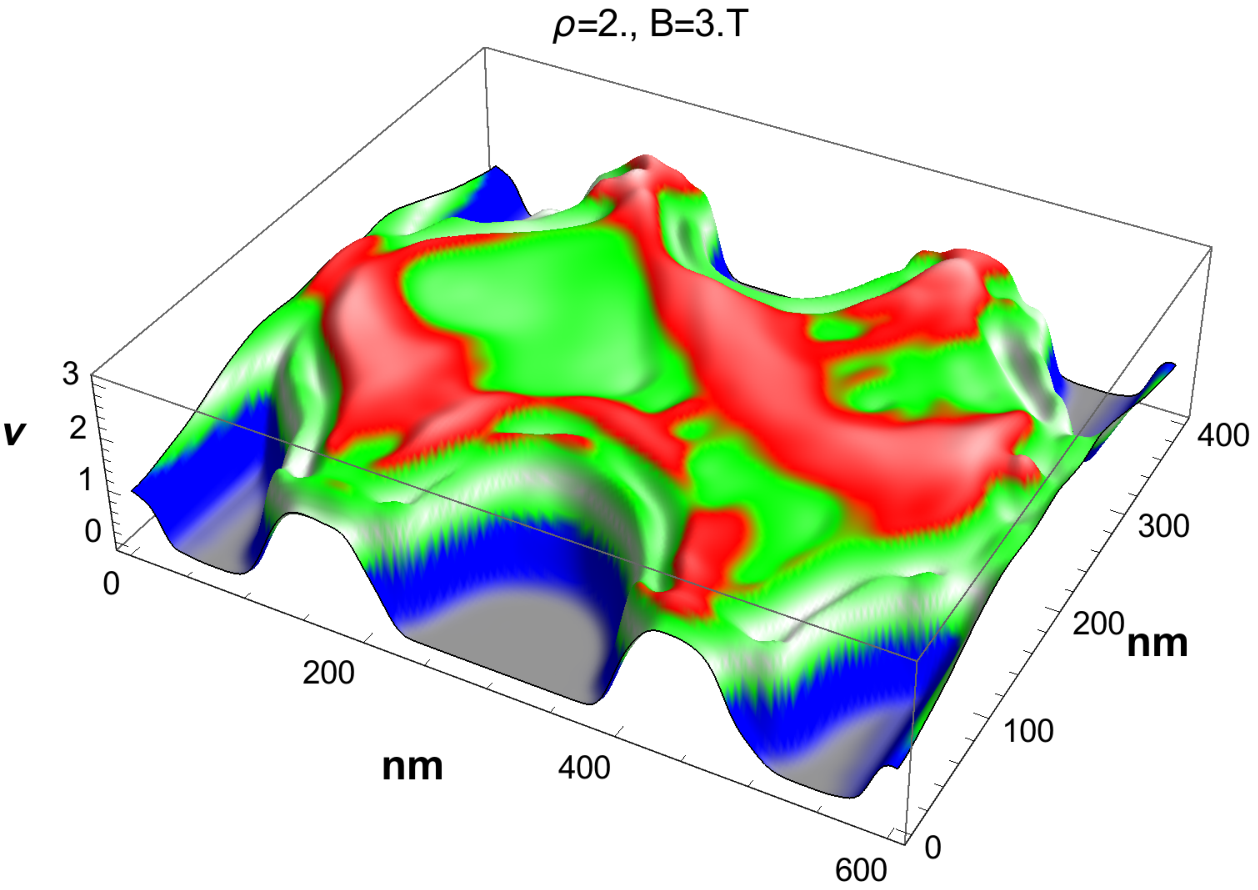}\\
%
%(b)
$\nu_\uparrow$\includegraphics[width=0.45\textwidth]{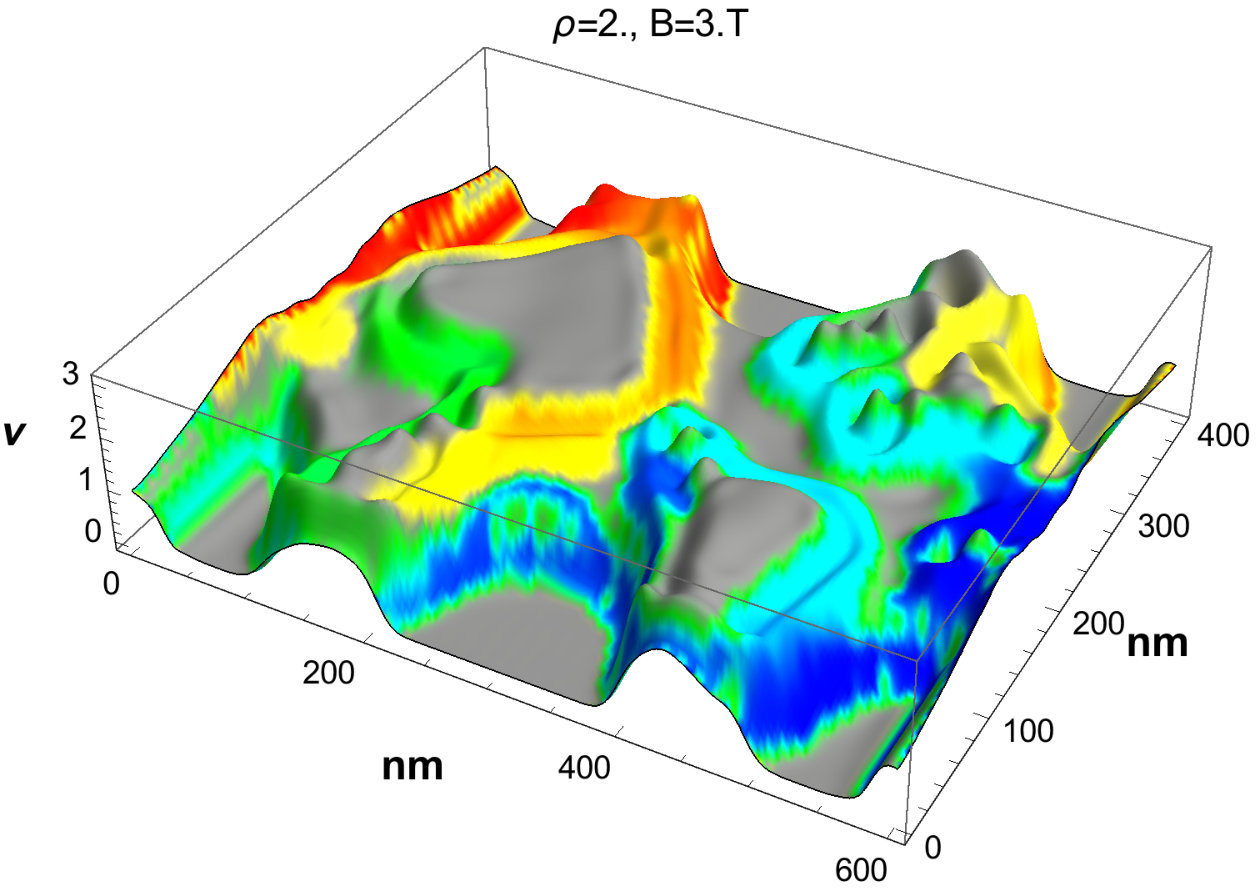}
$\nu_\downarrow$\includegraphics[width=0.45\textwidth]{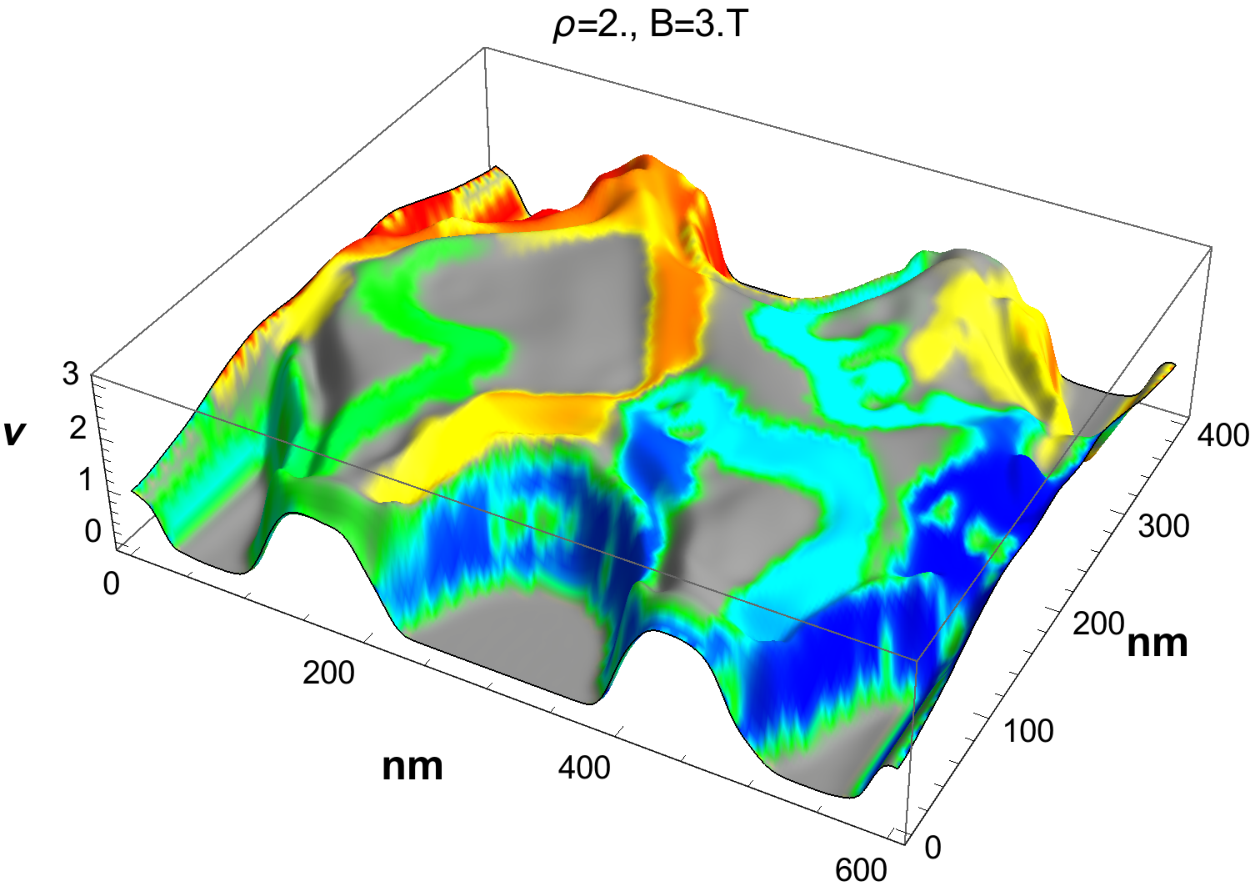}
% (c)
% \includegraphics[width=0.45\textwidth]{HF-CD_B311_n200_S1030_sp1.eps}
% \includegraphics[width=0.45\textwidth]{HF-CD_B311_n200_S1030_sp2.eps}
%
\caption{
\label{fig-CD-NNM}
%\label{CDS1x3_sp2}
%\label{fig-fillingfactor-down} 
Top row: Spatial distribution of filling factors $\nu_\uparrow$ (left) and $\nu_\downarrow$ (right). The colors represent the filling factor, where blue means the first LL for $\nu_{\uparrow,\downarrow} = 0 \rightarrow 1$, green the second LL for $\nu_{\uparrow,\downarrow} = 1 \rightarrow 2$ and red the third. The filling factor range close to $\nu_{\uparrow,\downarrow} = 1.5$  is highlighted in light gray in order to identify the stripes appearing close to the half filled top LL. They appear only for $\nu_{\uparrow}$.
Bottom row: Spatial distribution of non-equilibrium chemical potential $\mu$ (colors) shown on top of the corresponding $\nu_{\uparrow,\downarrow}$ distributions (grey heights) as in the top row. The colors represent $\mu$ in arbitrary units with overall clock-wise propagating potential reducing from the high potential supplied to the current contact on the left from red to orange and yellow while the low potential is supplied to the current contact on the right and is indicated as increasing from blue to cyan to green. 
The parameters in both rows are identical, i.e.\ constant carrier density $\rho= 2 \times \SI{e11}{cm^{-2}}$ and $B=\SI{3}{Tesla}$. The situation corresponds to the $\nu = 3 \rightarrow 4$ plateau transition of Fig.~\ref{fig-transport}. 
}
\end{figure*}

In Fig.\ \ref{fig-CD-NNM} we show the spatial distribution of the filling factors $\nu_\uparrow$ and $\nu_\downarrow$, with $\nu= \nu_\uparrow + \nu_\downarrow$, obtained from the HF calculation as well as the associated non-equilibrium chemical potentials $\mu$ resulting from the NNM. The data were computed for the disorder potential strength $V_\text{impurities}$ of maximally $\pm \SI{10}{mV}$ used to generate Fig.\ \ref{fig-transport}.
Our results for $\nu_\uparrow$ have been shown before \cite{OswaldPRB2017}, while the $\nu_\downarrow$ only appeared in the supplement. The direct comparison between $\nu_\uparrow$ and $\nu_\downarrow$ makes it clear that regions which are populated for $\nu_\downarrow$ are still largely unpopulated for $\nu_\uparrow$. 
Hence, instead of observing an overall, spatially homogeneous, increase or decrease of the carrier density, we find shrinking or growing clusters of fully filled spin-up LL at $ \nu_{\uparrow} =2$ and growing or shrinking areas of depleted spin-up LL at $ \nu_{\uparrow} =1$. On average this results in a continuous change of the spatially averaged $\nu_\uparrow$ and $\nu_\downarrow$. Therefore a combined $\nu_{\uparrow} =1.5$ is made up by half of the area taken up by clusters of $\nu_{\uparrow}=1$ and the other half taken up by $\nu_{\uparrow} =2$. This is of course similar for e.g. $\nu_{\downarrow}=1.5$ in the case of an average total filling factor $\nu=2.5$.
%%%%%%%%%%%%%%%%%%%%%%%%%%%%%%%%%%%%%%%%%%%%%%%%%%%%%%%%%%%%%%%%%%%%
%%%%%%%%%%%%%%%%%%%%%%%%%%%%%%%%%%%%%%%%%%%%%%%%%%%%%%%%%%%%%%%%%%%%
%%%%%%%%%%%%%%%%%%%%%%%%%%%%%%%%%%%%%%%%%%%%%%%%%%%%%%%%%%%%%%%%%%%%
\begin{figure*}[b]
%
%(b)
$\nu_\uparrow$\includegraphics[width=0.45\textwidth]{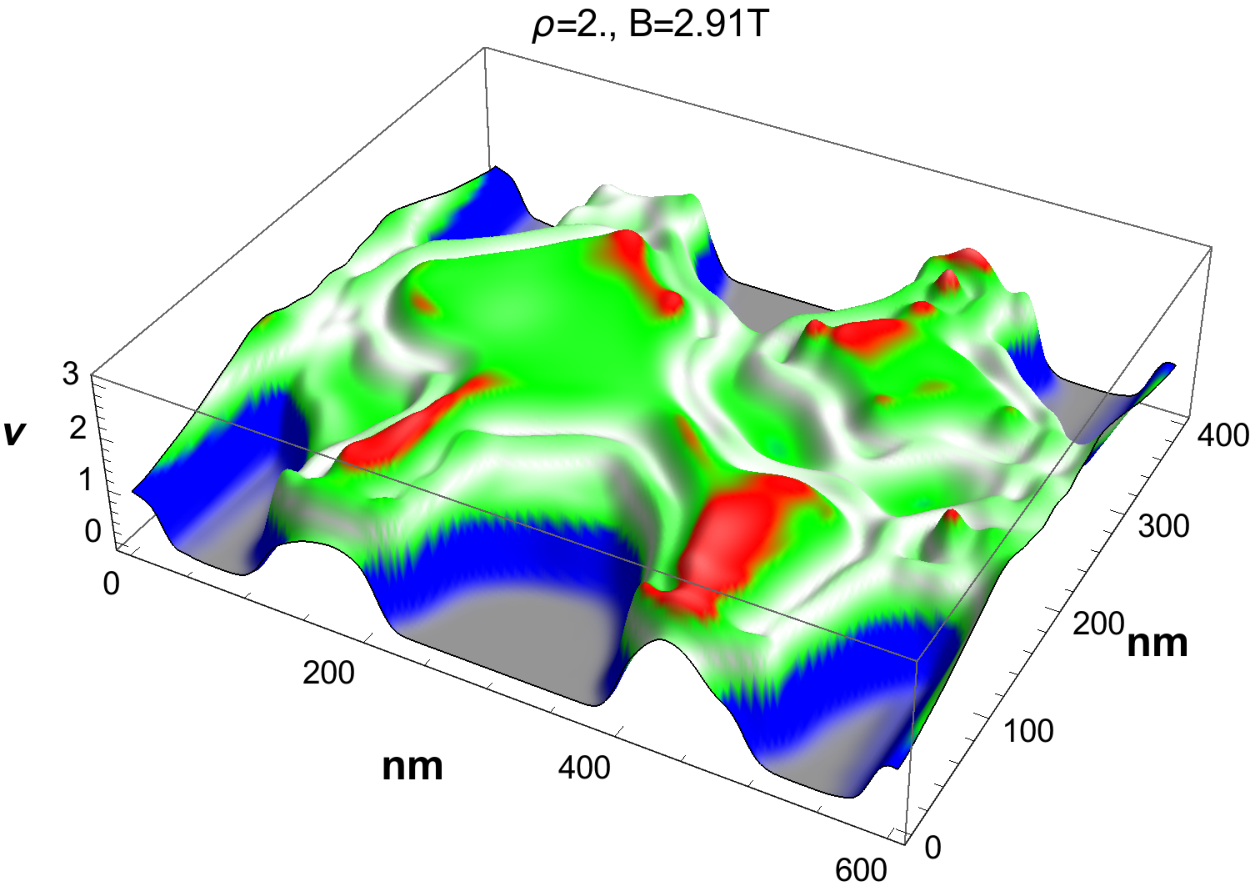}
$\nu_\downarrow$\includegraphics[width=0.45\textwidth]{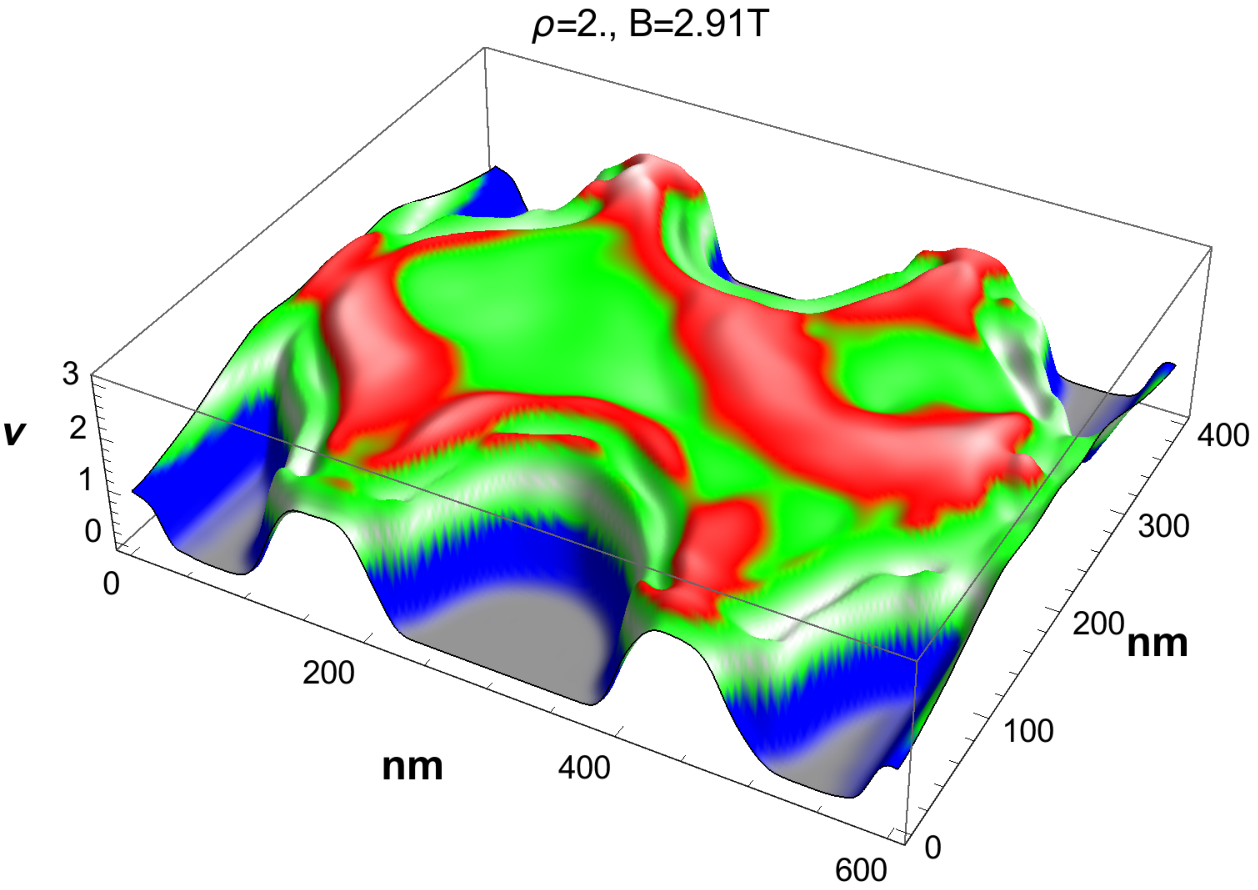}\\
%
% (a)
% \includegraphics[width=0.45\textwidth]{HF-CD_B300_n200_S1030_sp1.eps} \includegraphics[width=0.45\textwidth]{HF-CD_B300_n200_S1030_sp2.eps}\\
%
%(c)
$\nu_\uparrow$\includegraphics[width=0.45\textwidth]{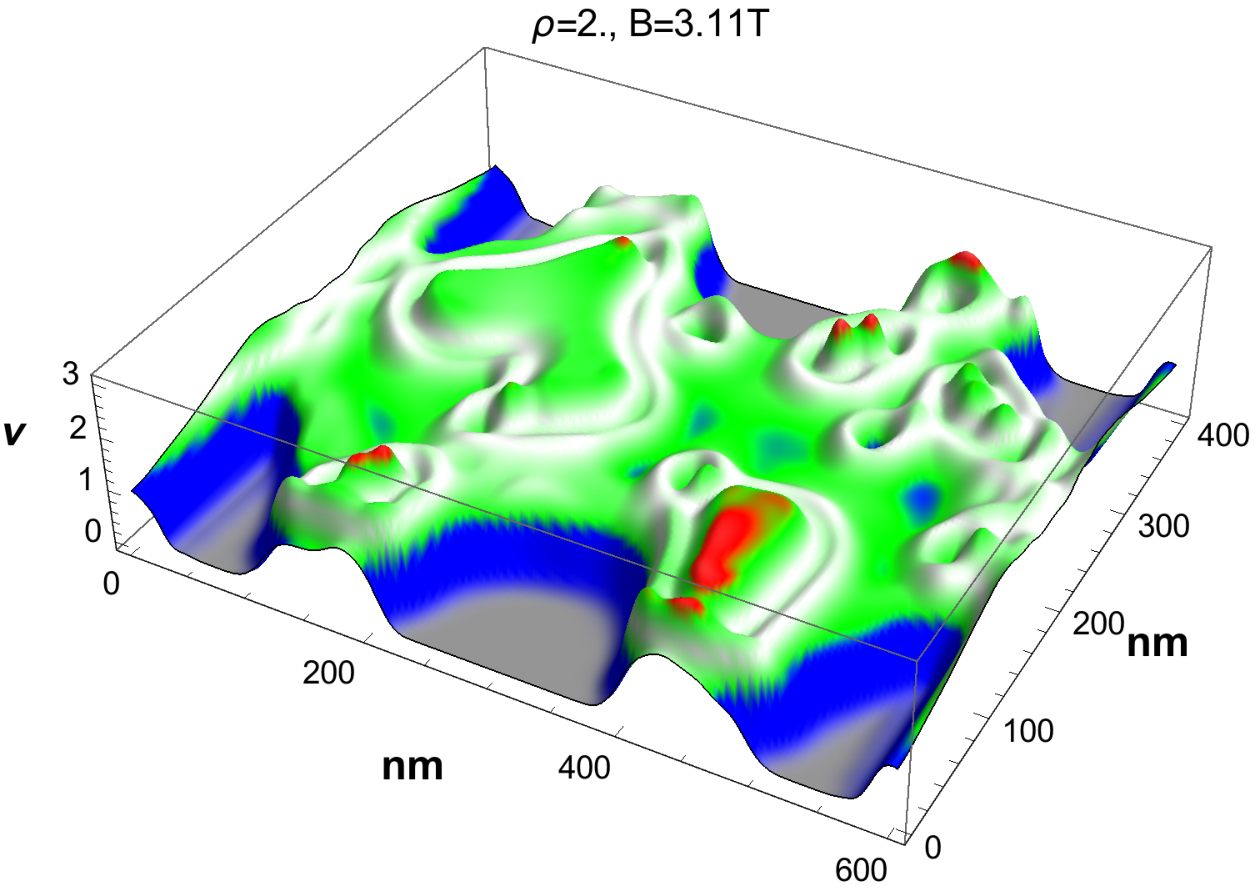}
$\nu_\downarrow$\includegraphics[width=0.45\textwidth]{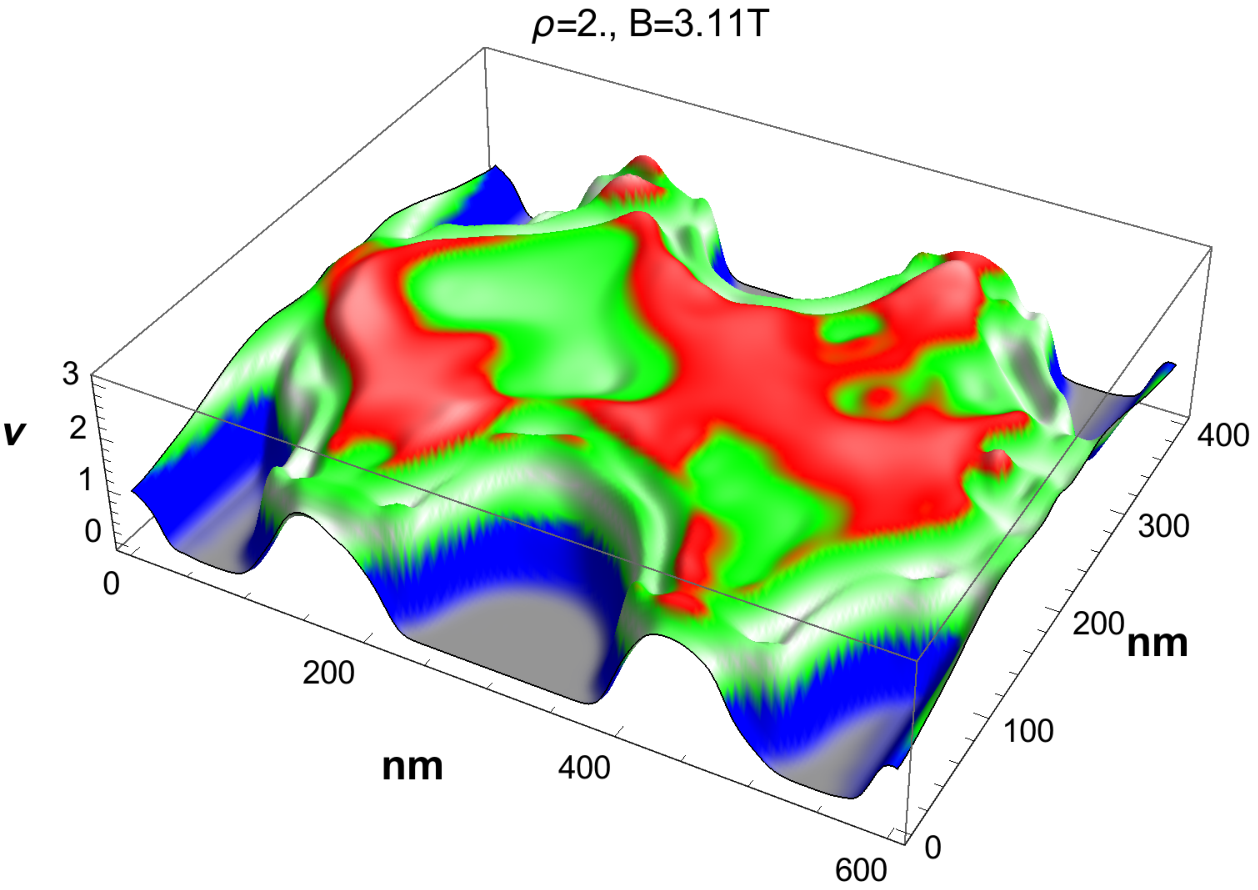}
\caption{
\label{fig-CD-mag}
%\label{CDS1x3_sp2}
%\label{fig-fillingfactor-down} 
Lateral distribution of filling factors $\nu_\uparrow$ (left) and $\nu_\downarrow$ (right) when (top row) decreasing or (bottom row) increasing the magnetic field $B$ at fixed $\rho= 2 \times \SI{e11}{cm^{-2}}$.
Colors are as in Fig.\ \ref{fig-CD-NNM}.
}
\end{figure*}
%%%%%%%%%%%%%%%%%%%%%%%%%%%%%%%%%%%%%%%%%%%%%%%%%%%%%%%%%%%%%%%%%%%%

\begin{figure*}[b]
%
%(d)
$\nu_\uparrow$\includegraphics[width=0.45\textwidth]{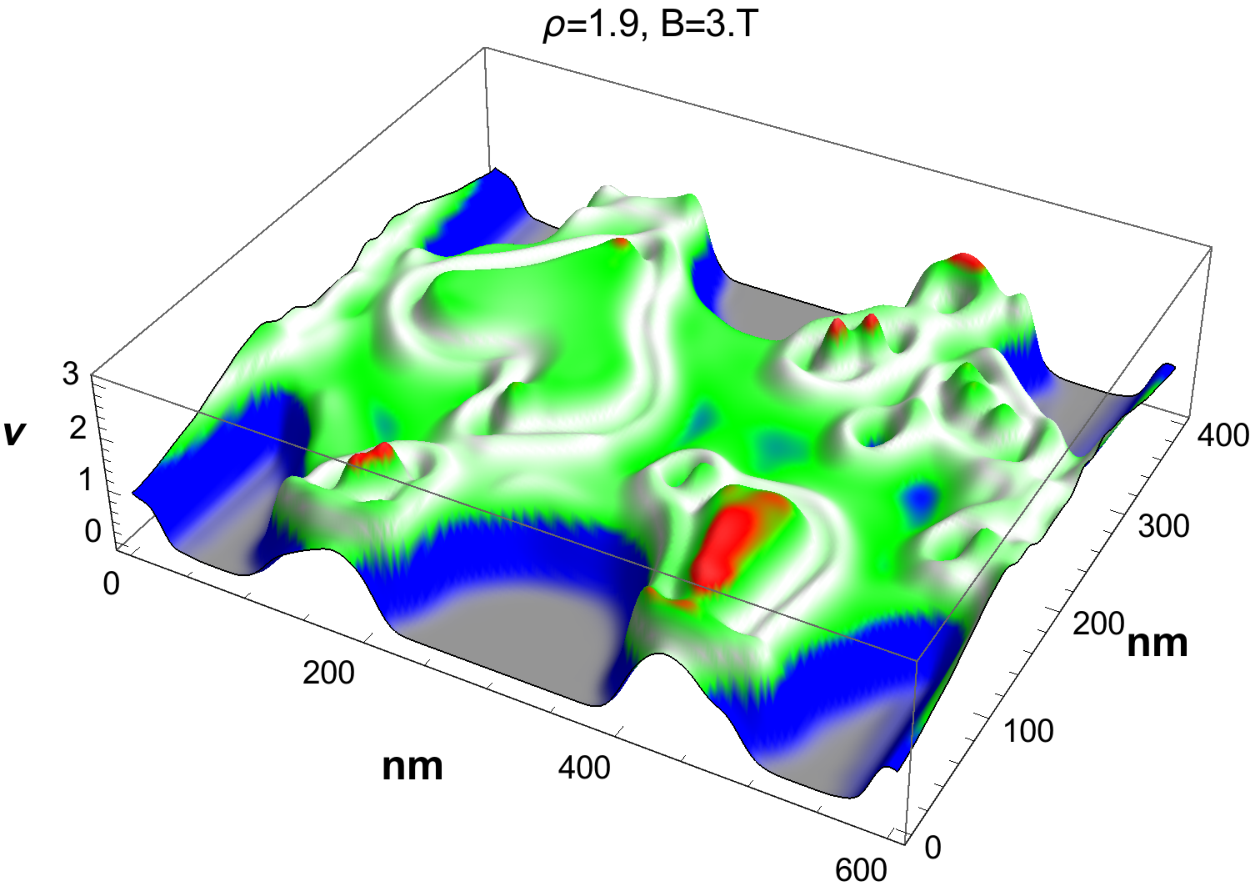}
$\nu_\downarrow$\includegraphics[width=0.45\textwidth]{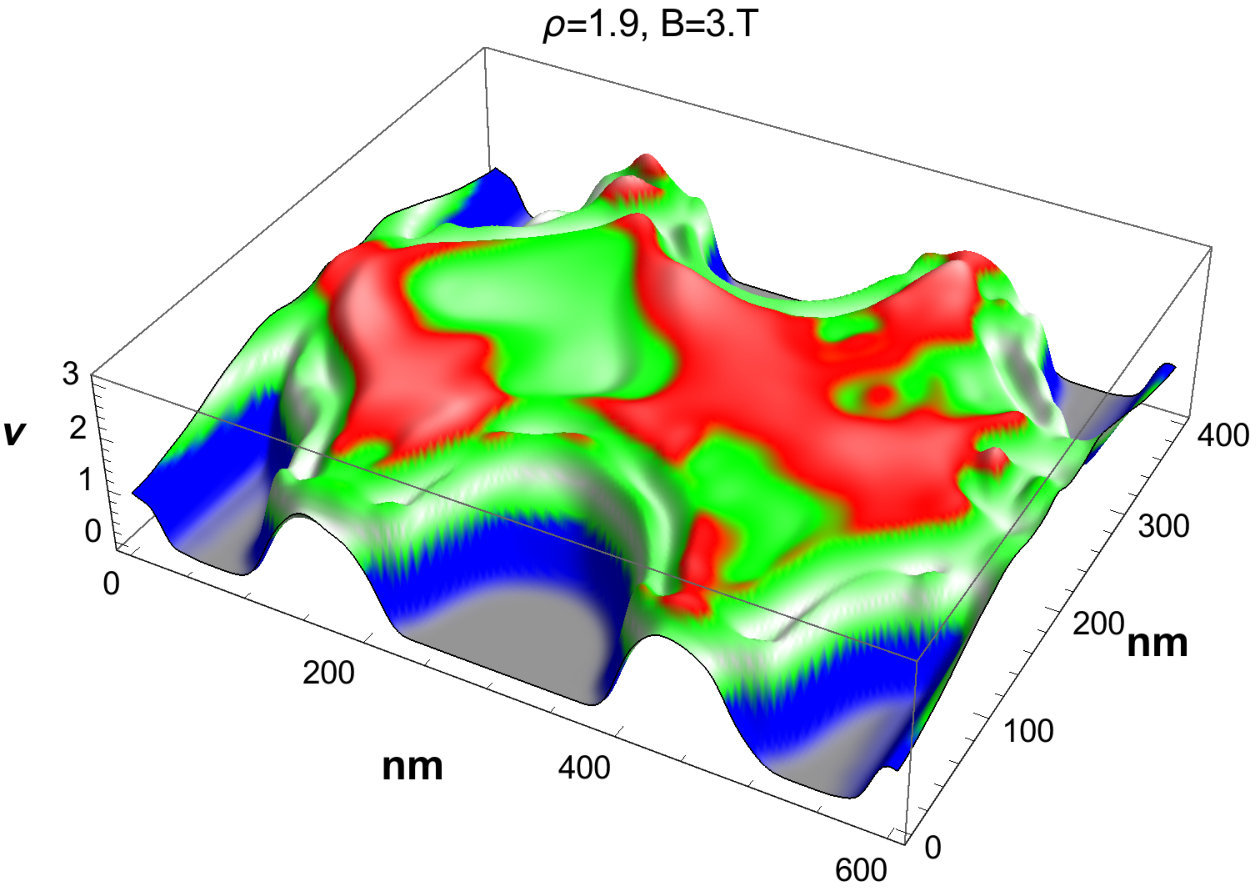}\\
% %
% (a)
% \includegraphics[width=0.45\textwidth]{HF-CD_B300_n200_S1030_sp1.eps} \includegraphics[width=0.45\textwidth]{HF-CD_B300_n200_S1030_sp2.eps}\\
%
%(e)
$\nu_\uparrow$\includegraphics[width=0.45\textwidth]{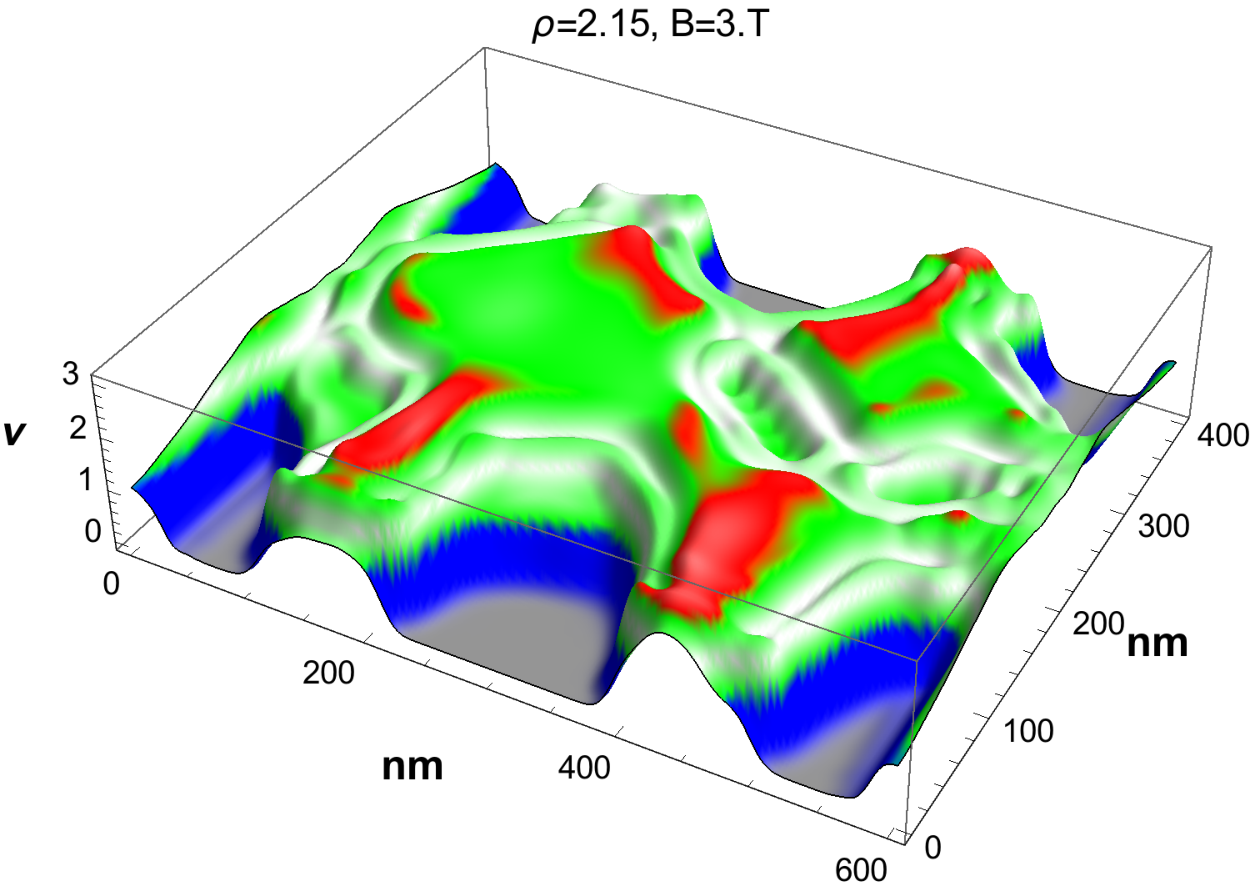}
$\nu_\downarrow$\includegraphics[width=0.45\textwidth]{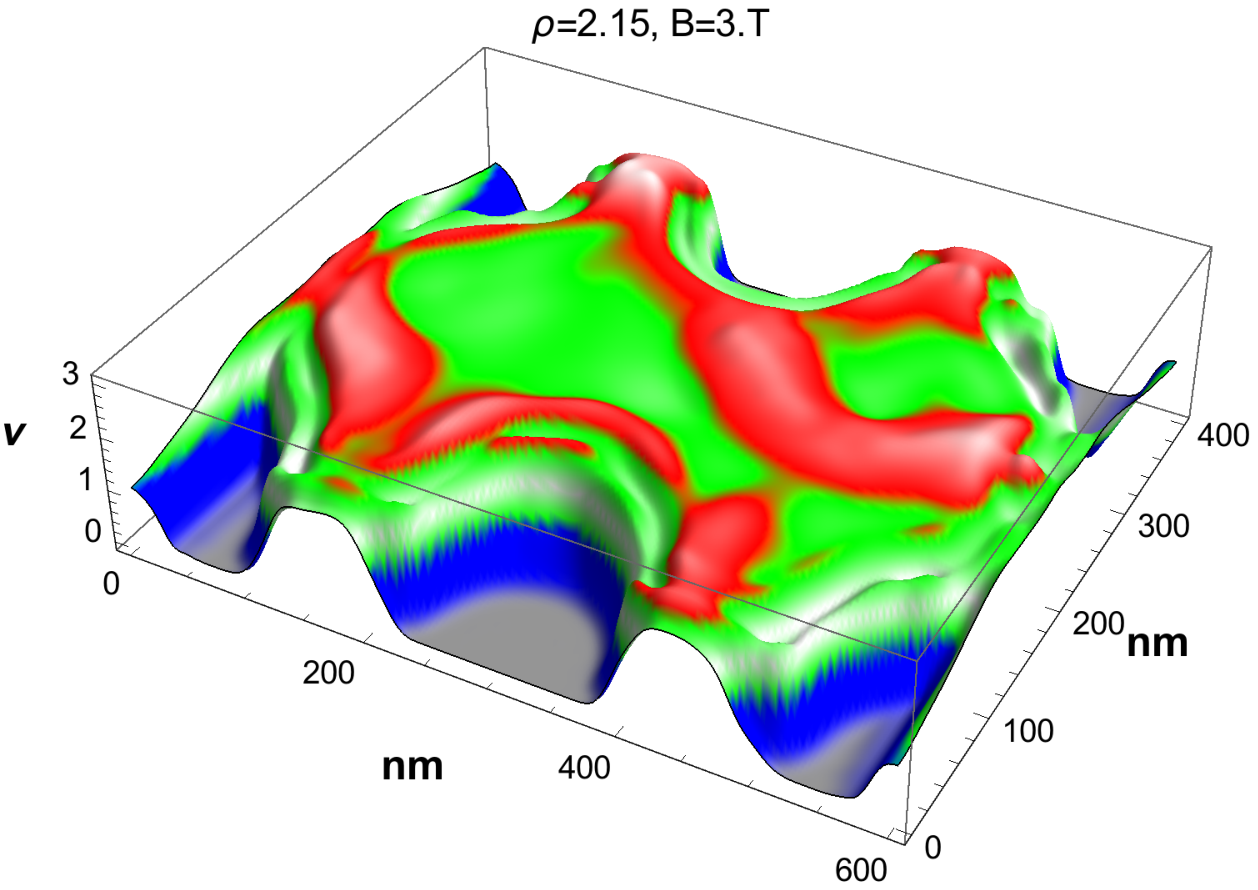}
\caption{
\label{fig-CD-density}
%\label{CDS1x3_sp2}
%\label{fig-fillingfactor-down} 
Lateral distribution of filling factors $\nu_\uparrow$ (left) and $\nu_\downarrow$ (right) when (top row) decreasing or (bottom row) increasing the density  $\rho$ at fixed $B=\SI{3}{Tesla}$.
Colors are as in Fig.\ \ref{fig-CD-NNM}.}
\end{figure*}

\begin{figure*}[b]
%
% (a)
% \includegraphics[width=0.45\textwidth]{HF-CD_B300_n200_S1030_sp1.eps} \includegraphics[width=0.45\textwidth]{HF-CD_B300_n200_S1030_sp2.eps}
%
%(f)
$\nu_\uparrow$\includegraphics[width=0.45\textwidth]{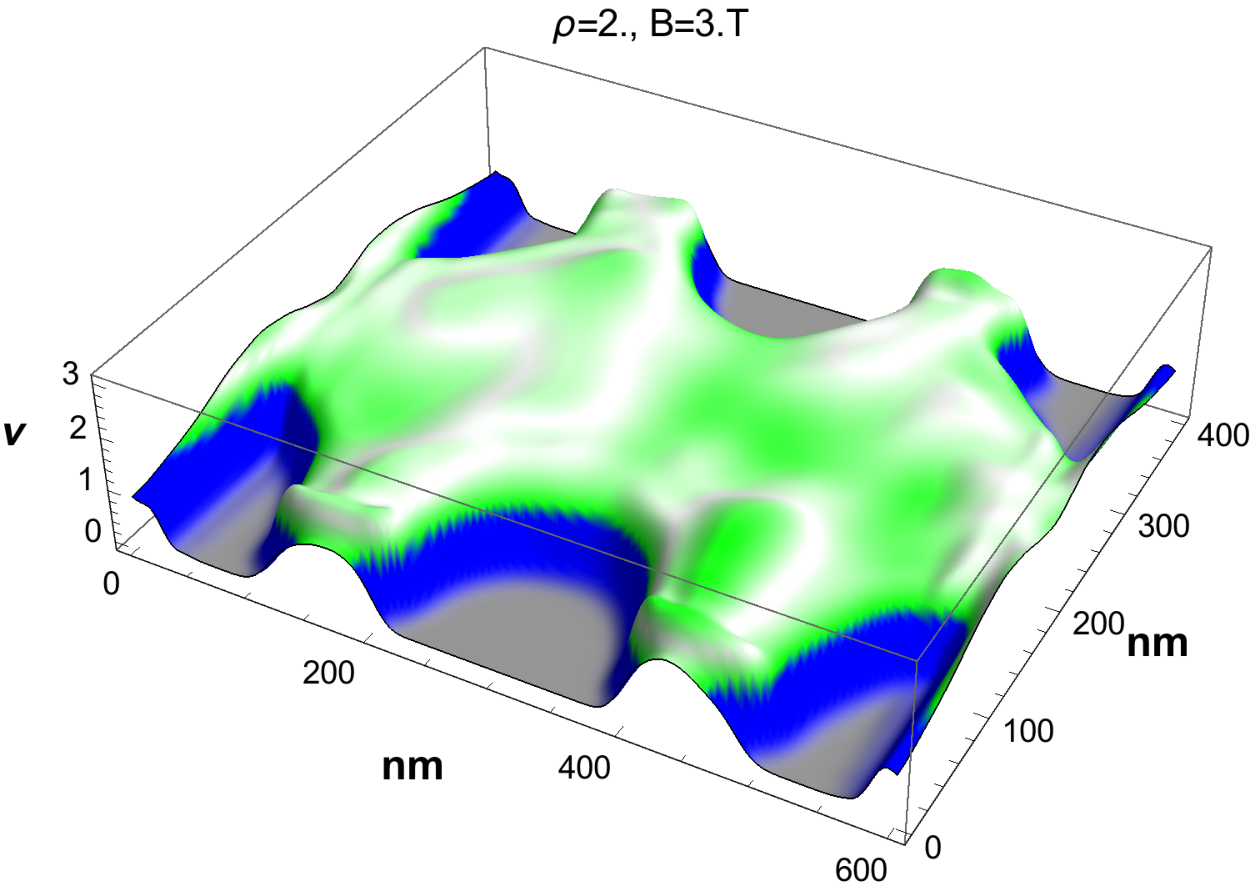}
$\nu_\downarrow$\includegraphics[width=0.45\textwidth]{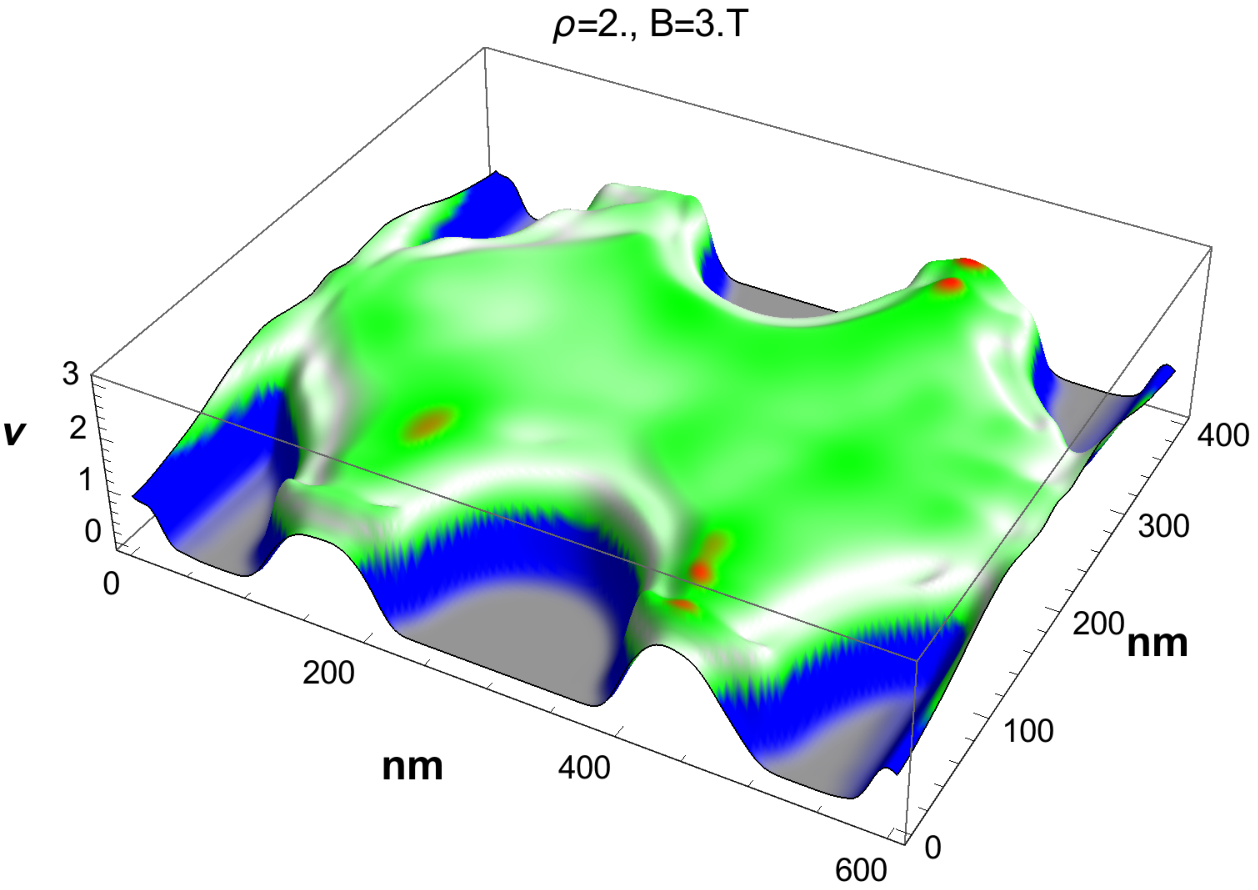}
\caption{
\label{fig-CD-temperature}
%\label{CDS1x3_sp2}
%\label{fig-fillingfactor-down} 
Distribution of filling factors $\nu_\uparrow$ (left) and $\nu_\downarrow$ (right) when increasing the temperature to $T=\SI{20}{Kelvin}$ ($\SI{2}{meV}$). 
Parameters are again $\rho= 2 \times \SI{e11}{cm^{-2}}$ and $B=\SI{3}{Tesla}$. We use the same colors scheme as in Fig.\ \ref{fig-CD-density}.
}
\end{figure*}

In Fig.\ \ref{fig-CD-mag} we show how $\nu_\uparrow$ (left) and $\nu_\downarrow$ (right) change when $B$ is varied while Fig.\ \ref{fig-CD-density} shows a situation analogous to Fig.\ \ref{fig-CD-mag}, but now instead of $B$, we vary the density $\rho$. The behaviour in these figures has already be discussed in detail in Refs.\ \cite{OswaldPRB2017,OswR17}. Here, we simply note that the aforementioned shrinking or growing clusters of fully filled and the implied growing or shrinking areas of depleted areas for the two spin directions in both figures follows the same exchange-enhanced $g$-factor "rules".
%as discussed with Fig.\ \ref{fig-CD-density}. ????????????????????????
%%%%%%%%%%%%%%%%%%%%%%%%%%%%%%%%%%%%%%%%%%%%%%%%%%%%%%%%%%%%%%%%%%%%
%%%%%%%%%%%%%%%%%%%%%%%%%%%%%%%%%%%%%%%%%%%%%%%%%%%%%%%%%%%%%%%%%%%%
Furthermore, we can study the temperature dependence, following from the broadening of the Fermi distribution, of $\nu_\uparrow$ and $\nu_\downarrow$ in Fig.\ \ref{fig-CD-temperature}. As should be expected,  results for $T=\SI{20}{Kelvin}$ show a much less structured distribution for $\nu_\uparrow$ and $\nu_\downarrow$ even for the HF results.
%%%%%%%%%%%%%%%%%%%%%%%%%%%%%%%%%%%%%%%%%%%%%%%%%%%%%%%%%%%%%%%%%%%%
%%%%%%%%%%%%%%%%%%%%%%%%%%%%%%%%%%%%%%%%%%%%%%%%%%%%%%%%%%%%%%%%%%%%

\begin{figure*}[tb]
%(a)
$\nu_\uparrow$\includegraphics[width=0.45\textwidth]{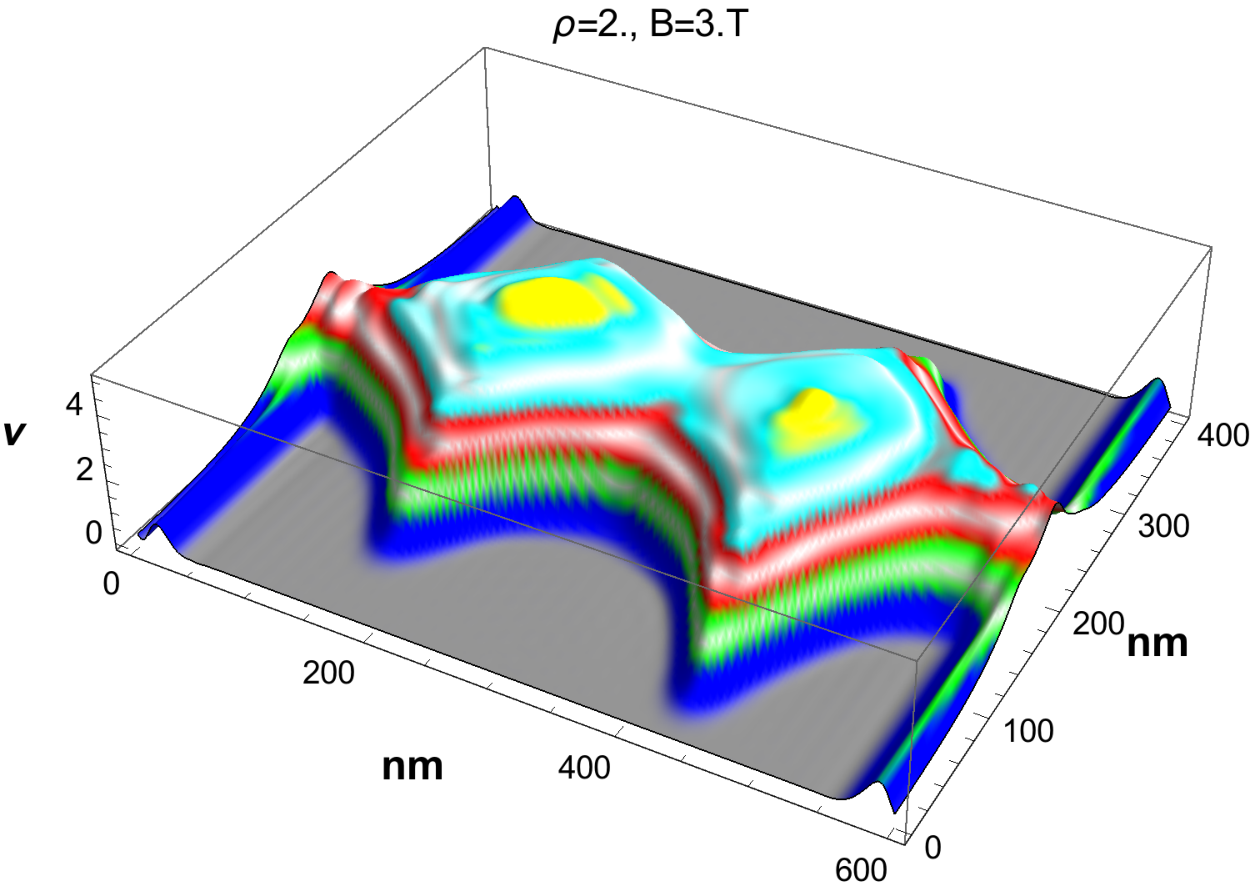} 
$\nu_\downarrow$\includegraphics[width=0.45\textwidth]{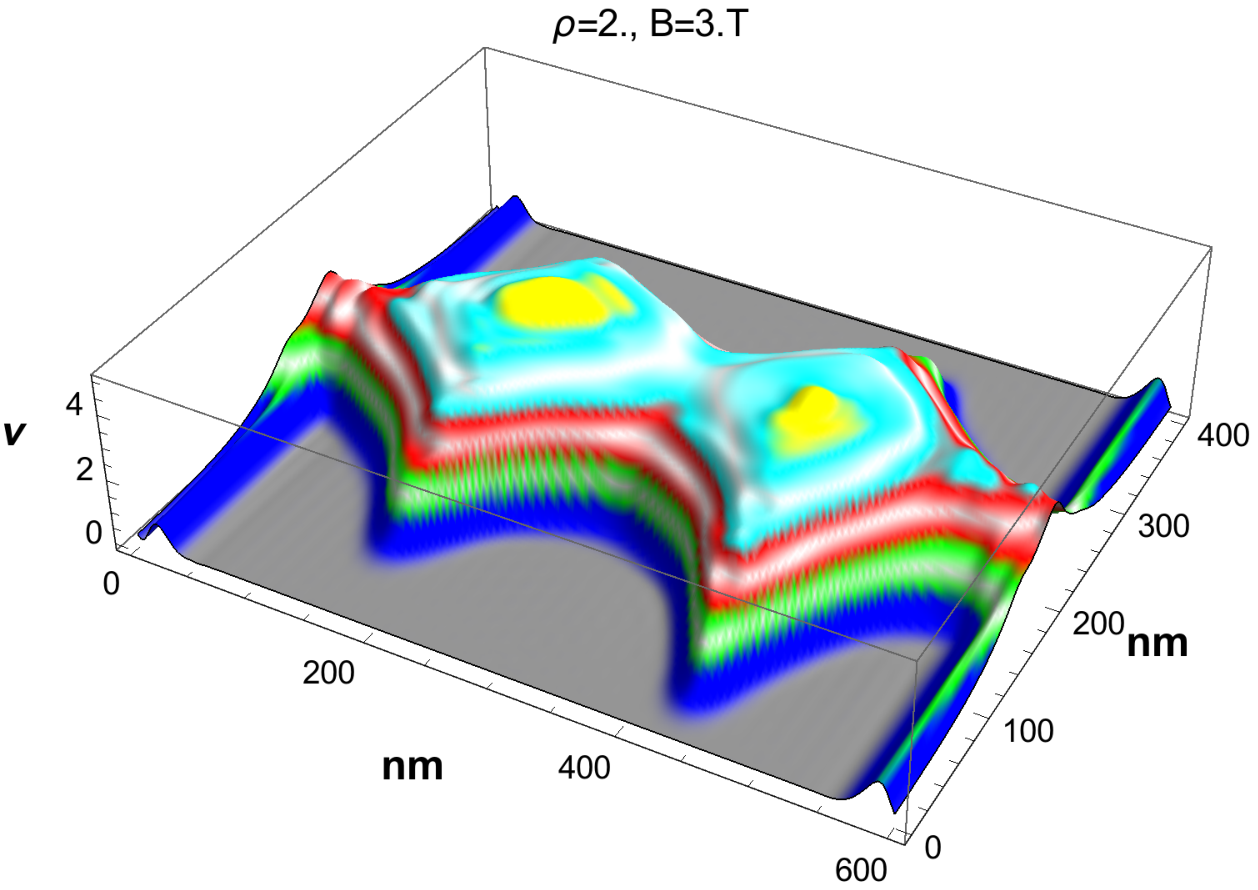}\\
%
%(b)
$\nu_\uparrow$\includegraphics[width=0.45\textwidth]{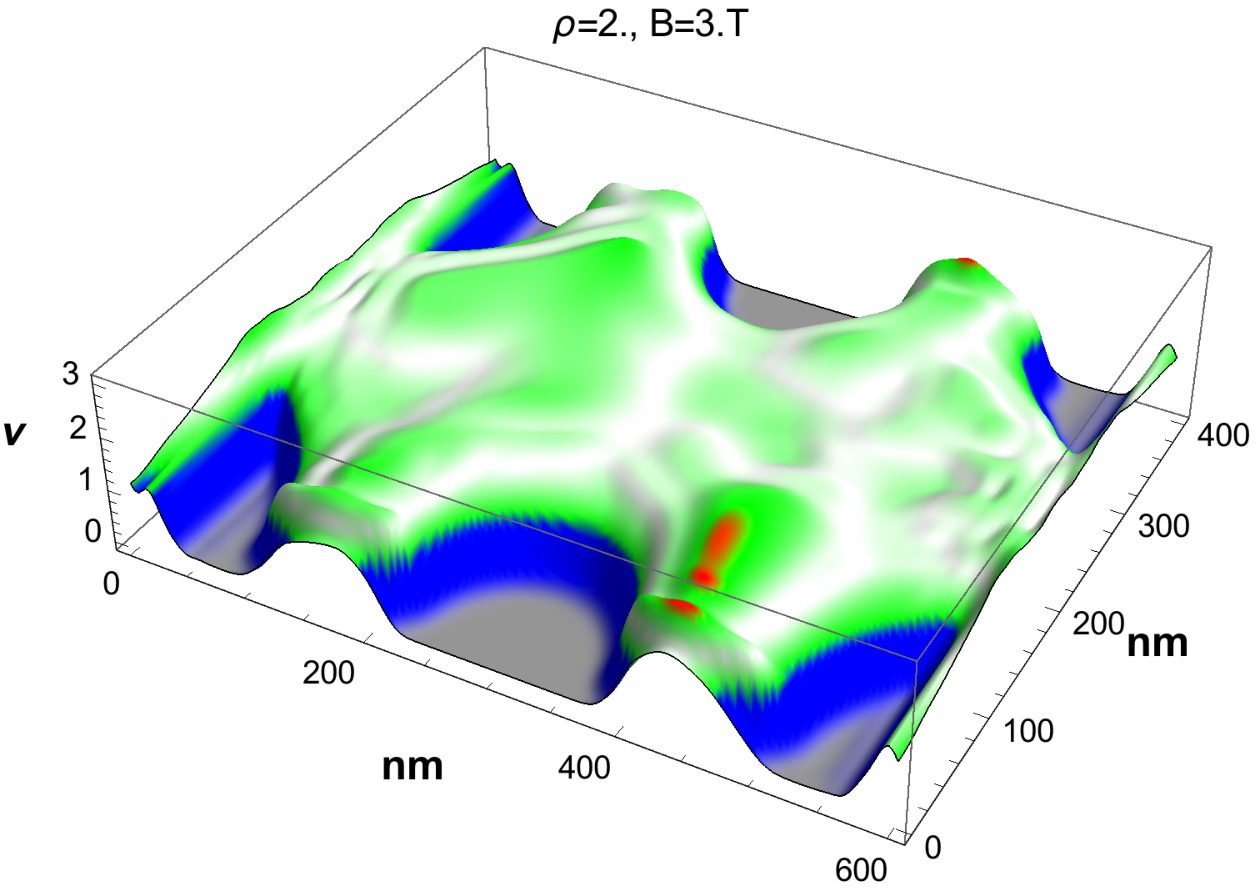} 
$\nu_\downarrow$\includegraphics[width=0.45\textwidth]{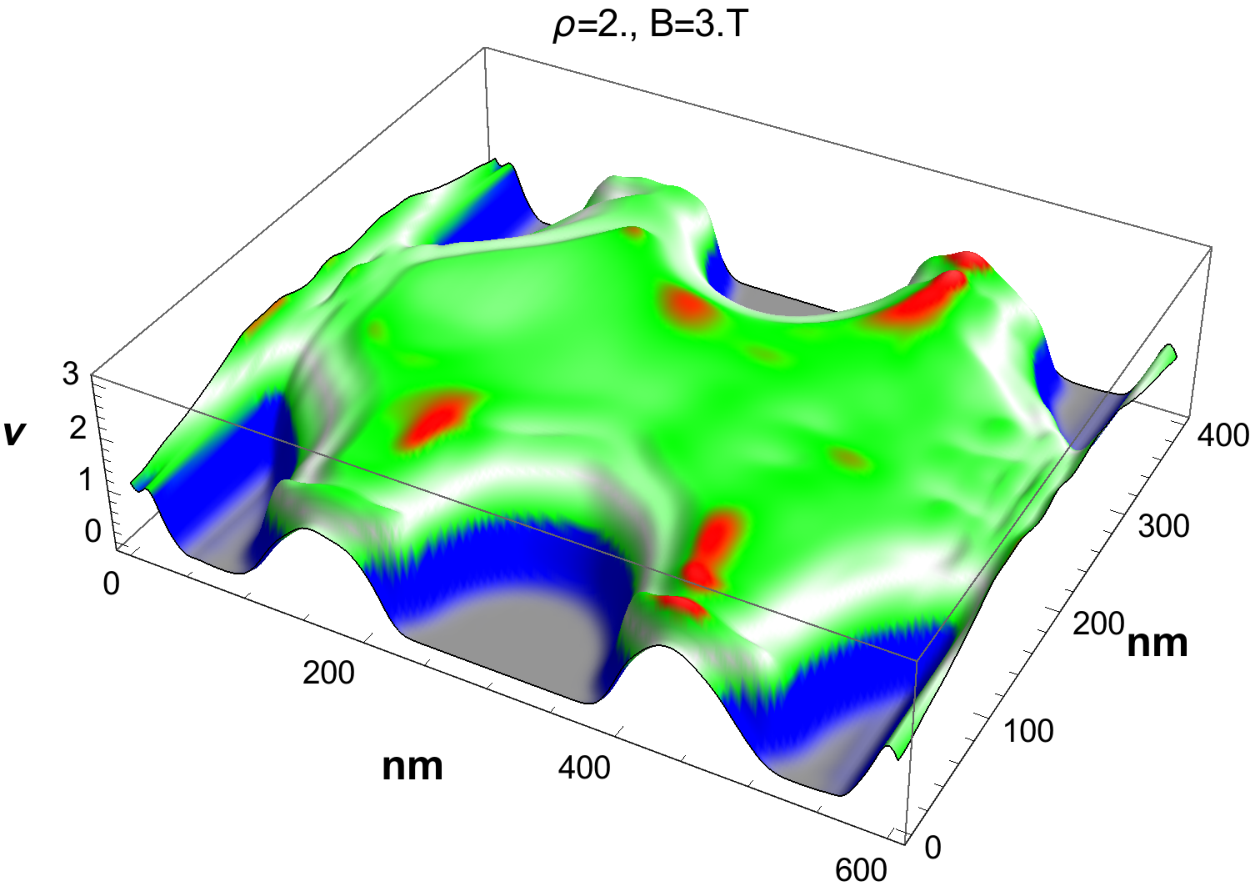} \\
%
%(c)
$\nu_\uparrow$\includegraphics[width=0.45\textwidth]{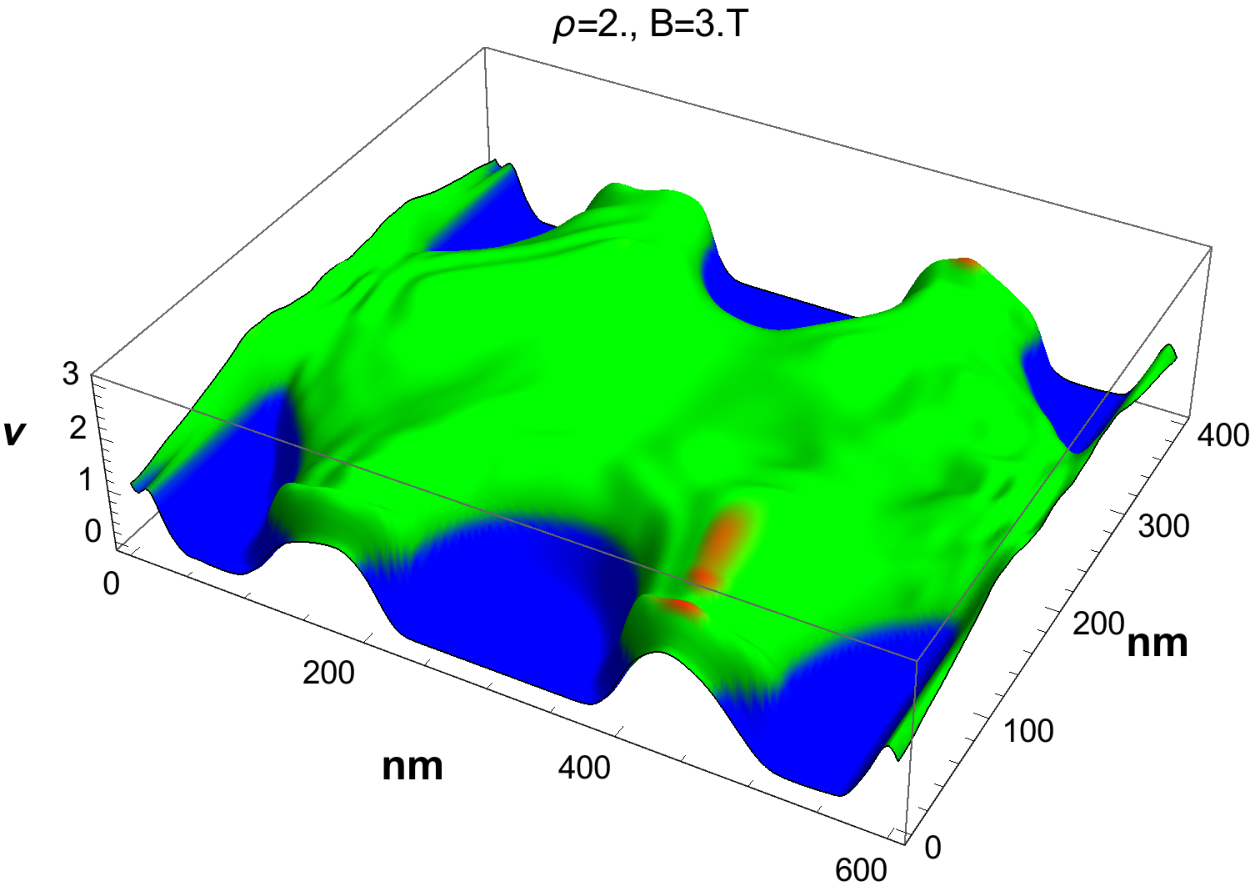} 
$\nu_\downarrow$\includegraphics[width=0.45\textwidth]{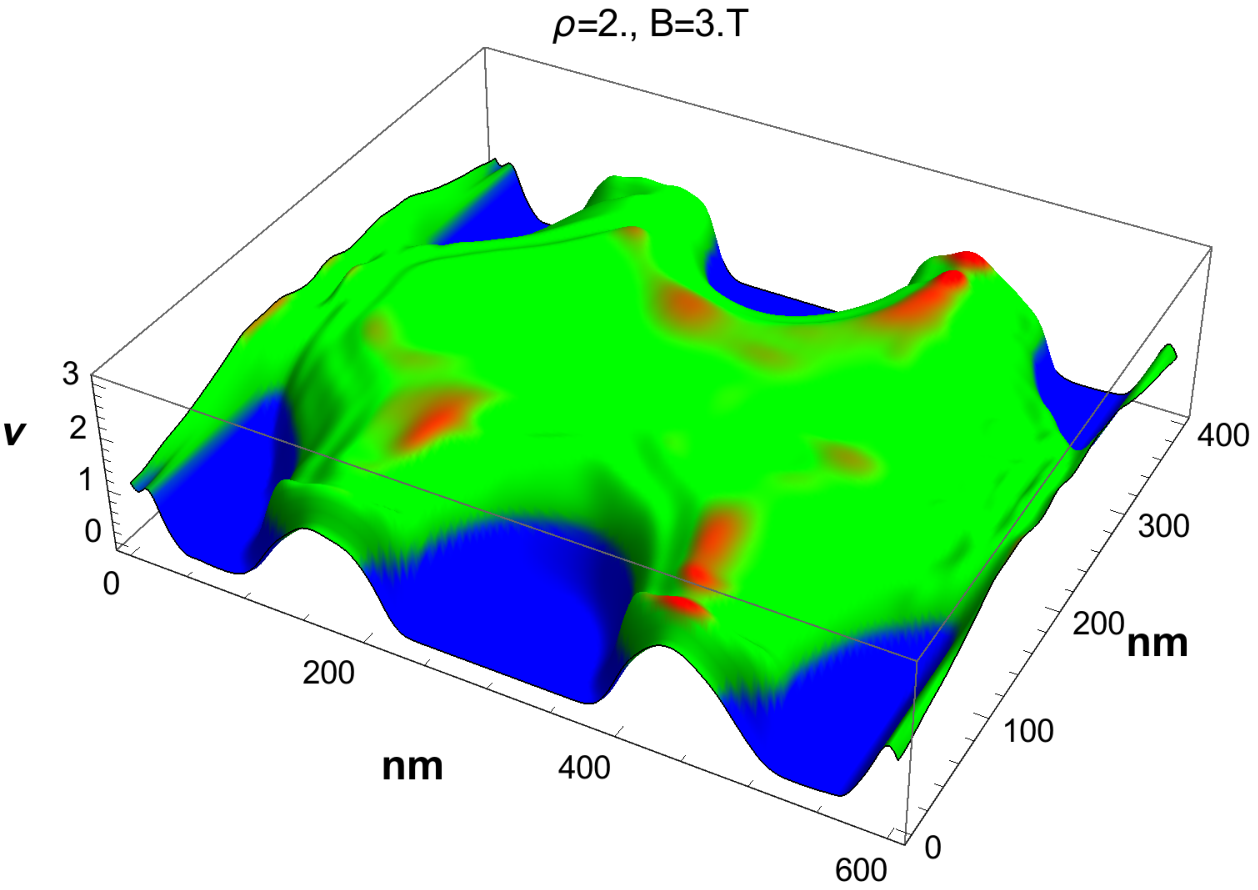} 
\caption{
\label{fig-CD-00-HH}
%\label{CDS1-n200-00-HH-HF}
%\label{fig-00-HH-HF-up} 
Distributions of $\nu_{\uparrow}$ and $\nu_\downarrow$ for (top row) the interaction-free single particle approximation, (centre row) the Hartree approximation and (bottom row) the same Hartee results, but now without color-highlighting of the half-odd integer filling.
The parameters correspond to $\rho= 2 \times \SI{e11}{cm^{-2}}$ and $B=\SI{3}{Tesla}$ as in Fig.\ \ref{fig-CD-NNM}.
The colors represent $\nu_{\uparrow}$ and $\nu_\downarrow$ as in Fig.\ \ref{fig-CD-NNM} with additional colors light blue and yellow denoting LLs $4$ and $5$. The filling factor range close to $\nu_\uparrow$, $\nu_\downarrow = $ half-odd integer is highlighted in light gray from LL2 onwards in order to identify possible stripes appearing close to the half filled top LL.
}
\end{figure*}

Last, we can also directly study the different responses of the electron system when changing the underlying interaction. 
In Fig.\ \ref{fig-CD-00-HH} we show the situation without interaction in the top row. There is no visible difference between the behaviour of $\nu_{\uparrow}$ and $\nu_\downarrow$ anymore since the chosen $g$-factor is too small and the exchange-interaction-based enhancement is gone. We also note that the spatial variations are much less pronounced  and for the same values of $\rho= 2 \times \SI{e11}{cm^{-2}}$ as before, the electron gas remains much more localized in the centre of the system, which leads to significantly larger local filling factors and hence, an additional conductance step in Fig.\ref{fig-transport}a.
The pure Hartree case is shown in the central and the bottom rows of Fig.\ \ref{fig-CD-00-HH}. The difference between $\nu_{\uparrow}$ and $\nu_\downarrow$ is still very small, but certainly visible. As for the non-interacting case, the spatial variations of $\nu$ are much less pronounced and the distinct features around half-odd integer fillings observed in Fig.\ \ref{fig-CD-NNM} are now hardly present. Indeed, if we were to plot $\nu_{\uparrow}$ and $\nu_\downarrow$ without the highlights for half-odd integer as done in the bottom panel of Fig.\ \ref{fig-CD-00-HH}, one would probably never even speculate about their existence in the HF case. Overall, the situation for pure-Hartee interaction, with very wide regions of gradually filling up to their integers, is, unsurprisingly, very reminiscent of the CSG picture \cite{Chklovskii1992ELECTROSTATICSCHANNELS}.

%%%%%%%%%%%%%%%%%%%%%%%%%%%%%%%%%%%%%%%%%%%%%%%%%%%%%%%%%%%%%%%%%%%%
%%%%%%%%%%%%%%%%%%%%%%%%%%%%%%%%%%%%%%%%%%%%%%%%%%%%%%%%%%%%%%%%%%%%

Hence, in partial summary, we have shown that our self-consistent HF modelling can reproduce the "classic" features associated with the IQHE such as the quantization of transport properties and the existence of network-like transport channels, along edges between plateau transitions and in the bulk directly at the transitions. In addition, we have full spatial resolution of local filling distributions $\nu(\vec{r})= \nu_{\uparrow}(\vec{r}) + \nu_\downarrow(\vec{r})$ and find a microscopic behaviour in line with exchange-enhanced CDWs. However, spatially resolved experimental information does not yet exist of these phases due to the intrinsic challenge of using local scanning probes in low temperatures for such remotely doped systems \cite{Hashimoto2008,HasCFS12,Friess2014}. Hence fully resolved modelling as shown here can provided a very valuable insight into the microscopic physics of the IQHE.

%%%%%%%%%%%%%%%%%%%%%%%%%%%%%%%%%%%%%%%%%%%%%%%%%%%%%%%%%%%%%%%%%%%%%%%%%
\subsection{Bubble and stripe phases}
\label{sec-SB}
%%%%%%%%%%%%%%%%%%%%%%%%%%%%%%%%%%%%%%%%%%%%%%%%%%%%%%%%%%%%%%%%%%%%%%%%%

Further examples of such microscopic detail are given by so-called "bubble" and "stripe" phases \cite{Fogler2002}. Their underlying geometric anisotropies have been identified, e.g., by transport experiments in higher Landau levels (LLs) of ultra-high mobility samples \cite{Lilly1999a,Du1999,Du2000c} and are characterized as stripes (strong transport anisotropies) or bubbles (reentrance effects). It is believed that the phases correspond to density modulations with characteristic geometric non-uniformities due to the interplay of Coulomb interaction and the wave functions in higher Landau levels. Recent theoretical modelling has mostly concentrated on transport signatures of these phases \cite{Ettouhami2006,Ettouhami2007,Cote2002,Cote2016} while the original papers either assumed uni-directional charge-density waves (CDWs) \cite{Koulakov1995,Fogler1996d} or showed the consistency of mean field treatments with anisotropic phases in a Fermi liquid \cite{Fradkin1999,Spivak2006}. 

\begin{figure*}[tb]
\includegraphics[width=0.95\textwidth]{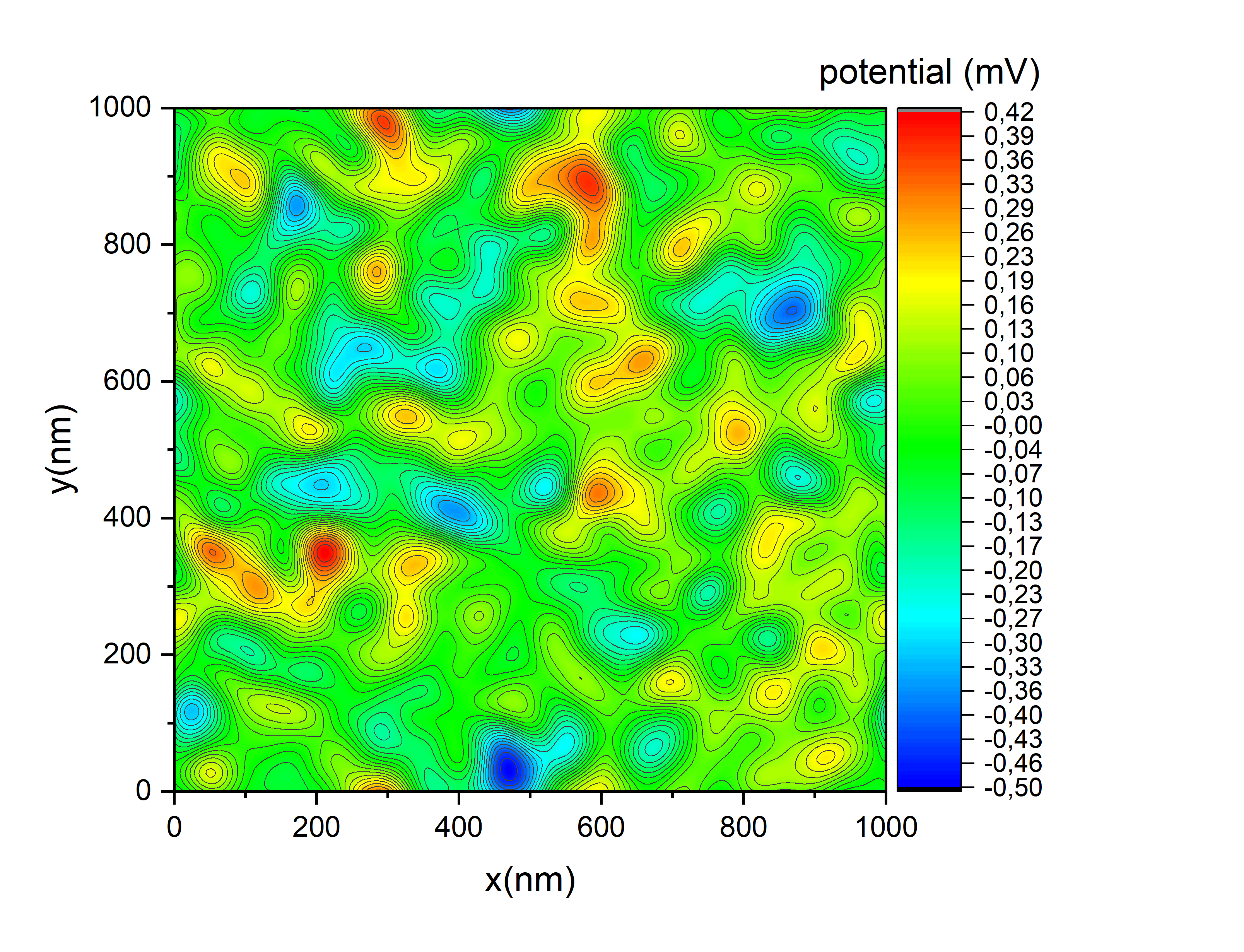}
\caption{
\label{fig-SB-disorder}
%\label{random_pot}
Lateral random disorder potential $V(\vec{r})$ visualized in a false color plot. The lines denote equipotentials while the potential energies are indicated by the colors as in the color scale provided.
}
\end{figure*}

\begin{figure*}[tb]
$\nu_\uparrow$\includegraphics[width=0.45\columnwidth]{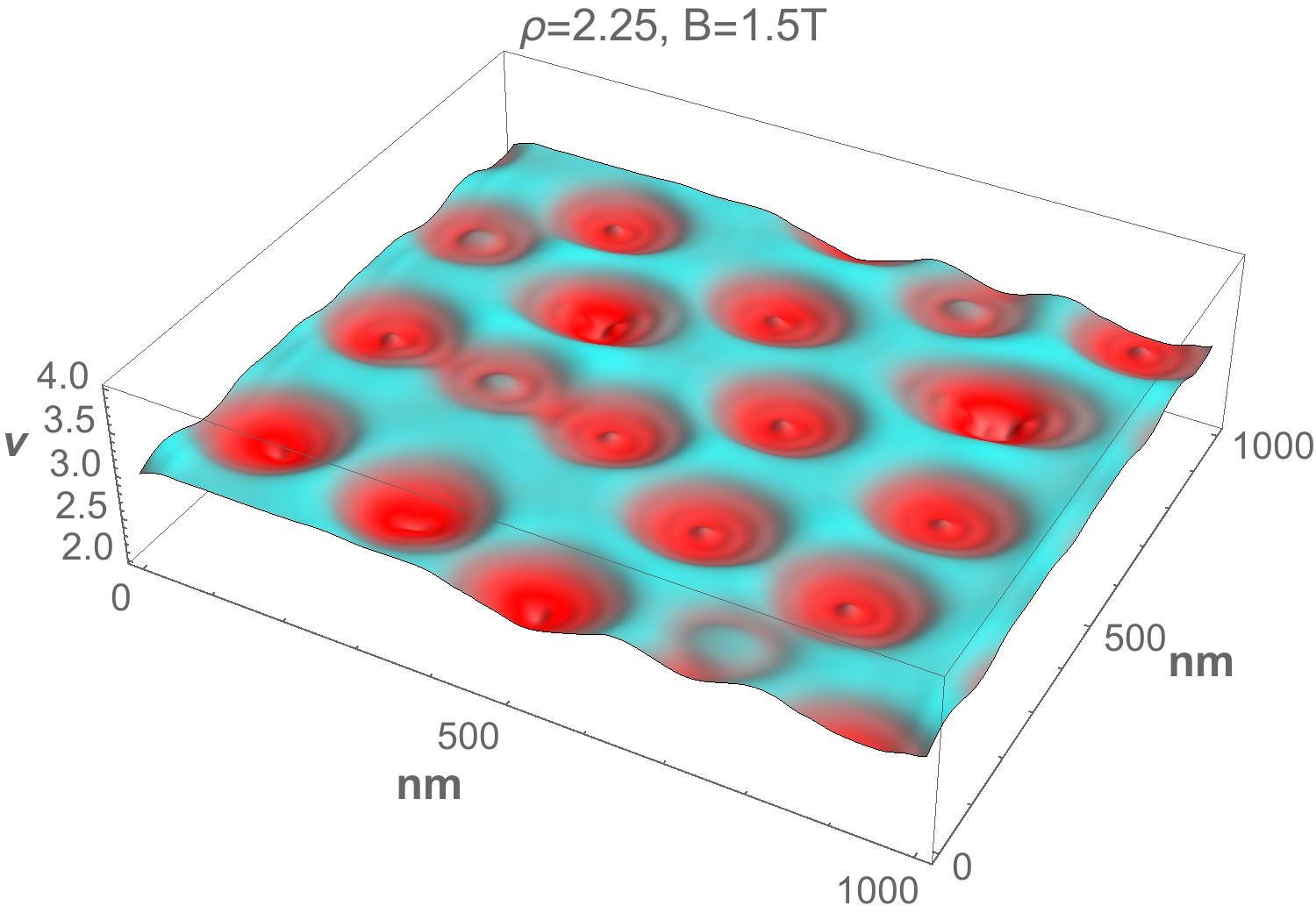}% Here is how to import EPS art
%\hspace*{-0.8\columnwidth} $\uparrow$ \hspace{0.8\columnwidth}$\downarrow$
$\nu_\downarrow$\includegraphics[width=0.45\columnwidth]{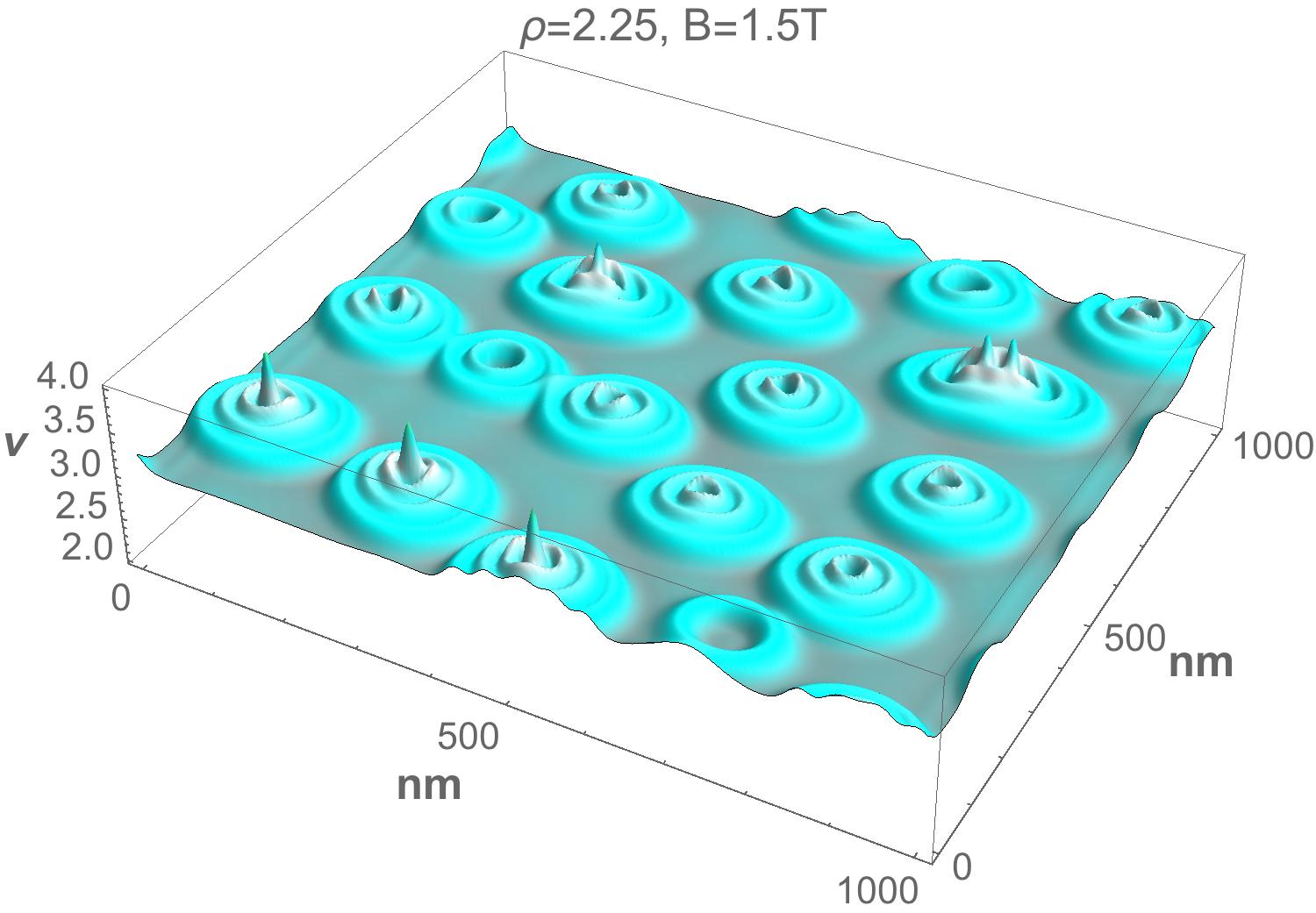}\\% Here is how to import EPS art
$\nu_\uparrow$\includegraphics[width=0.45\columnwidth]{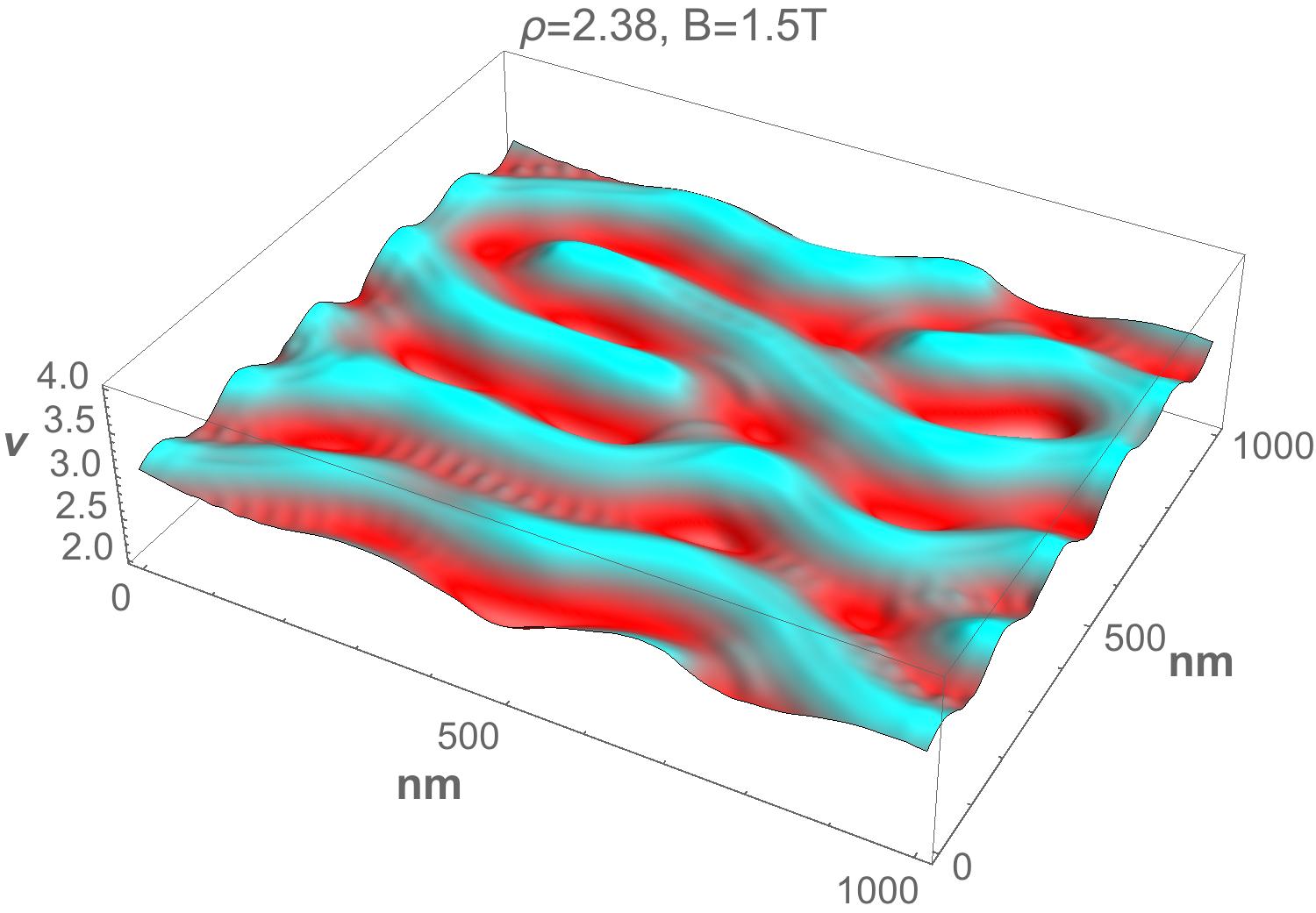}% Here is how to import EPS art
%\hspace*{-0.8\columnwidth} $\uparrow$ \hspace{0.8\columnwidth}$\downarrow$
$\nu_\downarrow$\includegraphics[width=0.45\columnwidth]{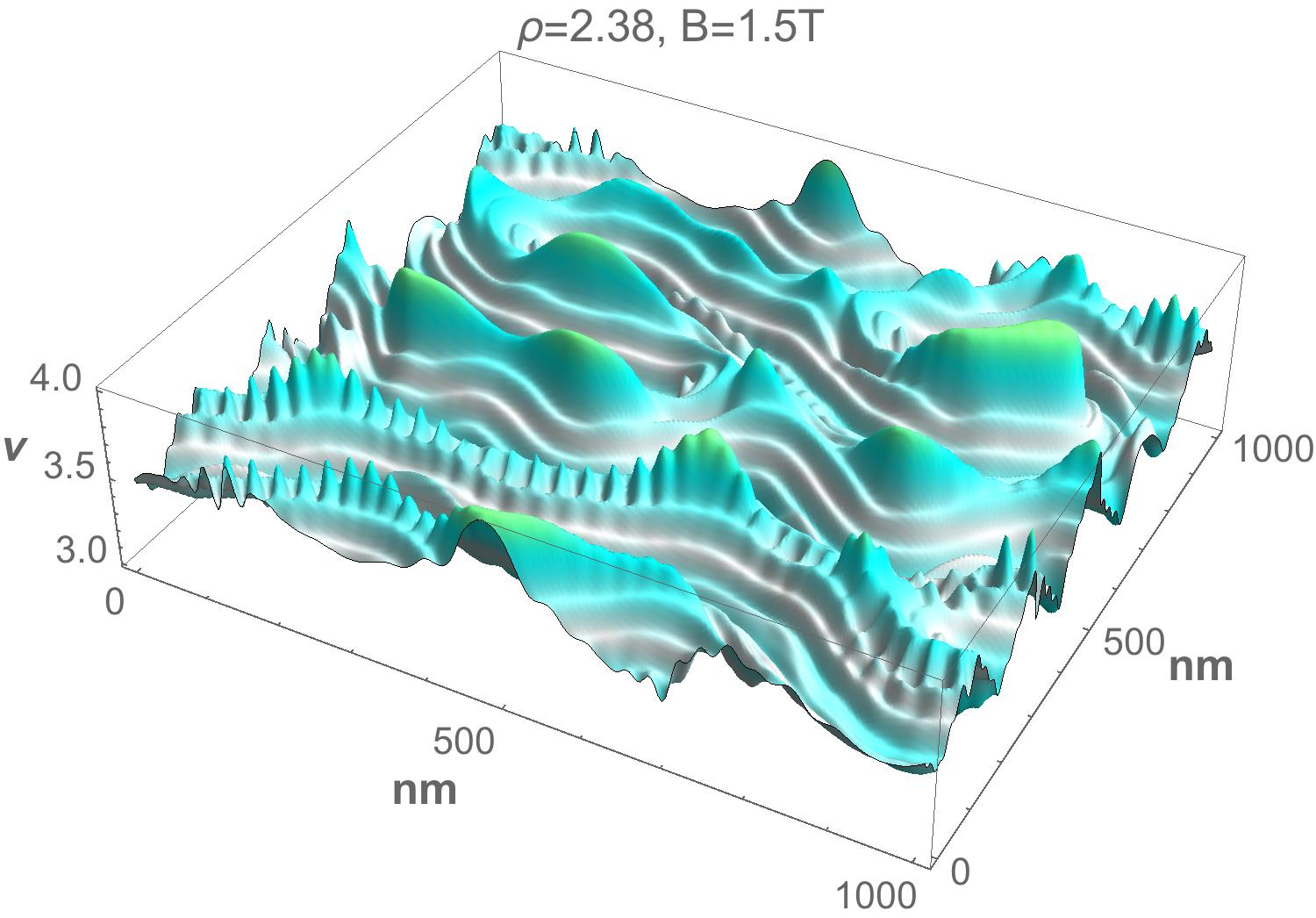}\\% Here is how to import EPS art
$\nu_\uparrow$\includegraphics[width=0.45\columnwidth]{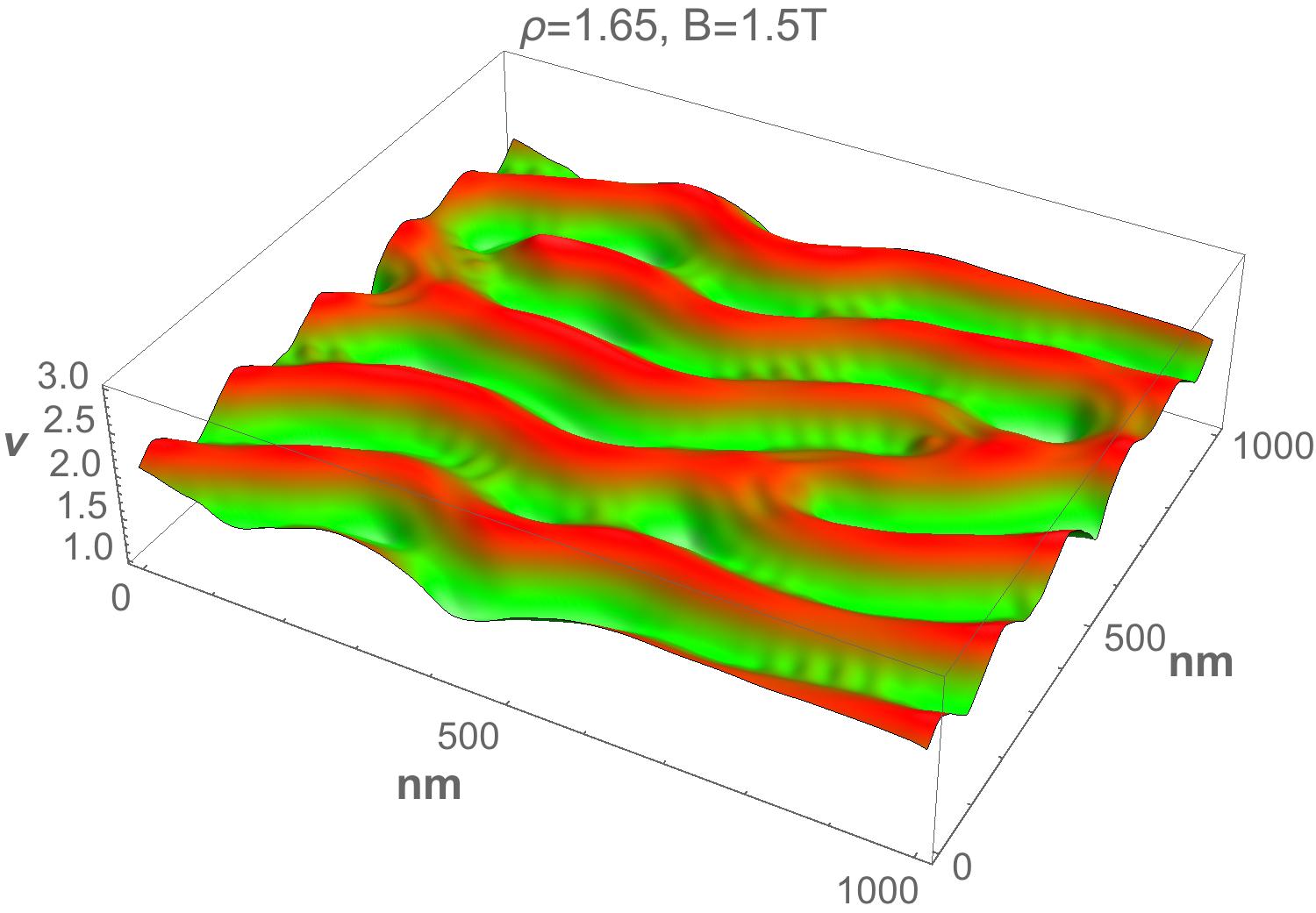}% Here is how to import EPS art
%\hspace*{-0.8\columnwidth} $\uparrow$ \hspace{0.8\columnwidth}$\downarrow$
$\nu_\downarrow$\includegraphics[width=0.45\columnwidth]{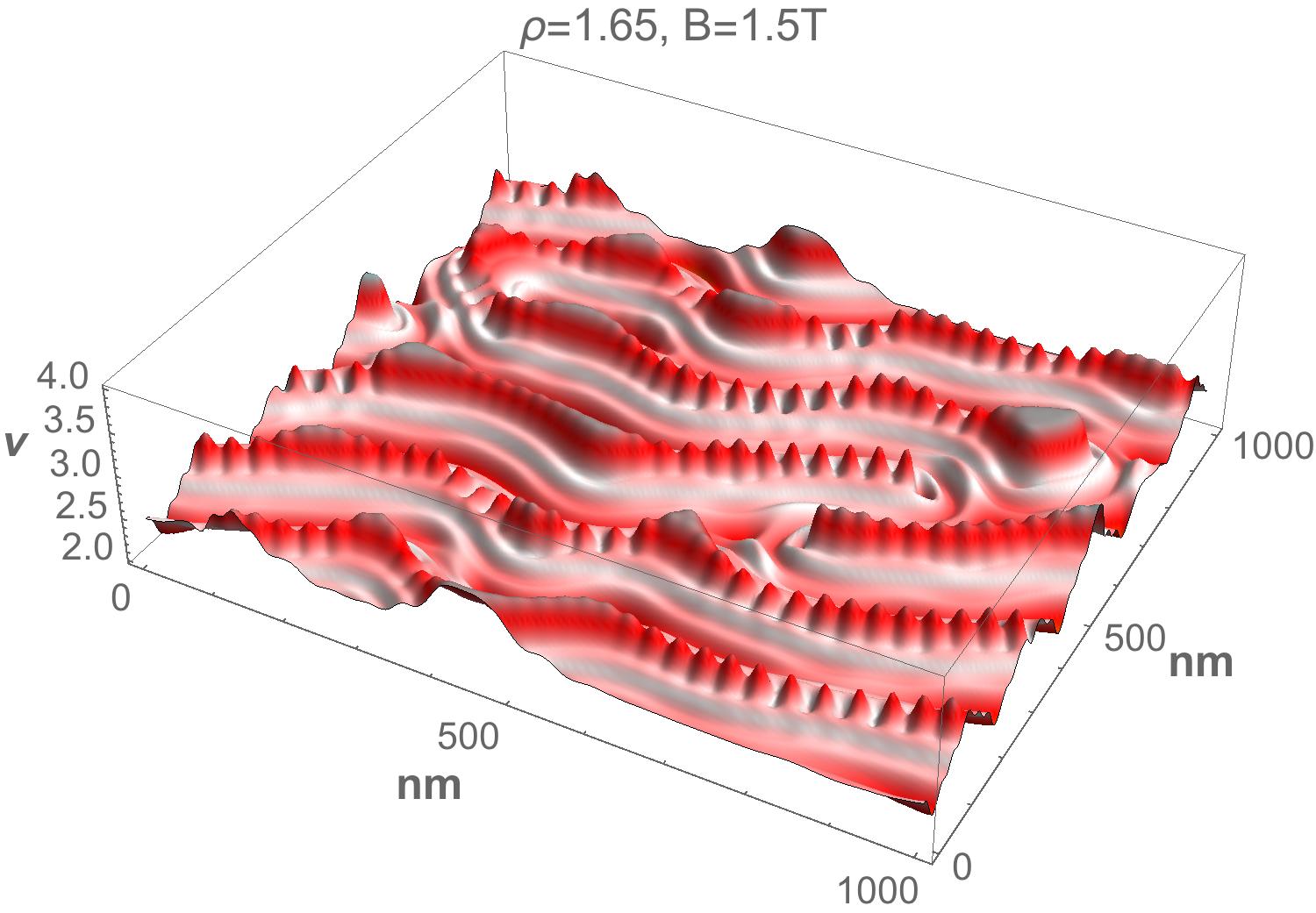}% Here is how to import EPS art
\caption{
\label{fig-SB-density}
%\label{fig-bubble_strips}
Distributions of $\nu_{\uparrow}$ and $\nu_\downarrow$ for (top row) $\nu = 6.20$, (central row) $\nu = 6.54$ and (bottom) $\nu = 4.54$ with the left column corresponding to $\nu_\uparrow$ and the right column to $\nu_\downarrow$ in all three rows. Colors are as given before, i.e.\ as in Fig.\ \ref{fig-CD-density} and others above. The parameters now correspond to $\rho=\SI{e11}{cm^{-2}}$ and $B=\SI{1.5}{Tesla}$.
}.
\end{figure*}

As we have shown in Ref.\ \cite{Oswald2020} the same HF approach used in the preceding sections for the strongly disordered QH liquid, also reproduces the stripe and bubble phases that emerge at weak disorder as similarly self-consistent solutions of the Hartree-Fock (HF) equations, i.e.\ in the experimentally relevant regime and without any ad hoc assumptions beyond a weak smooth disorder. For example, the quantitative prediction of the period of the stripes finds $d = \alpha R_c =\alpha \sqrt{(2n+1)\hbar B/e}$ with $\alpha=2.9 \pm 0.1$ \cite{Fogler1996d,Koulakov1995,Goerbig2004a,Kukushkin2011a,Friess2014,Oswald2020}.
This shows a perhaps still underappreciated strength of the HF approach.
To simulate ultra-high mobility samples, the random potential strength is kept low (cp.\ Fig.\ \ref{fig-SB-disorder}). 
%%%%%%%%%%%%%%%%%%%%%%%%%%%%%%%%%%%%%%%%%%%%%%%%%%%%%%%%%%%%%%%%%%%%%%%%%
%%%%%%%%%%%%%%%%%%%%%%%%%%%%%%%%%%%%%%%%%%%%%%%%%%%%%%%%%%%%%%%%%%%%%%%%%
The random potential is generated by Gaussian impurity potentials of radius $\SI{40}{nm}$, the number of impurities is $N=2000$, their random placement results in a fluctuating potential of $V_\text{max} = \SI{0.43}{mV}$ and $V_\text{min} = \SI{-0.50}{mV}$, considerably less than the $V_\text{impurities} \in [\SI{-10}{mV}, \SI{10}{mV}]$ used in the preceeding sections. At a total filling factor of, e.g., $\nu = 6.54$ ($= \nu_\downarrow + \nu_\uparrow = 3.54 + 3.0$) and $B = \SI{1.5}{Tesla}$, the present low-disorder situation still corresponds to more than $2000$ electrons.
The filling factor and the lateral system size are made as large as possible with respect to the available computing power in order not to unduly influence the stripe and bubble formation by possible boundary effects. Configurations up to $\SI{1}{\mu m^2}$ are achievable as shown in Fig.\ \ref{fig-SB-density} with spatial resolution of $\sim \SI{4.4}{nm}$ well below $l_B$ for a magnetic field $B$ varying from, e.g., $\SI{1}{Tesla}$ ($l_B\sim \SI{26}{nm}$) to $\SI{6.5}{Tesla}$ ($l_B\sim \SI{10}{nm}$).  
In Fig.\ \ref{fig-SB-density} we show the computed variation in $\nu(\vec{r})$ for three different densities $\rho$ at fixed magnetic field $B$
%. These results are identical to Fig.\ 1 of the main text 
for $\nu_{\downarrow}(\vec{r})$ but also $\nu_{\uparrow}(\vec{r})$ in the left column. The choice of colors is as in Refs.\ \cite{OswR17,OswaldPRB2017}.
Clearly, depending on the value of $\rho$, "bubble" and "stripe" phase have emerged self-consistenyl without prior intentional induction of any spatial anisotropy --- except of course that which is induced by the randomness of the disorder. "Bubble" and "stripe" phases are hence stable phases of the HF ground state, no more "interaction" is needed for their emergence. For a detailed discussion of the features shown in Fig.\ \ref{fig-SB-density}, we refer the reader to Ref.\ \cite{Oswald2020}. 
%%%%%%%%%%%%%%%%%%%%%%%%%%%%%%%%%%%%%%%%%%%%%%%%%%%%%%%%%%%%%%%%%%%%%%%%%%%%%%%%%
%%%%%%%%%%%%%%%%%%%%%%%%%%%%%%%%%%%%%%%%%%%%%%%%%%%%%%%%%%%%%%%%%%%%%%%%%%%%%%%%%

\begin{figure*}[tb]
no interaction: \hfill  \mbox{ }\\
%\hspace*{-0.8\columnwidth} $\uparrow$ \hspace{0.8\columnwidth}$\downarrow$
$\nu_\uparrow$\includegraphics[width=0.45\textwidth]{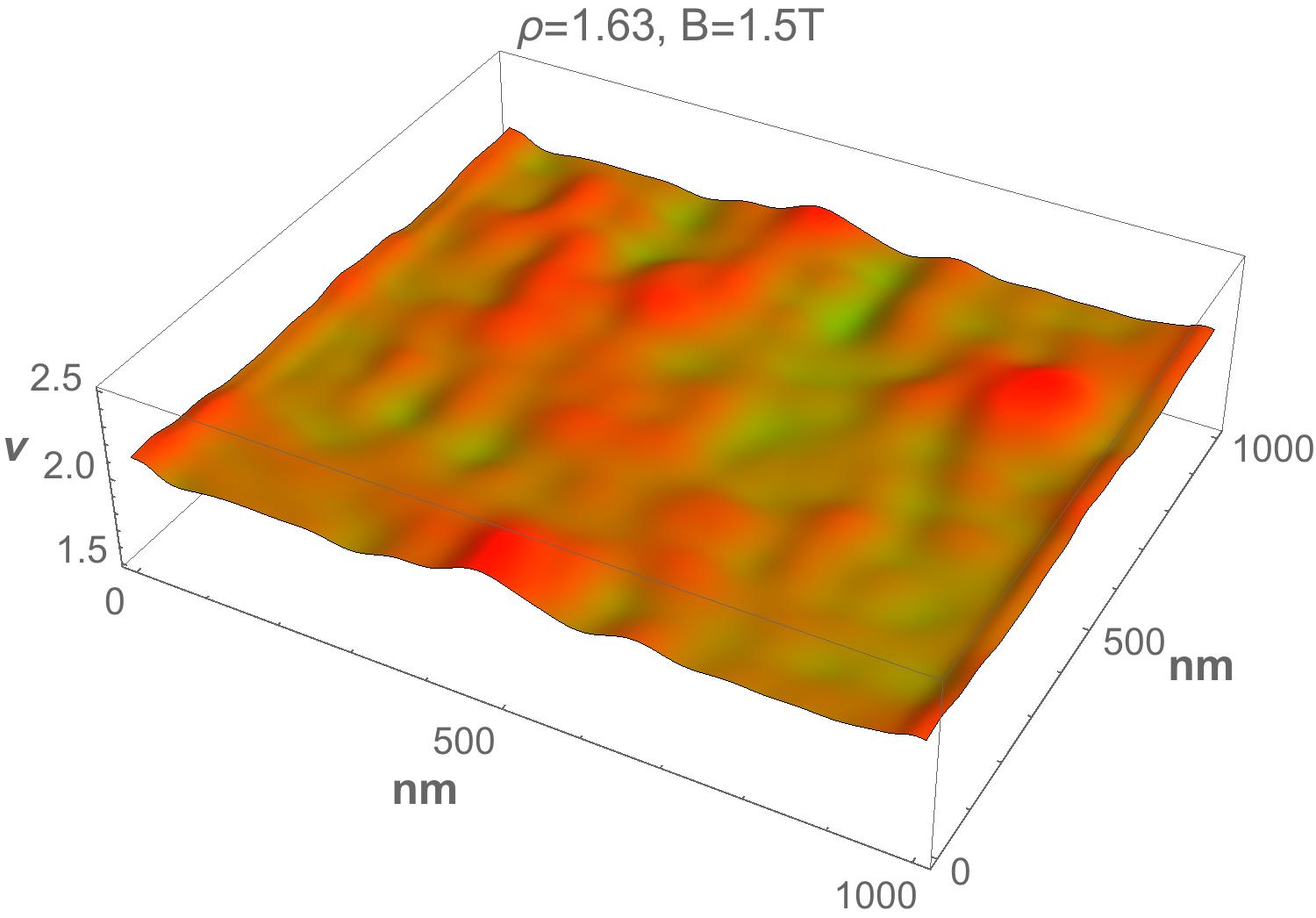}
$\nu_\downarrow$\includegraphics[width=0.45\textwidth]{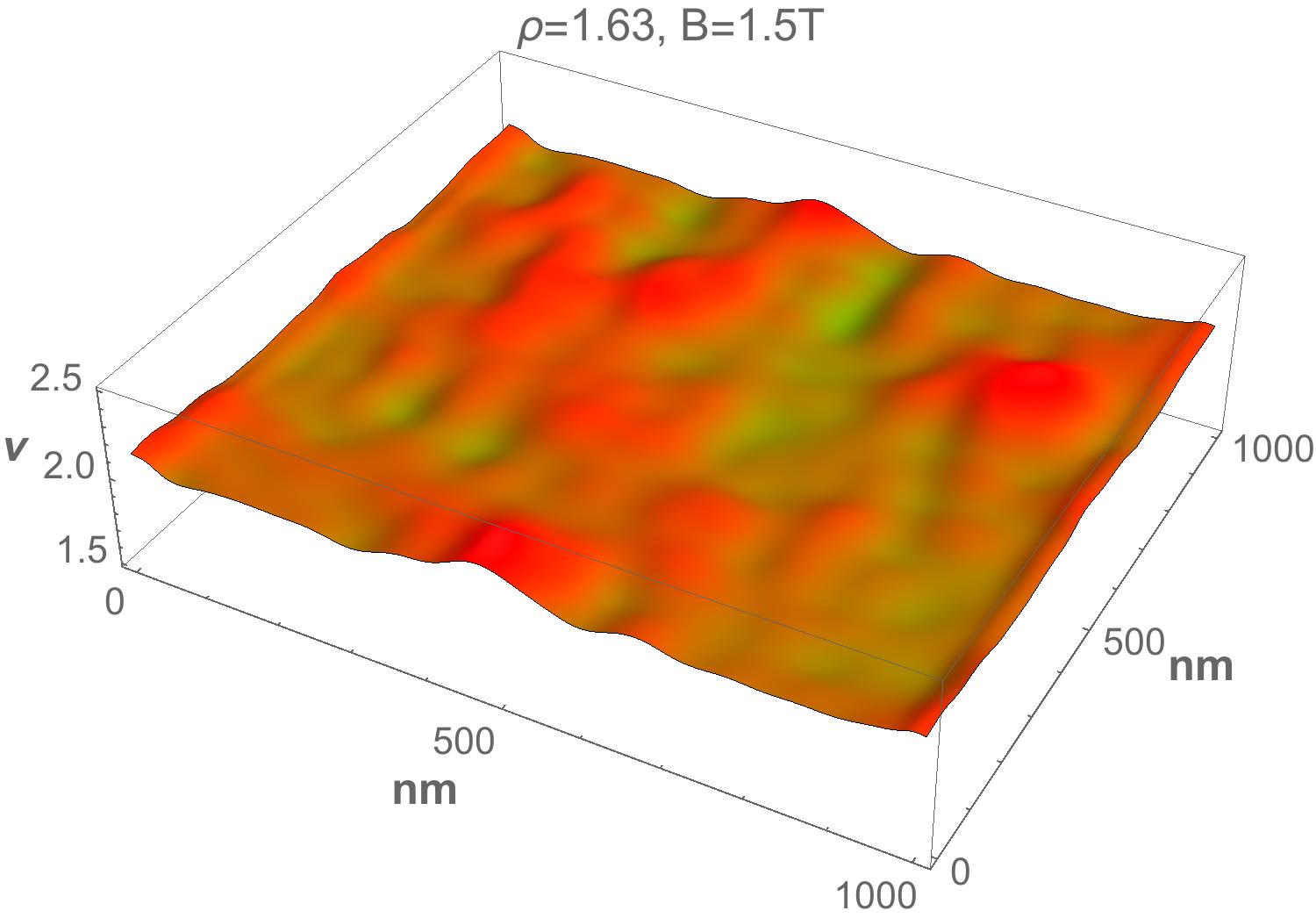}\\[3ex]
Hartree interaction: \hfill  \mbox{ }\\
$\nu_\uparrow$\includegraphics[width=0.45\textwidth,clip]{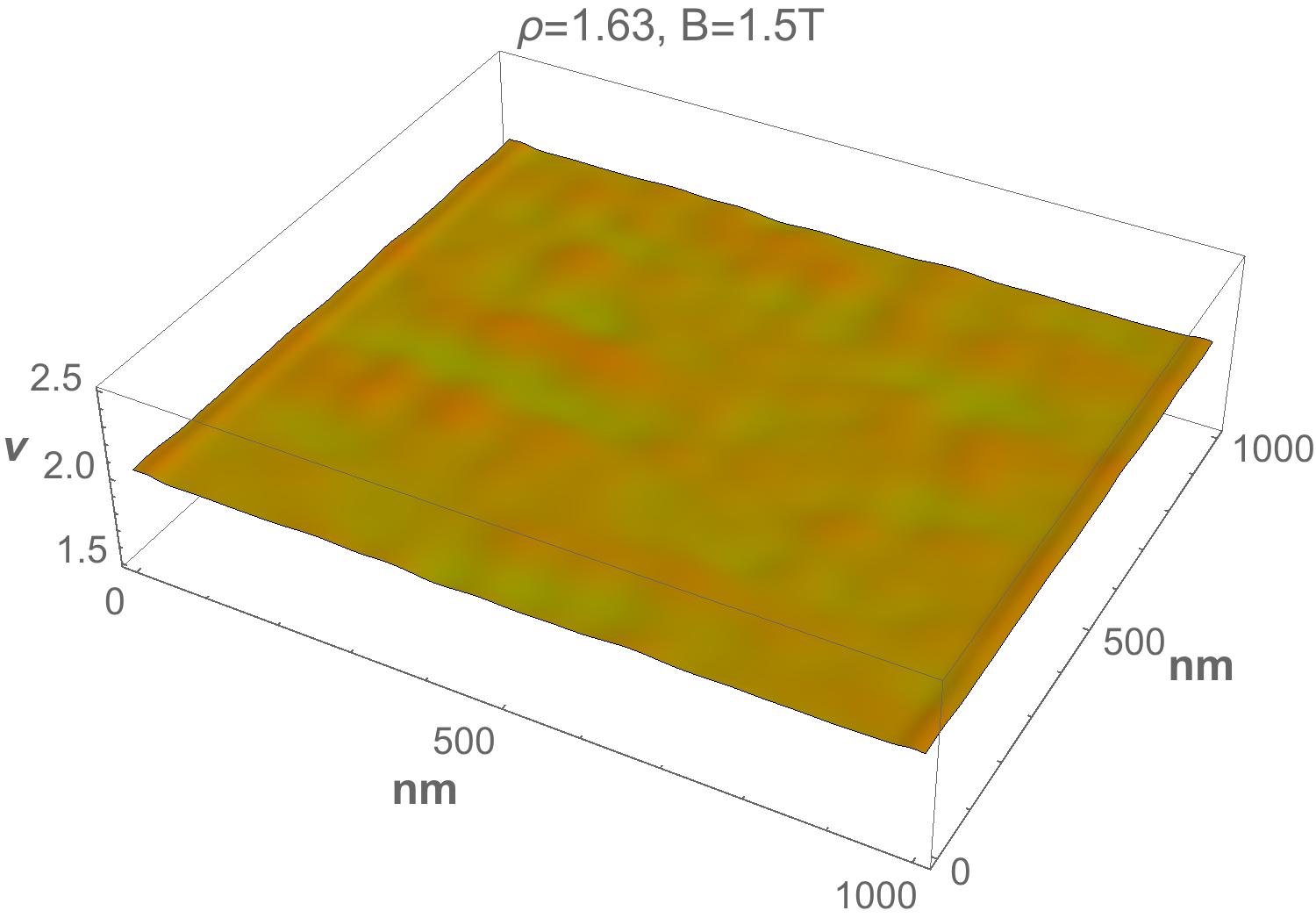}
$\nu_\downarrow$\includegraphics[width=0.45\textwidth]{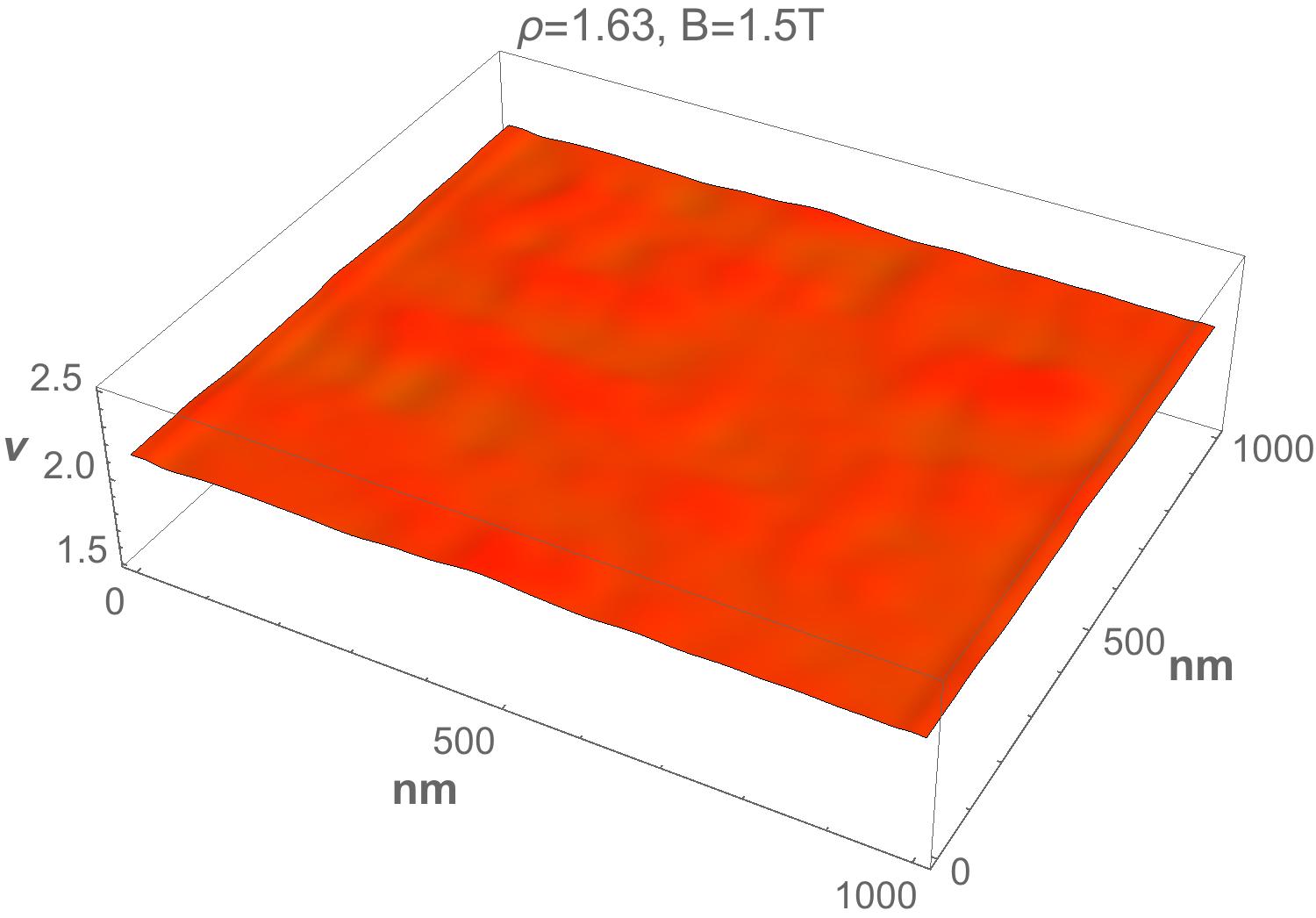}
\caption{
\label{fig-SB-HH-00}
%\label{CD_nu_450-Sx-HH-00}
Lateral charge density of the partly filled top Landau levels at total filling factor $\nu=4.5$. The top row shows results for the non-interacting system while the bottom row corresponds to a pure Hartree interaction with left column corresponding to $\nu_\uparrow$ and the right column to $\nu_\downarrow$ in all two rows. The color shades represent the filling factor $\nu(\vec{r})$, i.e.\ as in Fig.\ \ref{fig-CD-density} and other above.
}
\end{figure*}

In the top row of Fig.\ \ref{fig-SB-HH-00} we show the situation analogous to Fig.\ \ref{fig-SB-density}, but now computed without the Coulomb interaction. Clearly, there is no stripe formation. The $\nu(\vec{r})$ modulation rather closely follows the random potential of Fig.\ \ref{fig-SB-disorder}.  The charge densities in the spin-up and spin-down levels do not influence each other and follow nearly identically the disorder potential because of the missing interaction. As we have chosen to keep the same colors as in Fig.\ \ref{fig-SB-density}, it is clear that there is comparatively very little spatial variation for $\nu_{\uparrow}(\vec{r})$ and $\nu_\downarrow(\vec{r})$.
In Fig.\ \ref{fig-SB-HH-00} one can also see that for pure Hartree interaction there is no stripe formation. The density modulation in $\nu(\vec{r})$ is much less than in Fig.\ \ref{fig-SB-density} and roughly follows the random potential shown in Fig.\ \ref{fig-SB-disorder}. Furthermore, the charge density modulation in the spin-up and spin-down levels shlightly "repel" each other due to the Hartree interaction.

%%%%%%%%%%%%%%%%%%%%%%%%%%%%%%%%%%%%%%%%%%%%%%%%%%%%%%%%%%%%%%%%%%%%%%%%%
%%%%%%%%%%%%%%%%%%%%%%%%%%%%%%%%%%%%%%%%%%%%%%%%%%%%%%%%%%%%%%%%%%%%%%%%%

%%%%%%%%%%%%%%%%%%%%%%%%%%%%%%%%%%%%%%%%%%%%%%%%%%%%%%%%%%%%%%%%%%%%%%%%%
\subsection{Knight shift}
\label{sec-KS}
%%%%%%%%%%%%%%%%%%%%%%%%%%%%%%%%%%%%%%%%%%%%%%%%%%%%%%%%%%%%%%%%%%%%%%%%%

% \ref{fig-KS-test}
% \ref{fig-KS-density}
% \ref{fig-KS-NMR}

In Ref.\ \cite{Friess2014}, Friess et al.\ investigated the Knight shift $I_{\nu}(f)$ in nuclear magnetic resonance (NMR) spectra and modelled it with a semiclassical model \cite{Tiemann2012,Friess2014} based on superpositions of the \emph{single} electron densities obtained from the Landau basis functions. A periodic variation of the filling factor, due to the presence of stripes and bubbles, should result in characteristic peaks, due to the implied regions with different spin polarizations for $\nu_{\uparrow}(\vec{r})$ and $\nu_\downarrow(\vec{r})$. While their method provides direct information about the area fractions, there is no direct information about geometry and periodicity.

\begin{figure*}[tb]
\includegraphics[width=0.95\textwidth]{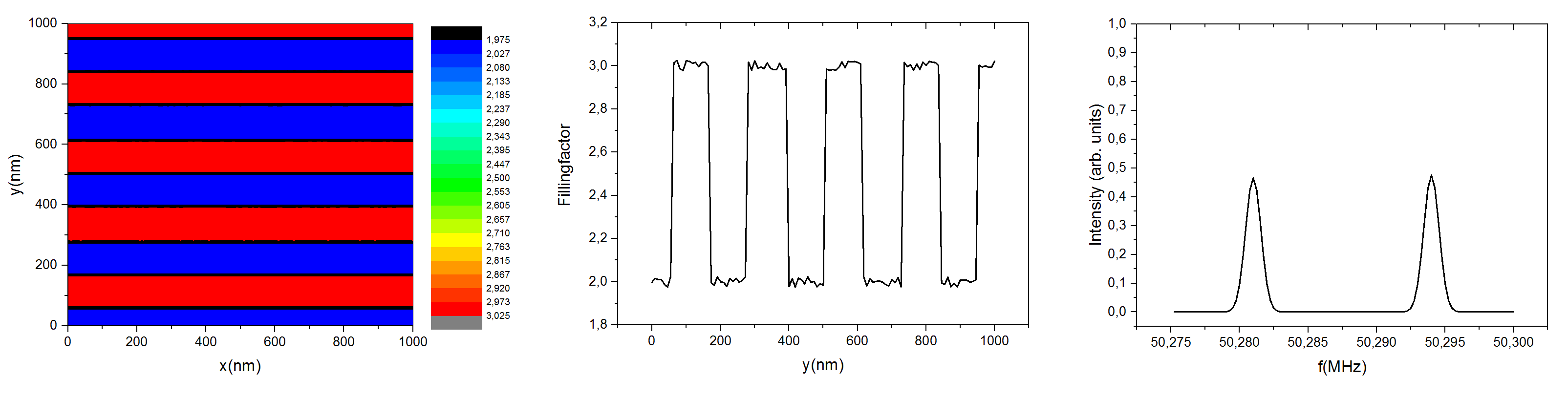}
\includegraphics[width=0.95\textwidth]{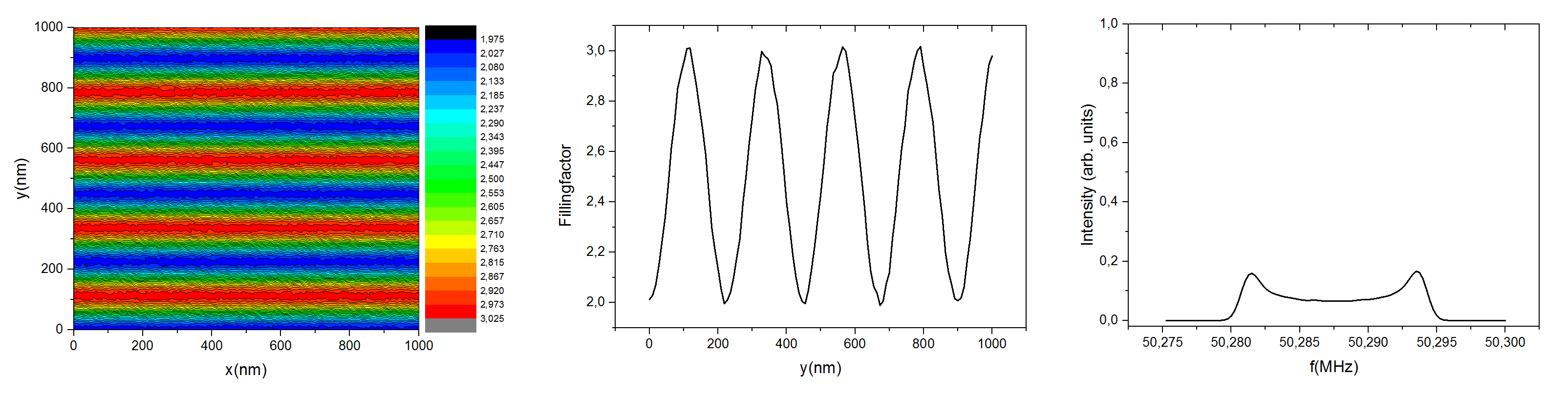}
\includegraphics[width=0.95\textwidth]{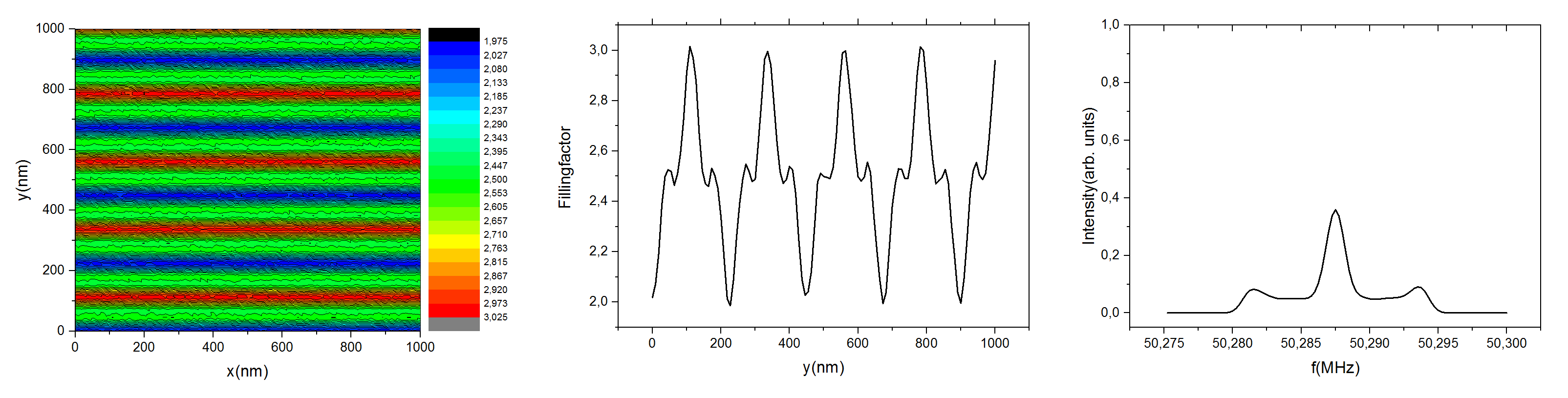}
\includegraphics[width=0.95\textwidth]{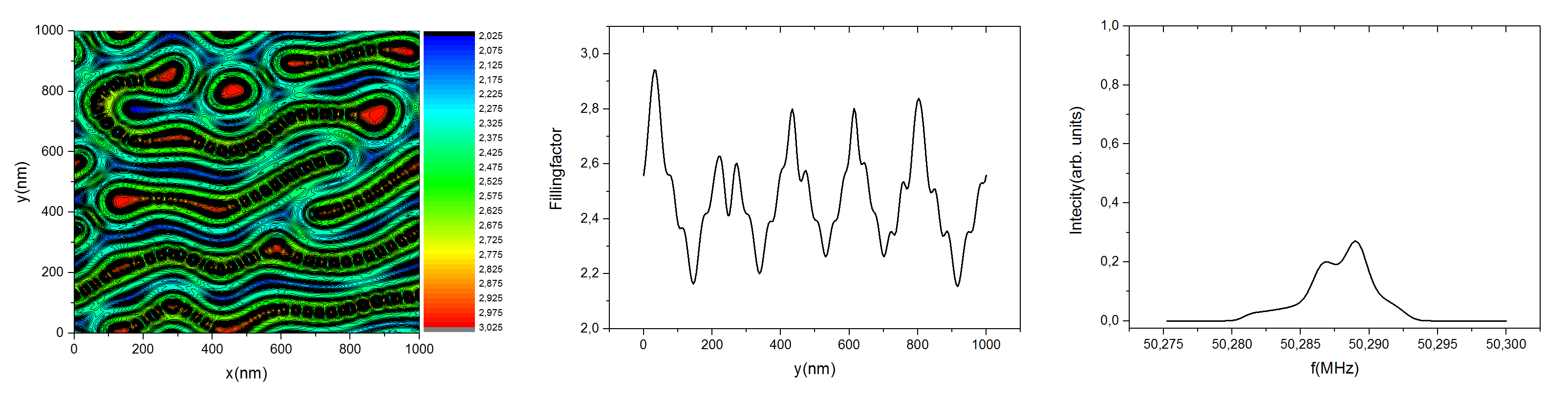}
\caption{
\label{fig-KS-test}
%\label{fig-NMR-test}
Calculations of the NMR intensity $I_{\nu}(f)$ for 4 different stripe-like variations of $\nu(\vec{r})$ at $\nu=4.5$. Rows 1-4 corresponds to (a) a simply square modulation, (b) a sinusoidal modulation, (c) a sinusoidal variations with left and right shoulders, (d) a modulation as computed from HF. The first column shows the spatial variations in $\nu(\vec{r})$ as given by the color scales. The second column represent a typical cross-section for each situation and the third column shows the estimated NMR intensity $I_{4.5}(f)$.
}.
\end{figure*}

In order to show that the patterns which we observe here within HF for $\nu_{\uparrow}(\vec{r})$ and $\nu_\downarrow(\vec{r})$, in particular the presence of the remnants of the LL wave functions around each stripe and bubble, are significant, we perform the calculations of $I_{\nu}(f)$ \cite{Oswald2020} also for three test patterns for $\nu(\vec{r})$. The results are given in Fig.\ \ref{fig-KS-test}. We find that the variation given by $\nu(\vec{r})$ as calculated with HF can reproduce global features of the experimental NMR results presented in Ref.\ \cite{Friess2014}, while all test density patterns have essential departures from the experimental results. 
%%%%%%%%%%%%%%%%%%%%%%%%%%%%%%%%%%%%%%%%%%%%%%%%%%%%%%%%%%%%%%%%%%%%%%%%%
%%%%%%%%%%%%%%%%%%%%%%%%%%%%%%%%%%%%%%%%%%%%%%%%%%%%%%%%%%%%%%%%%%%%%%%%%

\begin{figure*}[tb]
\mbox{ }  \hfill $\nu=2$--$3$ \hfill $\nu=4$--$5$ \hfill  \mbox{ }\\
(a)\includegraphics[width=0.47\textwidth]{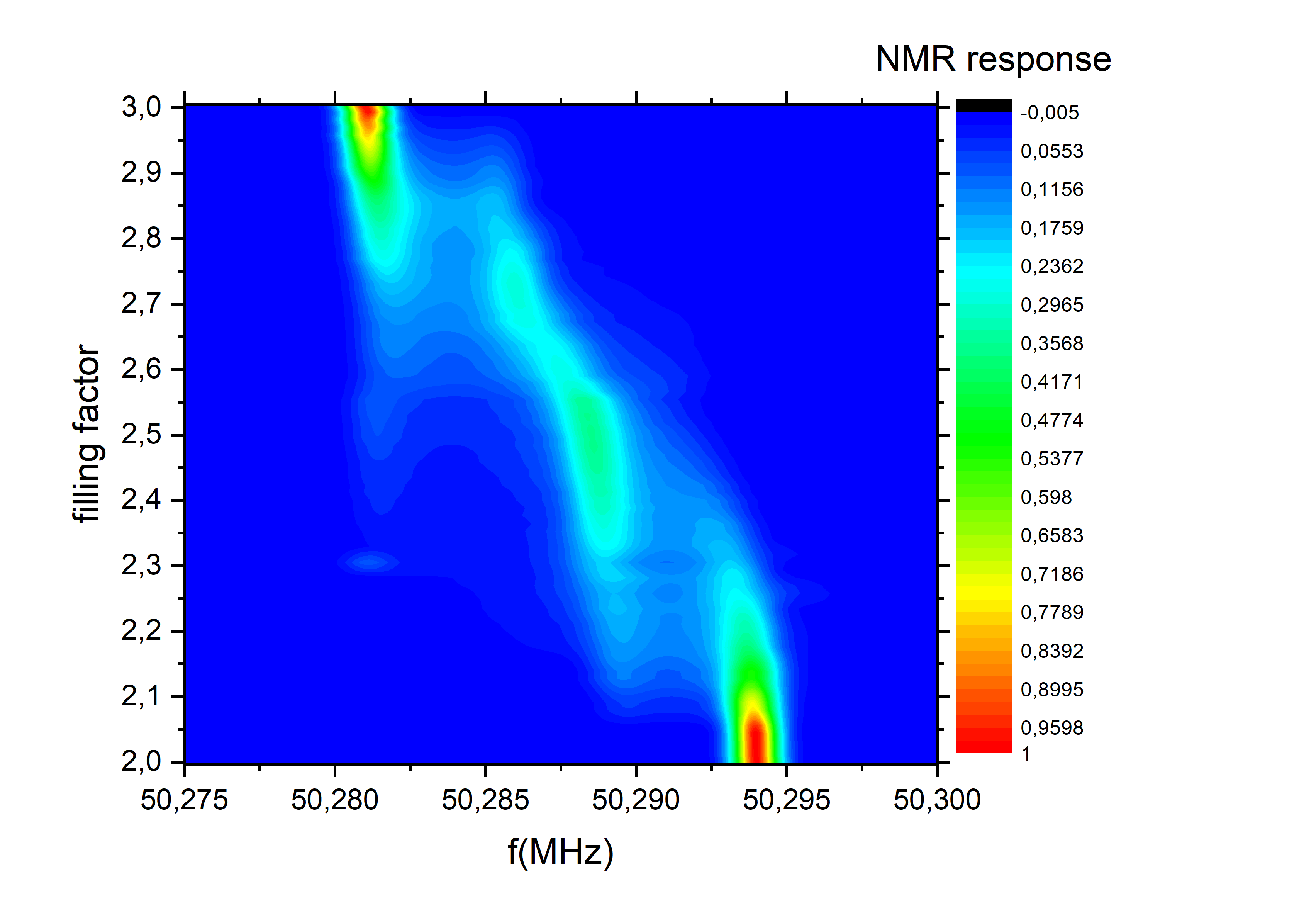}
(b)\includegraphics[width=0.47\textwidth]{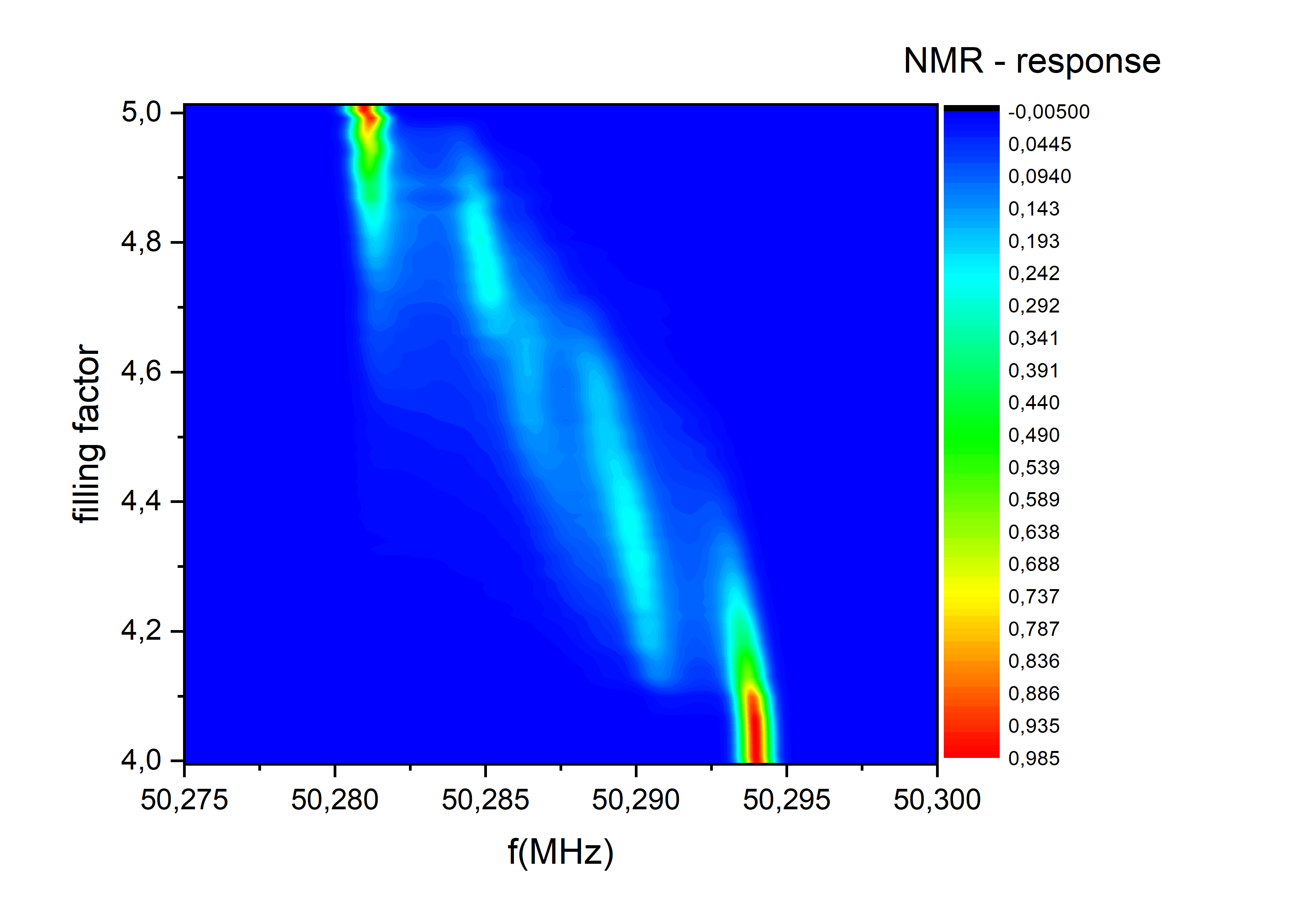}\\
\mbox{ }  \hfill $\nu=2.5$ \hfill $\nu=4.5$ \hfill \mbox{ }\\
(c)\includegraphics[width=0.47\textwidth]{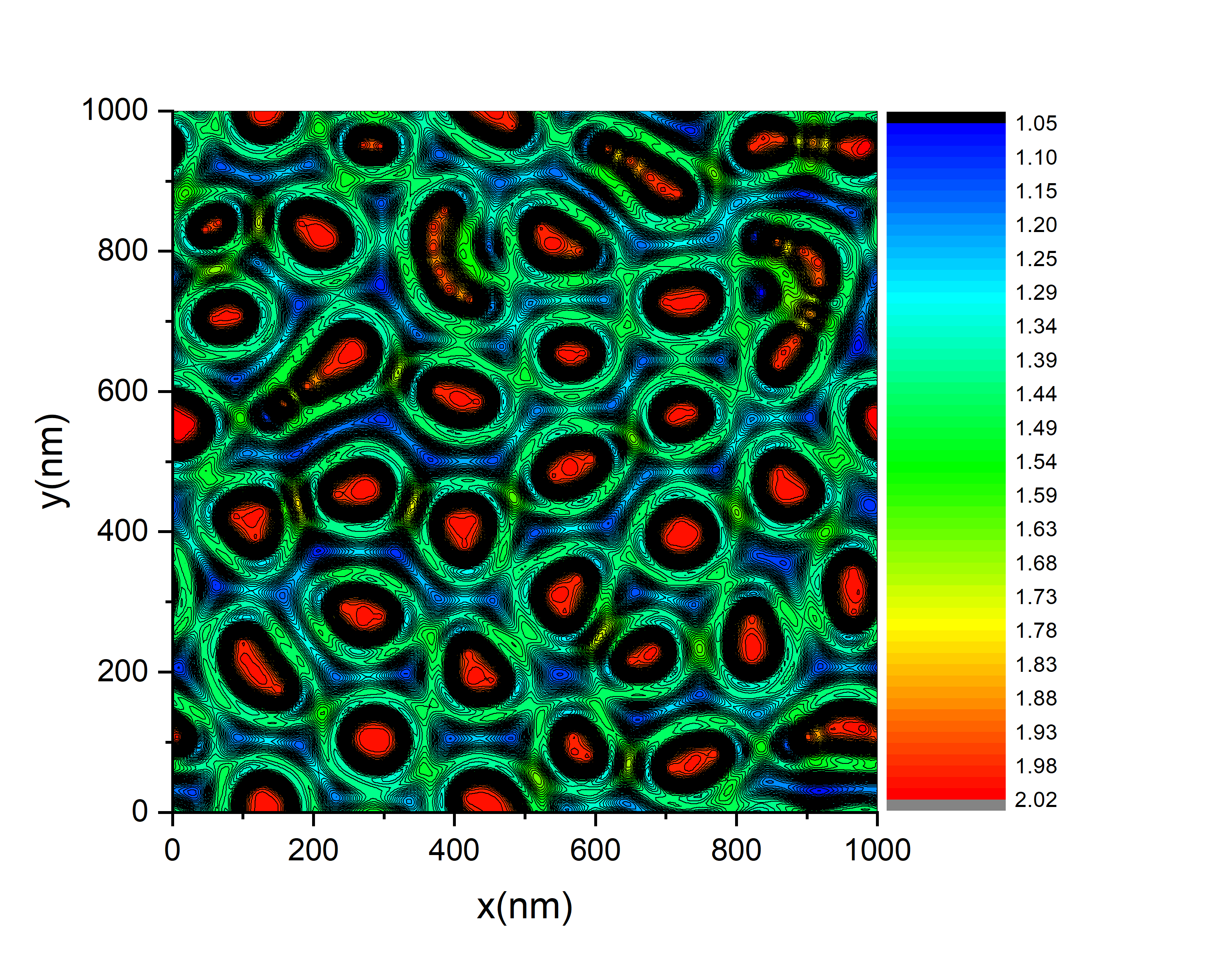}
(d)\includegraphics[width=0.47\textwidth]{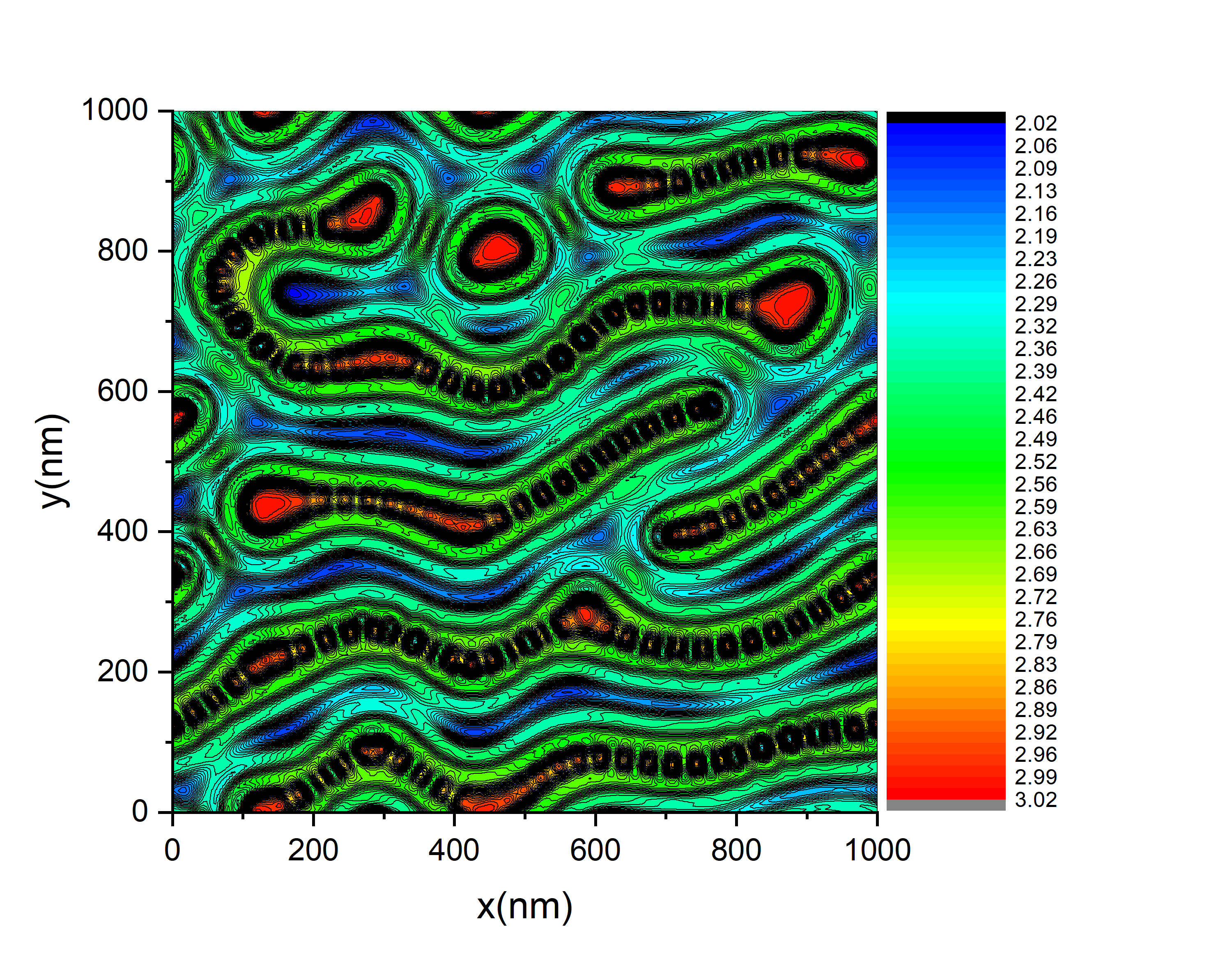}
\caption{
\label{fig-KS-density}
%\label{fig-NMR-stripes-bubbles}
NMR intensities $I_{\nu}(f)$ for (a) $\nu=2-3$ and (b) $\nu=4-5$ and local $\nu(\vec{r})$ at (c) $\nu=2.5$ and (d) $\nu=4.5$ in left and right columns, respectively. (b+d) The right column shows results as in the bottom row of Fig.\ \ref{fig-KS-test}. The color shades represent  $I_{\nu}(f)$ and $\nu(\vec{r})$ as given by the scales. Lines in (c+d) connect equal height in $\nu(\vec{r})$.
}
\end{figure*}

\begin{figure*}[tb]
\includegraphics[width=0.95\textwidth]{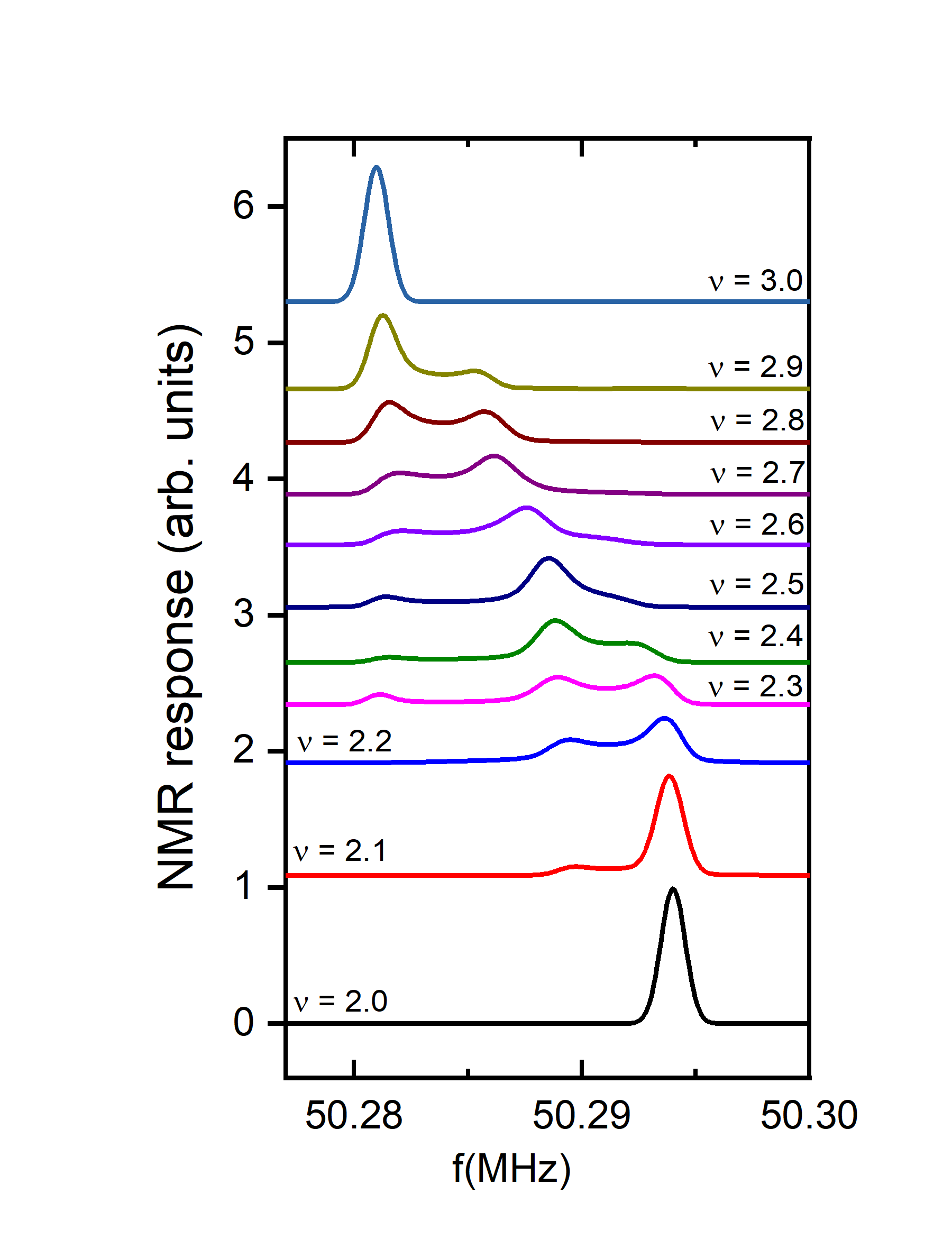}
\caption{
\label{fig-KS-NMR}
%\label{fig:NMRspec-2-3}
NMR intensities $I_{\nu}(f)$ as shown in Fig.\ \ref{fig-KS-test} for selected filling factors in range $\nu=2 - 3$. The style is similar to the one used in Ref.\ \cite{Friess2014}.
}
\end{figure*}

We note that normally stripes appear only starting with filling factor $\nu=4.5$. This is known also experimentally, but for experimental reasons the authors of Ref.\ \cite{Friess2014} could not go to that filling factor. Instead, they used filling factor $\nu=2.5$ and forced, by using an in-plane component of the magnetic field, the electron system to form a stripe pattern. Clearly, there is no need for our simulations to also model this experimental "trick". 
In order to compare the effect of stripe patterns on the NMR Knight shift, we therefore use  the stripe pattern in the "correct" range $\nu=4$--$5$. The result in Fig.\ \ref{fig-KS-density} (b) has striking similarities but seems indeed a bit richer in features than the experimental curve for $\nu=2$--$3$. 
However, since the Knight shift spectrum looses its local information due to the spatial integration, we can also evaluate the range $\nu = 2$--$3$ as shown in Fig.\ \ref{fig-KS-density}. Indeed the agreement with the experiments of Ref.\ \cite{Friess2014} becomes even better in this filling factor range, although no stripes at all are formed in our simulations in this filling factor range (see Fig.\ref{fig-KS-density}c). For example, in Fig.\ \ref{fig-KS-NMR} around $\nu\approx 2.5$, we can still see in total $3$ peaks, two of them clearly separated and a third one as a shoulder on the high frequency flank, just as shown for the experiments in Fig. 2b of Ref.\ \cite{Friess2014}.
%%%%%%%%%%%%%%%%%%%%%%%%%%%%%%%%%%%%%%%%%%%%%%%%%%%%%%%%%%%%%%%%%%%%%%%%%
%%%%%%%%%%%%%%%%%%%%%%%%%%%%%%%%%%%%%%%%%%%%%%%%%%%%%%%%%%%%%%%%%%%%%%%%%

% This makes the agreement almost perfect for the experiments as shown in Fig.\ \ref{fig:NMRspec-2-3} albeit not with the non-interacting modelling of Fig.\ 2c \cite{Friess2014}. Filling factor $6$--$7$ requires much more computing time and, although a most interesting question, there are currently no experiments available for comparison.
%%%%%%%%%%%%%%%%%%%%%%%%%%%%%%%%%%%%%%%%%%%%%%%%%%%%%%%%%%%%%%%%%%%%%%%%%
%%%%%%%%%%%%%%%%%%%%%%%%%%%%%%%%%%%%%%%%%%%%%%%%%%%%%%%%%%%%%%%%%%%%%%%%%

%%%%%%%%%%%%%%%%%%%%%%%%%%%%%%%%%%%%%%%%%%%%%%%%%%%%%%%%%%%%%%%%%%%%%%%%%
\section{Conclusions}
%%%%%%%%%%%%%%%%%%%%%%%%%%%%%%%%%%%%%%%%%%%%%%%%%%%%%%%%%%%%%%%%%%%%%%%%%

In conclusion, we have shown that a computational, fully-self consistent HF calculation reproduces much of the physics of the IQHE regime, across many LLs, disorder strengths, magnetic fields and temperatures. Near macroscopic system sizes of up to $\SI{1}{\mu m^2}$ can be reached. When coupled with a NNM, the approach is able to model the observed quantization of $R_{xy}$ at strong disorders as well as to reproduce the transport anisotropies for stripe and bubble phases at low disorders. This success relies in a fundamental way on the underlying basis set of LL wave functions. When these fail to be good descriptors of the physical situation, then the predictions of the HF approach become less reliable. This is clearly the case for $B \rightarrow 0$, i.e.\ when $l_c \sim L$. On the other hand, for strong magnetic fields at low disorders, one expects to see fractional quantum Hall physics to appear \cite{TsuSG82,Chakraborty1995QuantumBasics}. In principle, the LL basis should still be able to capture such behavior. However, our results up to $\nu=6$ suggest that this requires a much larger number of LLs than just those needed to accommodate all electrons. So far this appears to be beyond available computing resources.

%%%%%%%%%%%%%%%%%%%%%%%%%%%%%%%%%%%%%%%%%%%%%%%%%%%%%%%%%%%%%%%%%%%%%%%%%
% \bibliographystyle{elsarticle-num-names}
% %\bibliography{references_RAR.bib}
% \bibliography{QH_RAR.bib}

\begin{thebibliography}{40}
\expandafter\ifx\csname natexlab\endcsname\relax\def\natexlab#1{#1}\fi
\providecommand{\url}[1]{\texttt{#1}}
\providecommand{\href}[2]{#2}
\providecommand{\path}[1]{#1}
\providecommand{\DOIprefix}{doi:}
\providecommand{\ArXivprefix}{arXiv:}
\providecommand{\URLprefix}{URL: }
\providecommand{\Pubmedprefix}{pmid:}
\providecommand{\doi}[1]{\href{http://dx.doi.org/#1}{\path{#1}}}
\providecommand{\Pubmed}[1]{\href{pmid:#1}{\path{#1}}}
\providecommand{\bibinfo}[2]{#2}
\ifx\xfnm\relax \def\xfnm[#1]{\unskip,\space#1}\fi
%Type = Article
\bibitem[{Huckestein(1995)}]{Huckestein1995ScalingEffect}
\bibinfo{author}{B.~Huckestein},
\newblock \bibinfo{title}{{Scaling theory of the integer quantum Hall effect}},
\newblock \bibinfo{journal}{Reviews of Modern Physics} \bibinfo{volume}{67}
  (\bibinfo{year}{1995}) \bibinfo{pages}{357--396}. \URLprefix
  \url{https://link.aps.org/doi/10.1103/RevModPhys.67.357}.
  \DOIprefix\doi{10.1103/RevModPhys.67.357}.
%Type = Book
\bibitem[{Chakraborty and
  Pietil{\"{a}}inen(1995)}]{Chakraborty1995QuantumBasics}
\bibinfo{author}{T.~Chakraborty}, \bibinfo{author}{P.~Pietil{\"{a}}inen},
  \bibinfo{title}{{Quantum Hall Effect: The Basics}}, \bibinfo{year}{1995}.
  \URLprefix \url{http://link.springer.com/10.1007/978-3-642-79319-6_1}.
  \DOIprefix\doi{10.1007/978-3-642-79319-6{\_}1}.
%Type = Article
\bibitem[{Chalker and Coddington(1988)}]{ChaC88}
\bibinfo{author}{J.~T. Chalker}, \bibinfo{author}{P.~D. Coddington},
\newblock \bibinfo{title}{{Percolation, quantum tunnelling and the integer Hall
  effect}},
\newblock \bibinfo{journal}{Journal of Physics C: Solid State Physics}
  \bibinfo{volume}{21} (\bibinfo{year}{1988}) \bibinfo{pages}{2665--2679}.
  \URLprefix
  \url{http://0-iopscience.iop.org.pugwash.lib.warwick.ac.uk/article/10.1088/0022-3719/21/14/008/pdf
  http://stacks.iop.org/0022-3719/21/i=14/a=008?key=crossref.d766ea045e0a719ca54a9caa16f83734}.
  \DOIprefix\doi{10.1088/0022-3719/21/14/008}.
%Type = Article
\bibitem[{Puschmann et~al.(2019)Puschmann, Cain, Schreiber, and
  Vojta}]{Puschmann2019IntegerLattice}
\bibinfo{author}{M.~Puschmann}, \bibinfo{author}{P.~Cain},
  \bibinfo{author}{M.~Schreiber}, \bibinfo{author}{T.~Vojta},
\newblock \bibinfo{title}{{Integer quantum Hall transition on a tight-binding
  lattice}},
\newblock \bibinfo{journal}{Physical Review B} \bibinfo{volume}{99}
  (\bibinfo{year}{2019}) \bibinfo{pages}{121301}. \URLprefix
  \url{https://arxiv.org/pdf/1805.09958.pdf
  https://link.aps.org/doi/10.1103/PhysRevB.99.121301}.
  \DOIprefix\doi{10.1103/PhysRevB.99.121301}.
%Type = Article
\bibitem[{Chklovskii et~al.(1992)Chklovskii, Shklovskii, and
  Glazman}]{Chklovskii1992ELECTROSTATICSCHANNELS}
\bibinfo{author}{D.~B. Chklovskii}, \bibinfo{author}{B.~I. Shklovskii},
  \bibinfo{author}{L.~I. Glazman},
\newblock \bibinfo{title}{{ELECTROSTATICS OF EDGE CHANNELS}},
\newblock \bibinfo{journal}{Physical Review B} \bibinfo{volume}{46}
  (\bibinfo{year}{1992}) \bibinfo{pages}{4026--4034}. \URLprefix
  \url{http://link.aps.org/doi/10.1103/PhysRevB.46.4026}.
  \DOIprefix\doi{10.1103/PhysRevB.46.4026}.
%Type = Article
\bibitem[{Fogler et~al.(1996)Fogler, Koulakov, and Shklovskii}]{Fogler1996d}
\bibinfo{author}{M.~M. Fogler}, \bibinfo{author}{A.~A. Koulakov},
  \bibinfo{author}{B.~I. Shklovskii},
\newblock \bibinfo{title}{{Ground state of a two-dimensional electron liquid in
  a weak magnetic field}},
\newblock \bibinfo{journal}{Physical Review B} \bibinfo{volume}{54}
  (\bibinfo{year}{1996}) \bibinfo{pages}{1853--1871}. \URLprefix
  \url{https://link.aps.org/doi/10.1103/PhysRevB.54.1853}.
  \DOIprefix\doi{10.1103/PhysRevB.54.1853}.
%Type = Article
\bibitem[{Werner and Oswald(2020)}]{Werner2020SizeRegime}
\bibinfo{author}{D.~Werner}, \bibinfo{author}{J.~Oswald},
\newblock \bibinfo{title}{{Size scaling of the exchange interaction in the
  quantum Hall effect regime}},
\newblock \bibinfo{journal}{Physical Review B} \bibinfo{volume}{102}
  (\bibinfo{year}{2020}) \bibinfo{pages}{235305}. \URLprefix
  \url{https://link.aps.org/doi/10.1103/PhysRevB.102.235305}.
  \DOIprefix\doi{10.1103/PhysRevB.102.235305}.
%Type = Article
\bibitem[{Gusev et~al.(1998)Gusev, Gennser, Kleber, Maude, Portal, Lubyshev,
  Basmaji, de~Silva, Rossi, and Nastaushev}]{Gusev1998PercolationPotential}
\bibinfo{author}{G.~Gusev}, \bibinfo{author}{U.~Gennser},
  \bibinfo{author}{X.~Kleber}, \bibinfo{author}{D.~Maude},
  \bibinfo{author}{J.~Portal}, \bibinfo{author}{D.~Lubyshev},
  \bibinfo{author}{P.~Basmaji}, \bibinfo{author}{M.~de~Silva},
  \bibinfo{author}{J.~Rossi}, \bibinfo{author}{Y.~V. Nastaushev},
\newblock \bibinfo{title}{{Percolation network in a smooth artificial
  potential}},
\newblock \bibinfo{journal}{Physical Review B - Condensed Matter and Materials
  Physics} \bibinfo{volume}{58} (\bibinfo{year}{1998})
  \bibinfo{pages}{4636--4643}. \DOIprefix\doi{10.1103/PhysRevB.58.4636}.
%Type = Article
\bibitem[{Dolgopolov(2014)}]{Dolgopolov2014IntegerPhenomena}
\bibinfo{author}{V.~T. Dolgopolov},
\newblock \bibinfo{title}{{Integer quantum Hall effect and related phenomena}},
\newblock \bibinfo{journal}{Uspekhi Fizicheskih Nauk} \bibinfo{volume}{184}
  (\bibinfo{year}{2014}) \bibinfo{pages}{113--136}.
  \DOIprefix\doi{10.3367/ufnr.0184.201402a.0113}.
%Type = Article
\bibitem[{Von~Klitzing(2019)}]{vonKlitzing2019}
\bibinfo{author}{K.~Von~Klitzing},
\newblock \bibinfo{title}{{Essay: Quantum Hall Effect and the New International
  System of Units}},
\newblock \bibinfo{journal}{Physical Review Letters} \bibinfo{volume}{122}
  (\bibinfo{year}{2019}) \bibinfo{pages}{200001}. \URLprefix
  \url{https://doi.org/10.1103/PhysRevLett.122.200001}.
  \DOIprefix\doi{10.1103/PhysRevLett.122.200001}.
%Type = Phdthesis
\bibitem[{Sohrmann(2007)}]{Soh07}
\bibinfo{author}{C.~Sohrmann}, \bibinfo{title}{{Interactions in the integer
  quantum Hall effect}}, Ph.D. thesis, University of Warwick,
  \bibinfo{year}{2007}. \URLprefix
  \url{http://webcat.warwick.ac.uk/record=b2244242~S1}.
%Type = Article
\bibitem[{Aoki(1979)}]{Aok79}
\bibinfo{author}{H.~Aoki},
\newblock \bibinfo{title}{{Effect of coexistence of random potential and
  electron-electron interaction in two-dimensional systems: Wigner glass}},
\newblock \bibinfo{journal}{Journal of Physics C: Solid State Physics}
  \bibinfo{volume}{12} (\bibinfo{year}{1979}) \bibinfo{pages}{633--645}.
  \URLprefix
  \url{http://stacks.iop.org/0022-3719/12/i=4/a=010?key=crossref.0cfaf094bf5c608702848a4d9ee7629a}.
  \DOIprefix\doi{10.1088/0022-3719/12/4/010}.
%Type = Article
\bibitem[{MacDonald and Aers(1986)}]{MacA86}
\bibinfo{author}{A.~H. MacDonald}, \bibinfo{author}{G.~C. Aers},
\newblock \bibinfo{title}{{Size dependence in small-system calculations for
  fractional quantum Hall states}},
\newblock \bibinfo{journal}{Physical Review B} \bibinfo{volume}{34}
  (\bibinfo{year}{1986}) \bibinfo{pages}{2906--2909}. \URLprefix
  \url{http://link.aps.org/doi/10.1103/PhysRevB.34.2906}.
  \DOIprefix\doi{10.1103/PhysRevB.34.2906}.
%Type = Article
\bibitem[{Klitzing et~al.(1980)Klitzing, Dorda, and Pepper}]{KliDP80}
\bibinfo{author}{K.~v. Klitzing}, \bibinfo{author}{G.~Dorda},
  \bibinfo{author}{M.~Pepper},
\newblock \bibinfo{title}{{New Method for High-Accuracy Determination of the
  Fine-Structure Constant Based on Quantized Hall Resistance}},
\newblock \bibinfo{journal}{Physical Review Letters} \bibinfo{volume}{45}
  (\bibinfo{year}{1980}) \bibinfo{pages}{494--497}. \URLprefix
  \url{https://journals.aps.org/prl/pdf/10.1103/PhysRevLett.45.494
  https://link.aps.org/doi/10.1103/PhysRevLett.45.494}.
  \DOIprefix\doi{10.1103/PhysRevLett.45.494}.
%Type = Article
\bibitem[{Sohrmann and R{\"{o}}mer(2008)}]{Sohrmann2008a}
\bibinfo{author}{C.~Sohrmann}, \bibinfo{author}{R.~A. R{\"{o}}mer},
\newblock \bibinfo{title}{{Kubo conductivity in the IQHE regime within
  Hartree-Fock}},
\newblock \bibinfo{journal}{physica status solidi (c)} \bibinfo{volume}{5}
  (\bibinfo{year}{2008}) \bibinfo{pages}{842--847}. \URLprefix
  \url{http://doi.wiley.com/10.1002/pssc.200777586}.
  \DOIprefix\doi{10.1002/pssc.200777586}.
%Type = Article
\bibitem[{Oswald and Oswald(2006)}]{Oswald2006CircuitRegime}
\bibinfo{author}{J.~Oswald}, \bibinfo{author}{M.~Oswald},
\newblock \bibinfo{title}{{Circuit type simulations of magneto-transport in the
  quantum Hall effect regime}},
\newblock \bibinfo{journal}{Journal of Physics: Condensed Matter}
  \bibinfo{volume}{18} (\bibinfo{year}{2006}) \bibinfo{pages}{R101--R138}.
  \URLprefix
  \url{http://stacks.iop.org/0953-8984/18/i=7/a=R01?key=crossref.f8058110a954ef62d579c557eb5fdf89
  https://iopscience.iop.org/article/10.1088/0953-8984/18/7/R01}.
  \DOIprefix\doi{10.1088/0953-8984/18/7/R01}.
%Type = Incollection
\bibitem[{Sohrmann et~al.(2009)Sohrmann, Oswald, and R{\"{o}}mer}]{SohOR09}
\bibinfo{author}{C.~Sohrmann}, \bibinfo{author}{J.~Oswald},
  \bibinfo{author}{R.~A. R{\"{o}}mer},
\newblock \bibinfo{title}{{Quantum Percolation in the Quantum Hall Regime}},
\newblock in: \bibinfo{editor}{A.~K. Sen}, \bibinfo{editor}{K.~K. Bardhan},
  \bibinfo{editor}{B.~K. Chakrabarti} (Eds.), \bibinfo{booktitle}{Quantum and
  Semi-classical Percolation and Breakdown in Disordered Solids}, volume
  \bibinfo{volume}{762}, \bibinfo{publisher}{Springer Berlin Heidelberg},
  \bibinfo{address}{Heidelberg}, \bibinfo{year}{2009}, pp.
  \bibinfo{pages}{1--31}. \URLprefix
  \url{http://link.springer.com/10.1007/978-3-540-85428-9_6}.
  \DOIprefix\doi{10.1007/978-3-540-85428-9{\_}6}.
%Type = Article
\bibitem[{Polyakov and Shklovskii(1995)}]{PolS95}
\bibinfo{author}{D.~G. Polyakov}, \bibinfo{author}{B.~I. Shklovskii},
\newblock \bibinfo{title}{{Universal prefactor of activated conductivity in the
  quantum hall effect}},
\newblock \bibinfo{journal}{Physical Review Letters} \bibinfo{volume}{74}
  (\bibinfo{year}{1995}) \bibinfo{pages}{150}. \URLprefix
  \url{http://journals.aps.org/prl/abstract/10.1103/PhysRevLett.74.150
  papers2://publication/uuid/FC1D60DC-F2C7-4AB9-B176-C6FC66A93533}.
%Type = Incollection
\bibitem[{Oswald(2016)}]{Osw16}
\bibinfo{author}{J.~Oswald},
\newblock \bibinfo{title}{{Linking Non-equilibrium Transport with the Many
  Particle Fermi Sea in the Quantum Hall Regime}},
\newblock in: \bibinfo{editor}{P.~Bracken} (Ed.), \bibinfo{booktitle}{Recent
  Advances in Quantum Dynamics}, \bibinfo{publisher}{InTech},
  \bibinfo{address}{Rijeka}, \bibinfo{year}{2016}, pp.
  \bibinfo{pages}{131--163}. \URLprefix \url{http://dx.doi.org/10.5772/62926}.
  \DOIprefix\doi{10.5772/62926}.
%Type = Article
\bibitem[{Oswald and R{\"{o}}mer(2017{\natexlab{a}})}]{OswaldPRB2017}
\bibinfo{author}{J.~Oswald}, \bibinfo{author}{R.~A. R{\"{o}}mer},
\newblock \bibinfo{title}{{Manifestation of many-body interactions in the
  integer quantum Hall effect regime}},
\newblock \bibinfo{journal}{Physical Review B} \bibinfo{volume}{96}
  (\bibinfo{year}{2017}{\natexlab{a}}) \bibinfo{pages}{125128}. \URLprefix
  \url{http://arxiv.org/abs/1707.01364
  https://link.aps.org/doi/10.1103/PhysRevB.96.125128}.
  \DOIprefix\doi{10.1103/PhysRevB.96.125128}.
%Type = Article
\bibitem[{Oswald and R{\"{o}}mer(2017{\natexlab{b}})}]{OswR17}
\bibinfo{author}{J.~Oswald}, \bibinfo{author}{R.~A. R{\"{o}}mer},
\newblock \bibinfo{title}{{Exchange-mediated dynamic screening in the integer
  quantum Hall effect regime}},
\newblock \bibinfo{journal}{EPL (Europhysics Letters)} \bibinfo{volume}{117}
  (\bibinfo{year}{2017}{\natexlab{b}}) \bibinfo{pages}{57009}. \URLprefix
  \url{http://stacks.iop.org/0295-5075/117/i=5/a=57009?key=crossref.1e0ca141aec04144a11fb11f12039fb1
  https://iopscience.iop.org/article/10.1209/0295-5075/117/57009}.
  \DOIprefix\doi{10.1209/0295-5075/117/57009}.
%Type = Article
\bibitem[{Hashimoto et~al.(2008)Hashimoto, Sohrmann, Wiebe, Inaoka, Meier,
  Hirayama, R{\"{o}}mer, Wiesendanger, and Morgenstern}]{Hashimoto2008}
\bibinfo{author}{K.~Hashimoto}, \bibinfo{author}{C.~Sohrmann},
  \bibinfo{author}{J.~Wiebe}, \bibinfo{author}{T.~Inaoka},
  \bibinfo{author}{F.~Meier}, \bibinfo{author}{Y.~Hirayama},
  \bibinfo{author}{R.~A. R{\"{o}}mer}, \bibinfo{author}{R.~Wiesendanger},
  \bibinfo{author}{M.~Morgenstern},
\newblock \bibinfo{title}{{Quantum Hall Transition in Real Space: From
  Localized to Extended States}},
\newblock \bibinfo{journal}{Physical Review Letters} \bibinfo{volume}{101}
  (\bibinfo{year}{2008}) \bibinfo{pages}{256802}. \URLprefix
  \url{http://dx.doi.org/10.1103/PhysRevLett.101.256802
  http://link.aps.org/doi/10.1103/PhysRevLett.101.256802
  https://link.aps.org/doi/10.1103/PhysRevLett.101.256802}.
  \DOIprefix\doi{10.1103/PhysRevLett.101.256802}.
%Type = Article
\bibitem[{Hashimoto et~al.(2012)Hashimoto, Champel, Florens, Sohrmann, Wiebe,
  Hirayama, R{\"{o}}mer, Wiesendanger, Morgenstern, Roemer, Wiesendanger,
  Morgenstern, R{\"{o}}mer, Wiesendanger, and Morgenstern}]{HasCFS12}
\bibinfo{author}{K.~Hashimoto}, \bibinfo{author}{T.~Champel},
  \bibinfo{author}{S.~Florens}, \bibinfo{author}{C.~Sohrmann},
  \bibinfo{author}{J.~Wiebe}, \bibinfo{author}{Y.~Hirayama},
  \bibinfo{author}{R.~A. R{\"{o}}mer}, \bibinfo{author}{R.~Wiesendanger},
  \bibinfo{author}{M.~Morgenstern}, \bibinfo{author}{R.~A. Roemer},
  \bibinfo{author}{R.~Wiesendanger}, \bibinfo{author}{M.~Morgenstern},
  \bibinfo{author}{R.~A. R{\"{o}}mer}, \bibinfo{author}{R.~Wiesendanger},
  \bibinfo{author}{M.~Morgenstern},
\newblock \bibinfo{title}{{Robust Nodal Structure of Landau Level Wave
  Functions Revealed by Fourier Transform Scanning Tunneling Spectroscopy}},
\newblock \bibinfo{journal}{Physical Review Letters} \bibinfo{volume}{109}
  (\bibinfo{year}{2012}) \bibinfo{pages}{116805}. \URLprefix
  \url{http://link.aps.org/doi/10.1103/PhysRevLett.109.116805
  http://arxiv.org/abs/1201.2235
  http://dx.doi.org/10.1103/PhysRevLett.109.116805
  https://link.aps.org/doi/10.1103/PhysRevLett.109.116805}.
  \DOIprefix\doi{10.1103/PhysRevLett.109.116805}.
%Type = Article
\bibitem[{Friess et~al.(2014)Friess, Umansky, Tiemann, von Klitzing, and
  Smet}]{Friess2014}
\bibinfo{author}{B.~Friess}, \bibinfo{author}{V.~Umansky},
  \bibinfo{author}{L.~Tiemann}, \bibinfo{author}{K.~von Klitzing},
  \bibinfo{author}{J.~H. Smet},
\newblock \bibinfo{title}{{Probing the Microscopic Structure of the Stripe
  Phase at Filling Factor 5/2}},
\newblock \bibinfo{journal}{Physical Review Letters} \bibinfo{volume}{113}
  (\bibinfo{year}{2014}) \bibinfo{pages}{076803}. \URLprefix
  \url{https://link.aps.org/doi/10.1103/PhysRevLett.113.076803}.
  \DOIprefix\doi{10.1103/PhysRevLett.113.076803}.
%Type = Incollection
\bibitem[{Fogler(2002)}]{Fogler2002}
\bibinfo{author}{M.~M. Fogler},
\newblock \bibinfo{title}{{Stripe and Bubble Phases in Quantum Hall Systems}},
\newblock in: \bibinfo{editor}{C.~Berthier}, \bibinfo{editor}{L.~P. Levy},
  \bibinfo{editor}{G.~Martinez} (Eds.), \bibinfo{booktitle}{High Magnetic
  Fields}, volume \bibinfo{volume}{595}, \bibinfo{publisher}{Springer Berlin /
  Heidelberg}, \bibinfo{year}{2002}, pp. \bibinfo{pages}{98--138}. \URLprefix
  \url{http://arxiv.org/abs/cond-mat/0111001
  http://link.springer.com/10.1007/3-540-45649-X_4}.
  \DOIprefix\doi{10.1007/3-540-45649-X{\_}4}.
%Type = Article
\bibitem[{Lilly et~al.(1999)Lilly, Cooper, Eisenstein, Pfeiffer, and
  West}]{Lilly1999a}
\bibinfo{author}{M.~P. Lilly}, \bibinfo{author}{K.~B. Cooper},
  \bibinfo{author}{J.~P. Eisenstein}, \bibinfo{author}{L.~N. Pfeiffer},
  \bibinfo{author}{K.~W. West},
\newblock \bibinfo{title}{{Evidence for an Anisotropic State of Two-Dimensional
  Electrons in High Landau Levels}},
\newblock \bibinfo{journal}{Phys. Rev. Lett.} \bibinfo{volume}{82}
  (\bibinfo{year}{1999}) \bibinfo{pages}{394}. \URLprefix
  \url{https://link.aps.org/doi/10.1103/PhysRevLett.82.394}.
  \DOIprefix\doi{10.1103/PhysRevLett.82.394}.
%Type = Article
\bibitem[{Du et~al.(1999)Du, Tsui, Stormer, Pfeiffer, Baldwin, and
  West}]{Du1999}
\bibinfo{author}{R.~Du}, \bibinfo{author}{D.~Tsui},
  \bibinfo{author}{H.~Stormer}, \bibinfo{author}{L.~Pfeiffer},
  \bibinfo{author}{K.~Baldwin}, \bibinfo{author}{K.~West},
\newblock \bibinfo{title}{{Strongly anisotropic transport in higher
  two-dimensional Landau levels}},
\newblock \bibinfo{journal}{Solid State Communications} \bibinfo{volume}{109}
  (\bibinfo{year}{1999}) \bibinfo{pages}{389--394}. \URLprefix
  \url{http://10.0.3.248/s0038-1098(98)00578-x
  https://dx.doi.org/10.1016/S0038-1098(98)00578-X
  https://kp-pdf.s3.amazonaws.com/ab4f0d57-8dc4-42dd-b100-b3c87139b067.pdf?AWSAccessKeyId=AKIAUROH2NUQSIQZIEG4&Signature=tNXuolm%2FVScz0WWOQEwI2jEDH9Y%3D&Expires=157624}.
  \DOIprefix\doi{10.1016/S0038-1098(98)00578-X}.
%Type = Article
\bibitem[{Du et~al.(2000)Du, Pan, Stormer, Tsui, Pfeiffer, Baldwin, and
  West}]{Du2000c}
\bibinfo{author}{R.~Du}, \bibinfo{author}{W.~Pan},
  \bibinfo{author}{H.~Stormer}, \bibinfo{author}{D.~Tsui},
  \bibinfo{author}{L.~Pfeiffer}, \bibinfo{author}{K.~Baldwin},
  \bibinfo{author}{K.~West},
\newblock \bibinfo{title}{{Strongly anisotropic electronic transport in higher
  Landau levels}},
\newblock \bibinfo{journal}{Physica E: Low-dimensional Systems and
  Nanostructures} \bibinfo{volume}{6} (\bibinfo{year}{2000})
  \bibinfo{pages}{36--39}. \URLprefix
  \url{https://linkinghub.elsevier.com/retrieve/pii/S1386947799000934}.
  \DOIprefix\doi{10.1016/S1386-9477(99)00093-4}.
%Type = Article
\bibitem[{Ettouhami et~al.(2006)Ettouhami, Doiron, Klironomos,
  C{\^{o}}t{\'{e}}, and Dorsey}]{Ettouhami2006}
\bibinfo{author}{A.~M. Ettouhami}, \bibinfo{author}{C.~B. Doiron},
  \bibinfo{author}{F.~D. Klironomos}, \bibinfo{author}{R.~C{\^{o}}t{\'{e}}},
  \bibinfo{author}{A.~T. Dorsey},
\newblock \bibinfo{title}{{Anisotropic States of Two-Dimensional Electrons in
  High Magnetic Fields}},
\newblock \bibinfo{journal}{Physical Review Letters} \bibinfo{volume}{96}
  (\bibinfo{year}{2006}) \bibinfo{pages}{196802}. \URLprefix
  \url{https://link.aps.org/doi/10.1103/PhysRevLett.96.196802}.
  \DOIprefix\doi{10.1103/PhysRevLett.96.196802}.
%Type = Article
\bibitem[{Ettouhami et~al.(2007)Ettouhami, Doiron, and
  C{\^{o}}t{\'{e}}}]{Ettouhami2007}
\bibinfo{author}{A.~M. Ettouhami}, \bibinfo{author}{C.~B. Doiron},
  \bibinfo{author}{R.~C{\^{o}}t{\'{e}}},
\newblock \bibinfo{title}{{Fluctuations and topological transitions of quantum
  Hall stripes: Nematics as anisotropic hexatics}},
\newblock \bibinfo{journal}{Physical Review B} \bibinfo{volume}{76}
  (\bibinfo{year}{2007}) \bibinfo{pages}{161306}. \URLprefix
  \url{https://link.aps.org/doi/10.1103/PhysRevB.76.161306}.
  \DOIprefix\doi{10.1103/PhysRevB.76.161306}.
%Type = Article
\bibitem[{C{\^{o}}t{\'{e}} et~al.(2002)C{\^{o}}t{\'{e}}, Fertig, Bourassa, and
  Bouchiha}]{Cote2002}
\bibinfo{author}{R.~C{\^{o}}t{\'{e}}}, \bibinfo{author}{H.~A. Fertig},
  \bibinfo{author}{J.~Bourassa}, \bibinfo{author}{D.~Bouchiha},
\newblock \bibinfo{title}{{Commensurate-incommensurate transitions of quantum
  Hall stripe states in double quantum well systems}},
\newblock \bibinfo{journal}{Physical Review B} \bibinfo{volume}{66}
  (\bibinfo{year}{2002}) \bibinfo{pages}{205315}. \URLprefix
  \url{https://link.aps.org/doi/10.1103/PhysRevB.66.205315}.
  \DOIprefix\doi{10.1103/PhysRevB.66.205315}.
%Type = Article
\bibitem[{C{\^{o}}t{\'{e}} and Simoneau(2016)}]{Cote2016}
\bibinfo{author}{R.~C{\^{o}}t{\'{e}}}, \bibinfo{author}{A.~M. Simoneau},
\newblock \bibinfo{title}{{Resistively detected NMR spectra of the crystal
  states of the two-dimensional electron gas in a quantizing magnetic field}},
\newblock \bibinfo{journal}{Physical Review B} \bibinfo{volume}{93}
  (\bibinfo{year}{2016}) \bibinfo{pages}{075305}. \URLprefix
  \url{https://link.aps.org/doi/10.1103/PhysRevB.93.075305}.
  \DOIprefix\doi{10.1103/PhysRevB.93.075305}.
%Type = Article
\bibitem[{Koulakov et~al.(1996)Koulakov, Fogler, and Shklovskii}]{Koulakov1995}
\bibinfo{author}{A.~A. Koulakov}, \bibinfo{author}{M.~M. Fogler},
  \bibinfo{author}{B.~I. Shklovskii},
\newblock \bibinfo{title}{{Charge Density Wave in Two-Dimensional Electron
  Liquid in Weak Magnetic Field}},
\newblock \bibinfo{journal}{Physical Review Letters} \bibinfo{volume}{76}
  (\bibinfo{year}{1996}) \bibinfo{pages}{499--502}. \URLprefix
  \url{http://arxiv.org/abs/cond-mat/9508017
  https://link.aps.org/doi/10.1103/PhysRevLett.76.499}.
  \DOIprefix\doi{10.1103/PhysRevLett.76.499}.
%Type = Article
\bibitem[{Fradkin and Kivelson(1999)}]{Fradkin1999}
\bibinfo{author}{E.~Fradkin}, \bibinfo{author}{S.~A. Kivelson},
\newblock \bibinfo{title}{{Liquid-crystal phases of quantum Hall systems}},
\newblock \bibinfo{journal}{Physical Review B} \bibinfo{volume}{59}
  (\bibinfo{year}{1999}) \bibinfo{pages}{8065--8072}. \URLprefix
  \url{https://link.aps.org/doi/10.1103/PhysRevB.59.8065}.
  \DOIprefix\doi{10.1103/PhysRevB.59.8065}.
%Type = Article
\bibitem[{Spivak and Kivelson(2006)}]{Spivak2006}
\bibinfo{author}{B.~Spivak}, \bibinfo{author}{S.~A. Kivelson},
\newblock \bibinfo{title}{{Transport in two dimensional electronic
  micro-emulsions}},
\newblock \bibinfo{journal}{Annals of Physics} \bibinfo{volume}{321}
  (\bibinfo{year}{2006}) \bibinfo{pages}{2071--2115}.
  \DOIprefix\doi{10.1016/j.aop.2005.12.002}.
%Type = Article
\bibitem[{Oswald and R{\"{o}}mer(2020)}]{Oswald2020}
\bibinfo{author}{J.~Oswald}, \bibinfo{author}{R.~A. R{\"{o}}mer},
\newblock \bibinfo{title}{{Microscopic details of stripes and bubbles in the
  quantum Hall regime}},
\newblock \bibinfo{journal}{Physical Review B} \bibinfo{volume}{102}
  (\bibinfo{year}{2020}) \bibinfo{pages}{121305}. \URLprefix
  \url{http://arxiv.org/abs/2001.07542
  https://link.aps.org/doi/10.1103/PhysRevB.102.121305}.
  \DOIprefix\doi{10.1103/PhysRevB.102.121305}.
%Type = Article
\bibitem[{Goerbig et~al.(2004)Goerbig, Lederer, and Smith}]{Goerbig2004a}
\bibinfo{author}{M.~O. Goerbig}, \bibinfo{author}{P.~Lederer},
  \bibinfo{author}{C.~M. Smith},
\newblock \bibinfo{title}{{Competition between quantum-liquid and
  electron-solid phases in intermediate Landau levels}},
\newblock \bibinfo{journal}{Physical Review B} \bibinfo{volume}{69}
  (\bibinfo{year}{2004}) \bibinfo{pages}{115327}. \URLprefix
  \url{http://link.aps.org/doi/10.1103/PhysRevB.69.115327
  https://link.aps.org/doi/10.1103/PhysRevB.69.115327}.
  \DOIprefix\doi{10.1103/PhysRevB.69.115327}.
%Type = Article
\bibitem[{Kukushkin et~al.(2011)Kukushkin, Umansky, von Klitzing, and
  Smet}]{Kukushkin2011a}
\bibinfo{author}{I.~V. Kukushkin}, \bibinfo{author}{V.~Umansky},
  \bibinfo{author}{K.~von Klitzing}, \bibinfo{author}{J.~H. Smet},
\newblock \bibinfo{title}{{Collective Modes and the Periodicity of Quantum Hall
  Stripes}},
\newblock \bibinfo{journal}{Physical Review Letters} \bibinfo{volume}{106}
  (\bibinfo{year}{2011}) \bibinfo{pages}{206804}. \URLprefix
  \url{https://link.aps.org/doi/10.1103/PhysRevLett.106.206804}.
  \DOIprefix\doi{10.1103/PhysRevLett.106.206804}.
%Type = Article
\bibitem[{Tiemann et~al.(2012)Tiemann, Gamez, Kumada, and Muraki}]{Tiemann2012}
\bibinfo{author}{L.~Tiemann}, \bibinfo{author}{G.~Gamez},
  \bibinfo{author}{N.~Kumada}, \bibinfo{author}{K.~Muraki},
\newblock \bibinfo{title}{{Unraveling the spin polarization of the {$\nu$} =
  5/2 fractional quantum hall state}},
\newblock \bibinfo{journal}{Science} \bibinfo{volume}{335}
  (\bibinfo{year}{2012}) \bibinfo{pages}{828--831}.
  \DOIprefix\doi{10.1126/science.1216697}.
%Type = Article
\bibitem[{Tsui et~al.(1982)Tsui, Stormer, and Gossard}]{TsuSG82}
\bibinfo{author}{D.~C. Tsui}, \bibinfo{author}{H.~L. Stormer},
  \bibinfo{author}{A.~C. Gossard},
\newblock \bibinfo{title}{{Two-Dimensional Magnetotransport in the Extreme
  Quantum Limit}},
\newblock \bibinfo{journal}{Physical Review Letters} \bibinfo{volume}{48}
  (\bibinfo{year}{1982}) \bibinfo{pages}{1559--1562}. \URLprefix
  \url{http://dx.doi.org/10.1103/PhysRevLett.48.1559
  papers2://publication/doi/10.1103/PhysRevLett.48.1559
  https://link.aps.org/doi/10.1103/PhysRevLett.48.1559}.
  \DOIprefix\doi{10.1103/PhysRevLett.48.1559}.

\end{thebibliography}

\end{document}